\documentclass[twocolumn,footinbib,superscriptaddress,floatfix,noeprint,longbibliography]{revtex4-1}
\usepackage{graphicx,subfigure,epsfig}
\usepackage{dcolumn}
\usepackage{amssymb}
\usepackage{times}
\usepackage{amsmath}
\usepackage{amsfonts}
\usepackage{mathrsfs}
\usepackage{setspace}
\usepackage{latexsym}
\usepackage{bbm}
\usepackage{float}
\usepackage{flafter}
\usepackage{bm}
\usepackage{epstopdf}
\usepackage{color}
\usepackage{multirow}
\usepackage{physics} 
\usepackage[export]{adjustbox}

\usepackage{braket}

\usepackage{soul} 

\def \be {\begin{eqnarray}}
\def \ee {\end{eqnarray}}

\newcommand{\bk}{{\bf k}}

\newcommand{\bq}{{\bf q}}
\newcommand{\bx}{{\bf x}}

\newcommand{\bX}{{\bf X}}

\newcommand{\ba}{{\bf a}}

\newcommand{\bb}{{\bf b}}

\newcommand{\bd}{{\bf d}}

\newcommand{\br}{{\bf r}}
\newcommand{\bR}{{\bf R}}

\newcommand{\bL}{{\bf L}}

\newcommand{\bM}{{\bf M}}

\usepackage{xcolor}

\newcommand{\bea}{\begin{equation} \begin{aligned}}
\newcommand{\eea}{\end{aligned} \end{equation} }

\newcommand{\bpm}{\begin{pmatrix}}
\newcommand{\epm}{\end{pmatrix}}

\newcommand{\mbf}[1]{\mathbf{#1}}

\renewcommand{\Tr}{\text{Tr}}

\newcommand{\bsl}[1]{\boldsymbol{#1}} 

\newcommand{\cc}[1]{\langle{#1}\rangle_c}

\usepackage{xr}
\usepackage[pdftex,breaklinks,bookmarks=false,colorlinks,linkcolor=blue,citecolor=blue,urlcolor=blue,hypertexnames=false]{hyperref}

\usepackage[normalem]{ulem}

\usepackage{cleveref}
\crefname{appendix}{App.}{Apps.}
\crefname{equation}{Eq.}{Eqs.}
\crefname{figure}{Fig.}{Figs.}
\crefname{table}{Tab.}{Tabs.}
\crefname{section}{Sec.}{Secs.}
\creflabelformat{appendix}{#2#1#3}

\newcommand{\eqa}[1]{\begin{align}\begin{split} #1 \end{split}\end{align}}

\newcommand{\eq}[1]{\begin{equation} #1 \end{equation}}

\newcommand{\dsZ}{\mathbb{Z}}

\begin{document}



\title{Corner Charge Fluctuation as an Observable for Quantum Geometry \\ and Entanglement
in Two-dimensional Insulators}

\author{Pok Man Tam}\email{pt4447@princeton.edu}
\affiliation{Princeton Center for Theoretical Science, Princeton University, Princeton, NJ 08544, USA}

\author{Jonah Herzog-Arbeitman}
\affiliation{Department of Physics, Princeton University, Princeton, NJ 08544, USA}

\author{Jiabin Yu}
\affiliation{Department of Physics, Princeton University, Princeton, NJ 08544, USA}
\affiliation{Department of Physics, University of Florida, Gainesville, FL, USA}

\begin{abstract}
Measuring bipartite fluctuations of a conserved charge, such as the particle number, is a powerful approach to understanding quantum systems. When the measured region has sharp corners, the bipartite fluctuation receives an additional contribution known to exhibit a universal angle-dependence in 2D isotropic and uniform systems. Here we establish that, for generic \textit{lattice} systems of \textit{interacting} particles, the corner charge fluctuation is directly related to quantum geometry. We first provide a practical scheme to isolate the corner contribution on lattices, and analytically prove that its angle-dependence in the \textit{small-angle limit} measures exclusively the integrated quantum metric. A model of a compact obstructed atomic insulator is introduced to illustrate this effect analytically, while numerical verification for various Chern insulator models further demonstrate the experimental relevance of the corner charge fluctuation in a finite-size quantum simulator as a probe of quantum geometry. Last but not least, for free fermions, we unveil an intimate connection between quantum geometry and quantum information through the lens of corner entanglement entropies. 

\end{abstract}
\maketitle

\noindent {\color{blue}\emph{Introduction.}} Quantum geometry has emerged as a new theme in the study of quantum matter by characterizing the manifold of ground states through the quantum geometric tensor (QGT) \cite{Torma2023, resta2011insulating, provost1980riemannian}. The imaginary part of the QGT is the well-known Berry curvature related to the phase difference between quantum states \cite{berry1984quantal}, which upon integration over the parameter space gives the Chern number characterizing the system's topology \cite{TKNN1982}. The real part gives the Fubini-Study \textit{quantum metric} that provides a measure of distance between two quantum states represented by projectors $P(\bk_1)$ and $P(\bk_2)$, respectively, via $D^2_{12} = 1-\tr[P(\bk_1)P(\bk_2)]$. Here ``$\tr$" is the trace over the Hilbert space, and $\bk$ is a set of parameters labeling a specific state on the manifold. For an infinitesimal parameter change, the distance can be expanded as $D^2_{12} = g_{ij}(\bk) dk^i dk^j$, defining the quantum metric
\begin{equation}\label{eq: metric_definition}
    g_{ij}(\bk) = \frac{1}{2}\tr[\partial_i P(\bk) \partial_j P(\bk)].
\end{equation}
This general discussion can be applied to band insulators where the manifold of interest corresponds to the occupied Bloch bands and the parameter $\bk$ is the momentum in the Brillouin zone (BZ). For generic interacting systems, one can consider a manifold of many-body ground states obeying twisted periodic boundary conditions, with the parameter $\bk$ labeling the twist phase \cite{SWM2000}. The metric determines the localization properties of many-body systems and provides a measure differentiating between metals and insulators  \cite{Kivelson1982,Marzari1997,SWM2000,resta1999}. 

The integrated quantum metric defined as
\begin{equation}
    \mathcal{G}_{ij} \equiv \int [d\bk]\; g_{ij}(\bk)
\end{equation}
is of physical significance. Here the measure is $[d\bk] \equiv d^D\bk/(2\pi)^D$ and integration is over the entire parameter space. For this work we focus on $D=2$. In band insulators, the metric trace $ \mathcal{G} \equiv \sum_{i=1}^D \mathcal{G}_{ii}$ is known to determine the gauge-invariant part of the Wannier spread functional \cite{Marzari1997}, and is lower bounded by many kinds of band topology \cite{Roy2014, peotta2015superfluidity, Xie2020, ozawa2021, Yu2022,Jonah2022b}. Particularly, $2\pi \mathcal{G} \geq \abs{C}$, where $C$ is the Chern number \cite{Roy2014}, and this bound has been recently generalized to the many-body context \cite{onishi2024topologicalboundstructurefactor}.
\begin{figure}[t!]
    \centering
    \includegraphics[width=0.75\columnwidth]{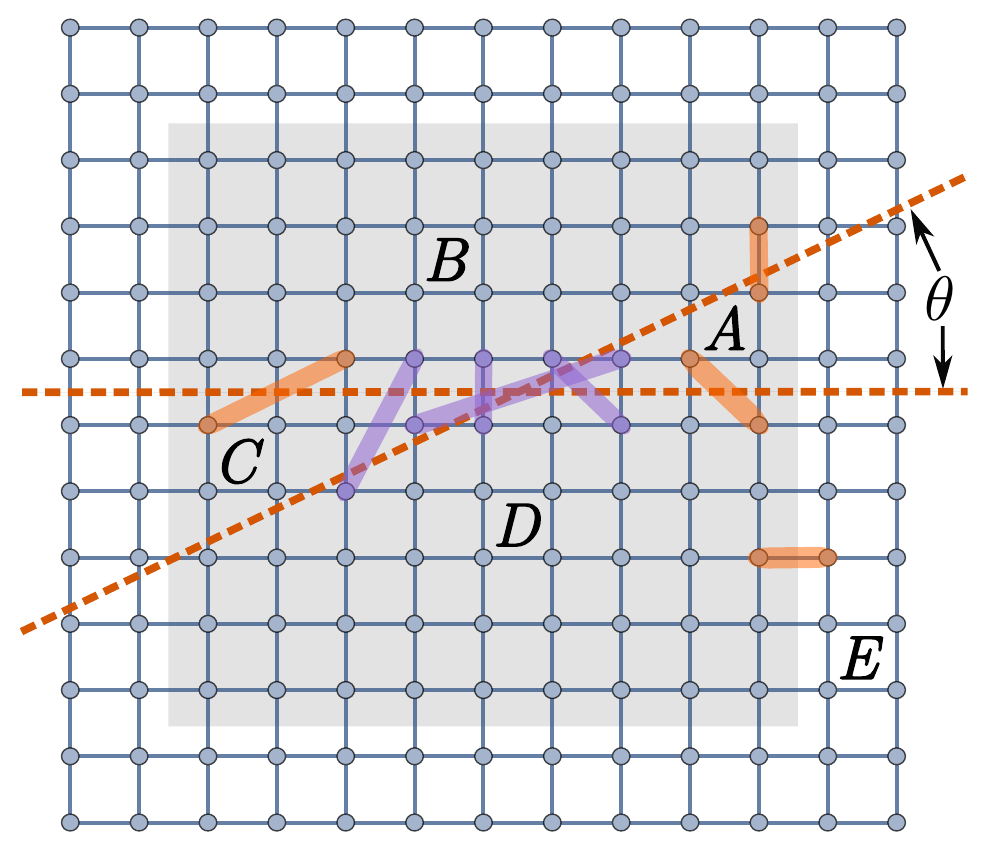}
    \caption{Partition of a square lattice defining the bipartite charge fluctuations and entanglement entropies, with partition boundaries (dashed) intersecting at the center of a plaquette at an angle $\theta$. Vertices represent unit cell positions. Subregions $A$, $B$, $C$ and $D$, which are relevant to the calculation in Eq. \eqref{eq: def_corner_fluc_lattice}, are chosen in the shaded bulk to eliminate edge effects. Purple (orange) bonds represent the corner (boundary) contribution. Refer also to the partition scheme described in the text.}
    \label{fig:setup}
\end{figure}
While ideas from quantum geometry have proved useful in understanding fractional Chern insulators \cite{Parameswaran2012, Regnault2013, Roy2014, jackson2015geometric,Claassen2015,Wang2021, Ledwith2023,LIU2024515}, flat-band superconductivity \cite{peotta2015superfluidity,torma2022superconductivity,Jonah2022,2022arXiv220900007H,PhysRevB.102.201112,PhysRevLett.130.226001,KTLaw2024}, nonlinear Hall effect \cite{gao2023quantum, Kaplan2024, fang2023quantumgeometryinducednonlinear}, excitons \cite{PhysRevLett.132.236001}, ferromagnets \cite{2024arXiv240207171K},  as well as electron-phonon coupling~\cite{Yu2024EPC}, direct experimental measurement of the quantum metric remains challenging. While many phenomena contains quantum geometric contributions, only a few types of observables are known to probe \textit{exclusively} the quantum metric in condensed matter systems \cite{SWM2000, Neupert2013, Ozawa2018, gianfrate2020measurement, ahn2022riemannian,deSousa2023,onishi2024fundamental, komissarov2023quantumgeometricorigincapacitance,kruchkov2023noise, kruchkov2023spectral, onishi2024quantum,2024arXiv240307052V}. 

Furthermore, while one expects connections between quantum geometry and quantum entanglement due to their shared relation to wavefunction localization \cite{Ryu2006, nisarga2024}, a precise and quantitative connection is yet to be established. Here we aim to address these issues by studying the \textit{corner} contributions to the bipartite fluctuation of particle number (henceforth referred to as ``charge") and the closely related entanglement entropies \cite{klich2009, calabrese2012exact, song2012, Rachel2012}. Charge fluctuation has demonstrated its use in extracting universal aspects of quantum critical systems \cite{Rachel2012, song2012, Herviou2019, wu2021universal, Meng2021, estienne2022cornering, wu2024bipartite}, and for probing the topology of metals \cite{tamkane2022, tamkane2024}. We now add the quantum geometry of two-dimensional (2D) insulators to that list. Our key result is a direct relation between the integrated quantum metric and the corner charge fluctuation (see Eqs. (\ref{eq:key_result}, \ref{eq: key result_band insulator})), which is applicable to generic interacting systems, while the relation to corner entanglement entropies is restricted to free fermions. 

\noindent {\color{blue}\emph{Corner charge fluctuation.}} 
We first introduce the central quantity of interest.
For a region $A$ whose shape contains a corner of angle $\theta$, as depicted in Fig. \ref{fig:setup}, its bipartite charge fluctuation behaves in the continuum as
\begin{equation}\label{eq: def_corner_fluc_continuum}
    \langle Q^2_A \rangle_c \equiv \langle Q^2_A \rangle -\langle Q_A \rangle^2 = \alpha \abs{\partial A} - b(\theta)+...,
\end{equation}
where $Q_A$ is the particle number operator for $A$, and $\langle\cdot\rangle$ refers to the ground-state expectation (with the subscript $c$ indicating the connected part). The dominant term is the boundary-law contribution scaling with the length of the boundary $\partial A$, the subdominant constant term $b(\theta)$ is the corner contribution arising from the singular shape of $A$,  and the ellipses represent terms vanishing in the thermodynamic limit. Since the boundary-law coefficient $\alpha$ is dimensionful, it is non-universal and not expected to capture the dimensionless integrated quantum metric in 2D. A natural place to hunt for the quantum geometric effect is thus the corner term. For a large class of systems with correlations decaying fast enough in space, the corner fluctuation is known to exhibit a universal angle-dependence \cite{estienne2022cornering}:
\begin{equation}\label{eq:Krempa_angle_dependence}
    b(\theta)=\gamma \beta(\theta),\quad \beta(\theta)=\frac{1+(\pi-\theta)\cot\theta}{4\pi^2},
\end{equation}
with the corner coefficient $\gamma= \pi\nabla^2_\bq S_\bq\vert_{\bq=0}$ related to the quadratic coefficient of the static structure factor at wavevector $\bq$. While initially obtained in conformally invariant critical systems \cite{Herviou2019, wu2021universal, Meng2021}, the universal angle-dependence actually holds for both gapless and gapped interacting systems in the \textit{uniform} and \textit{isotropic} limit \cite{estienne2022cornering}, but is expected to fail for lattice systems. 

On a lattice, $b(\theta)$ defined via Eq. \eqref{eq: def_corner_fluc_continuum} is ambiguous due to the intrinsically rough partitioning boundary. Instead, we define the corner contribution by the following combination of bipartite charge fluctuations based on Fig. \ref{fig:setup}:
\begin{equation}\label{eq: def_corner_fluc_lattice}
\begin{split}
\mathcal{C}^{(Q)}(\theta) & \equiv \frac{1}{2}\Big[-\langle Q^2_A \rangle_c-\langle Q^2_B \rangle_c-\langle Q^2_C \rangle_c-\langle Q^2_D \rangle_c  \\
&+\langle Q^2_{AB} \rangle_c +\langle Q^2_{CD} \rangle_c+\langle Q^2_{BC}\rangle_c+\langle Q^2_{AD} \rangle_c \\
&-\langle Q^2_{ABCD} \rangle_c \Big].
\end{split}
\end{equation}
Any boundary contribution arising from the correlation of two sites in neighboring regions (i.e., the orange pairs in Fig. \ref{fig:setup}) is canceled exactly in the above combination, which leaves us with the correlated pairs that connect regions sharing only the ``corner" (i.e., the purple pairs). More precisely, 
\begin{equation}\label{eq: fluctuation_simplified}
    -\cc{Q_A^2} = \braket{Q_A Q_{\bar{A}}}_c = \sum_{\bR \in A, \bR' \in \bar{A}} \cc{\rho(\bR)\rho(\bR')},
\end{equation}
where $\rho(\bR) = \sum_\sigma c^\dagger_{\bR,\sigma} c_{\bR,\sigma}$ is the density operator for a unit-cell positioned at $\bR$ and $c_{\bR,\sigma}$ is the annihilation operator for the $\sigma$th orbital. In the first equality we used the conservation of total charge and in the second equality we recognize that the bipartite charge fluctuation is simply summing the density-density \textit{connected} correlator between $A$ and its complement $\bar{A}=BCDE$. Importantly, for a generic multi-orbital systems with a given choice of unit cell, we have stipulated the following \underline{partition scheme}: all orbitals of a unit cell at $\bR$ (labeled by a vertex in Fig. \ref{fig:setup}) are assigned to the same region to which $\bR$ belongs. 

Noticing that terms associated with $\{\bR \in A, \bR'\in B\}$ contribute equally to $\braket{Q^2_A}_c$,$\braket{Q^2_B}_c$, $\braket{Q^2_{BC}}_c$ and $\braket{Q^2_{AD}}_c$, the prescribed combination in Eq. \eqref{eq: def_corner_fluc_lattice} eliminates such contributions. Thus we are left with the correlation between corner-sharing regions, which simplifies Eq. \eqref{eq: def_corner_fluc_lattice} to
\begin{equation}\label{eq:simplified_CQ}
    \mathcal{C}^{(Q)}(\theta) =  -\cc{Q_A Q_C} - \cc{Q_B Q_D}.
\end{equation}
We further define the lattice corner coefficient 
\begin{equation}\label{eq:def_corner_coefficient}
    \gamma^{(Q)}(\theta) = \frac{\mathcal{C}^{(Q)}(\theta)}{\beta(\theta)+\beta(\pi-\theta)},
\end{equation}
with $\beta(\theta)$ introduced in Eq. \eqref{eq:Krempa_angle_dependence}. For isotropic and uniform systems, $\gamma^{(Q)}(\theta)$ should be $\theta$-independent \cite{estienne2022cornering}, but as we will see with explicit, analytical examples, this is no longer true on lattices. The corner charge fluctuation is also generally dependent on the orientation of cuts. For the partition in Fig. \ref{fig:setup} with one cut lying along the $x$-direction, the corresponding quantities are denoted as $\mathcal{C}^{(Q)}_x$ and $\gamma^{(Q)}_x$, respectively.

\noindent {\color{blue}\emph{Structure factor and quantum geometry.}}
Our key result is to relate the corner charge fluctuation to the structure factor in lattice systems, which is in turn related to the quantum geometry for both band insulators and interacting many-body systems (this latter relation is highlighted recently in Ref. \cite{onishi2024quantum}, under the concept of ``quantum weight"). 

As we will see, the quantum metric to be extracted from the corner charge fluctuation (under the partition scheme specified under Eq. \eqref{eq: fluctuation_simplified}) is related to the momentum space density operator 
\begin{equation}\label{eq: origin_embedding FT}
    \widetilde{\rho}_\bq  = \sum_{\bR} e^{-i\bq\cdot \bR}\rho(\bR)
\end{equation}
with $\bR$ a Bravais lattice vector. Various conventions, called orbital embeddings, can be adopted for the Fourier transform and are known to affect the Berry curvature and quantum metric on a lattice ~\cite{haldane2014attachment, Lim2015, Simon2020,Jonah2022}. We refer to the specific convention in Eq. \eqref{eq: origin_embedding FT} as the \emph{origin orbital embedding} and we highlight all quantities related to the origin orbital embedding with a tilde symbol \footnote{Another commonly used orbital embedding is to Fourier transform as $\rho_\bq = \sum_{\bR,\sigma} e^{-i\bq\cdot (\bR+\br_\sigma)} c^\dagger_{\bR,\sigma} c_{\bR,\sigma}$, where $\bR+\br_\sigma$ is the physical position of the orbital. This is not used in the main text, so we shall discuss it only in the supplementary.}.

For generic groundstates with or without interactions, the static structure factor $\widetilde{S}_\bq = \mathcal{A}^{-1} \cc{\widetilde{\rho}_\bq \widetilde{\rho}_{-\bq}}$ (with $\mathcal{A}$ the system area) is always related to the \textit{many-body} quantum geometry as
\begin{equation}\label{eq:relating S to G}
    \frac{1}{2}\partial_i\partial_j \widetilde{S}_\bq \vert_{\bq=0} = \int [d\bk] \frac{1}{2} \tr[\partial_i \widetilde{P}(\bk)\partial_j \widetilde{P}(\bk)] \equiv \widetilde{\mathcal{G}}_{ij},
\end{equation}
where $\widetilde{P}(\bk)$ is the many-body projector $\widetilde{P}(\bk) = e^{-i\bk\cdot \widetilde{\bX}} \ket{\Psi_\bk} \bra{\Psi_\bk} e^{i\bk\cdot \widetilde{\bX}} $, with $\ket{\Psi_\bk}$ a many-body state obeying a twisted boundary condition specified by $\bk$ \footnote{To be concrete, the twisted periodic boundary condition corresponds to $\Psi_\bk(\br_1,..., \br_a+\bL, ..., \br_N) = e^{i\bk\cdot \bL} \Psi_\bk(\br_1,..., \br_a, ..., \br_N)$, with $k_i \in [0,2\pi/L_i)$ and $L_i$ is the period of the system in the $i$-th direction.} and $\widetilde{\bX} = \sum_{\mbf{R}\sigma} \mbf{R} c^\dag_{\mbf{R},\sigma}c_{\mbf{R},\sigma}$ is the many-body position operator.
To show \cref{eq:relating S to G}, we first derive from Eq. \eqref{eq: origin_embedding FT} that
\begin{equation}\label{eq: relating S to localization}
    \frac{1}{2}\partial_i\partial_j \widetilde{S}_\bq \vert_{\bq=0} = \frac{1}{\mathcal{A}} \cc{\widetilde{X}_i \widetilde{X}_j}.
\end{equation}
%
On the right hand side we have the localization tensor, which is known to be equal to the integrated many-body quantum metric as discussed in Refs. \cite{SWM2000, resta2011insulating}, thus leading to Eq. \eqref{eq:relating S to G}.
The relation between $\cc{\widetilde{X}_i \widetilde{X}_j}$  and the many-body quantum metric was first derived by Souza-Wilkens-Martin in Ref. \cite{SWM2000} by assuming Kohn's conjecture about wavefunction localization in insulators \cite{Kohn1964}, while we provide an alternative derivation in the supplementary material \cite{supp} without relying on this assumption.

In the non-interacting limit of band insulators, Eq. \eqref{eq:relating S to G} naturally reduces to its single-particle version where $\bk$ is the Bloch momentum and $\widetilde{P}(\bk) \to \sum_{m\in occ} \widetilde{U}_m(\bk) \widetilde{U}_m^\dagger(\bk)$ becomes the single-particle projector onto the occupied bands with eigenvectors $\widetilde{U}_m(\bk)$ \footnote{A simple proof of this statement is included in the Supplementary \cite{supp} for completeness. See also Refs. \cite{Regnault2013,onishi2024quantum}}. We remark that the origin orbital embedding can be adopted in any system with generic orbitals, yielding a quantum geometric observable. We now show how to physically measure this quantity from the corner charge fluctuation.

\noindent {\color{blue}\emph{Main result.}} For a generic lattice system (either interacting or non-interacting), we now prove:
\begin{equation}\label{eq:key_result}
    \lim_{\theta \rightarrow 0 } \theta \cdot \mathcal{C}_i^{(Q)}(\theta)= \frac{1}{2} \partial_j^2 \widetilde{S}_\bq\vert_{\bq=0}, \;\;\; \text{for}\;\; i \perp j\;.
\end{equation}
It is useful to first recognize $\widetilde{S}_\bq$ as
\eqa{
\label{eq:structure_factor_origin_embedding}
    \widetilde{S}_\bq =\frac{1}{\mathcal{A}_{\text{cell}}}\sum_{\bR-\bR'} e^{-i\bq\cdot(\bR-\bR')}\cc{\rho(\bR-\bR')\rho(0)},
}
where $\mathcal{A}_{\text{cell}}$ is the area of the unit cell.
%
From Eq. \eqref{eq:simplified_CQ}, we note that for lattice partitions with a small $\theta$ (refer to Fig. \ref{fig:setup}) the unit-cell positions $\bR \in A$ and $\bR'\in C$ must be far separated. By the short-range nature of insulator, we can ignore the first term and focus on the second term in Eq. \eqref{eq:simplified_CQ}. For a fixed bond separation $\bR-\bR'$, the number of bonds satisfying $\bR \in B$ and $\bR'\in D$ is counted as 
\begin{equation}\label{eq:exact_counting}
    \frac{1}{\mathcal{A}_{\text{cell}}}\big[(\bR-\bR')_y \cot{\theta} - (\bR-\bR')_x\big] (\bR-\bR')_y, 
\end{equation}
with $\mathcal{A}_{\text{cell}}$ the area of a unit cell. This counting is \textit{exact} on the square lattice provided that (i) the angle is chosen to satisfy $\cot{\theta}\in 2\mathbb{N}$ and (ii) all subregions are non-empty, containing at least one site. This counting is generalized to any Bravais lattice in the Supplementary \cite{supp}. Combined with the small-angle limit,
\begin{equation}\label{eq: intermediate step 1}
    \lim_{\theta \rightarrow 0 } \mathcal{C}_x^{(Q)}(\theta) = -\frac{\cot\theta}{2 \mathcal{A}_{\text{cell}}} \sum_{\bR-\bR'} (\bR-\bR')_y^2 \cc{\rho(\bR-\bR')\rho(0)},
\end{equation}
where lattice translation symmetry is used to express $\cc{\rho(\bR)\rho(\bR')}=\cc{\rho(\bR-\bR')\rho(0)}$, and the $\frac{1}{2}$ factor arises from having $\bR-\bR'$ to be summed over the entire lattice. Taking two derivatives of \cref{eq:structure_factor_origin_embedding} and comparing with Eq. \eqref{eq: intermediate step 1}, we arrive at our main result in \cref{eq:key_result}. 

The above argument only assumes lattice translation symmetry and is applicable to any two-dimensional Bravais lattices, with partition oriented along a crystal axis. In the Supplementary \cite{supp}, we further elaborate on the case of triangular lattice. In terms of the integrated quantum metric and the corner coefficient (in anticipation of numerical studies for generic $\theta$), our main result can be recast as 
\begin{equation}\label{eq: key result_band insulator}
\lim_{\theta\rightarrow 0 } \gamma^{(Q)}_i(\theta) = 4\pi \widetilde{\mathcal{G}}_{jj}, \;\;\; \text{for}\;\; i \perp j.
\end{equation}

\begin{figure}[t]
    \centering
    \includegraphics[width=1.\linewidth]{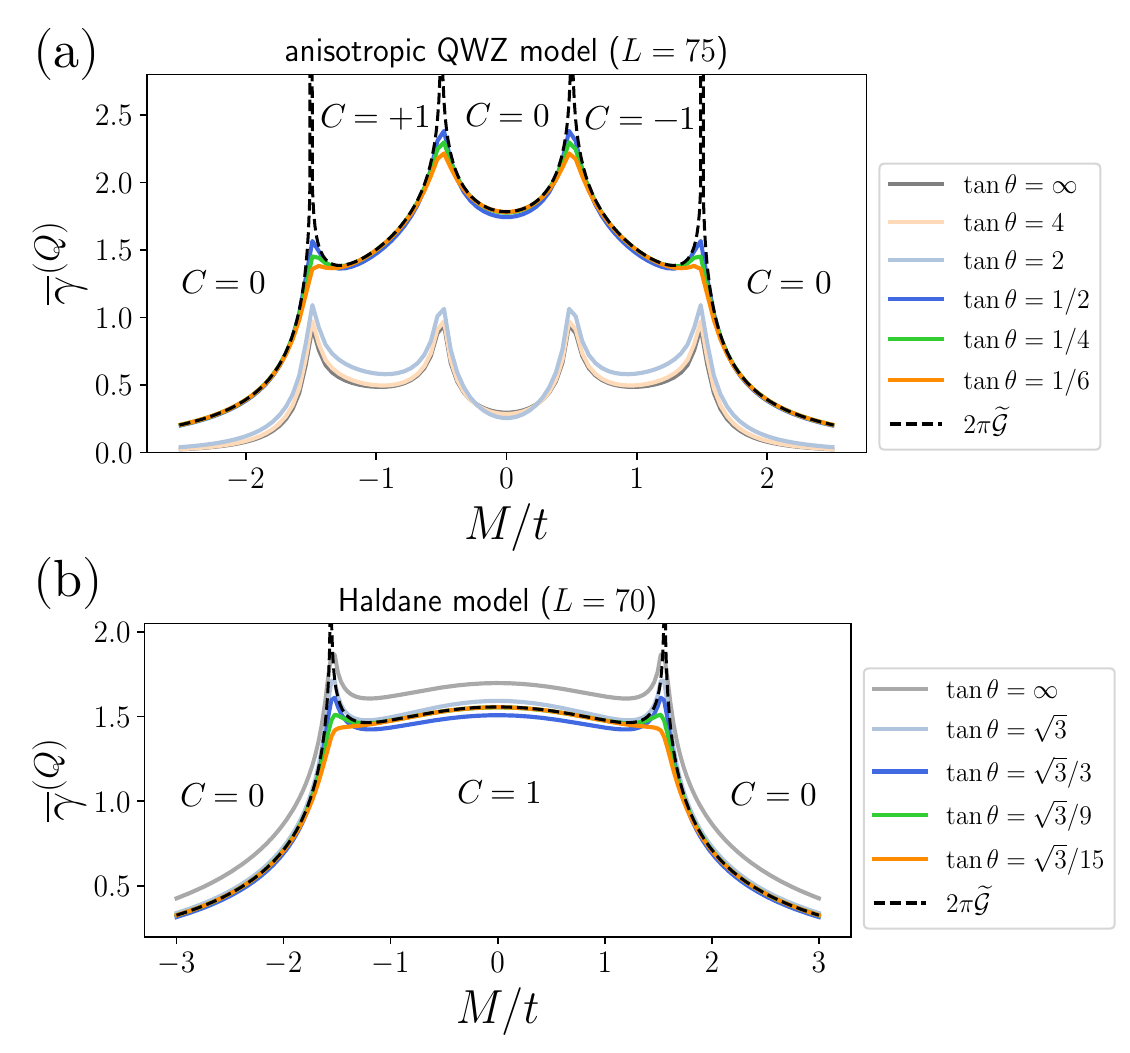}
    \caption{Corner charge fluctuation in various Chern insulator models, with the Chern number $C$ indicated in each phase. Topology and quantum geometry of the insulator are varied by tuning the sublattice mass $M$ (rescaled by hopping $t$). For charge fluctuation, a universal angle-dependence arises for small $\theta$, where the average corner coefficient $\overline{\gamma}^{(Q)} \equiv \frac{1}{2}(\gamma^{(Q)}_x +\gamma^{(Q)}_y)$ equals to the trace of integrated quantum metric $2\pi \widetilde{\mathcal{G}} \geq \abs{C}$, in consistence with Eq. \eqref{eq: key result_band insulator}.}
    \label{fig:numerics_Fluc}
\end{figure}

\begin{figure}[b]
    \centering
    \includegraphics[width=0.85\linewidth]{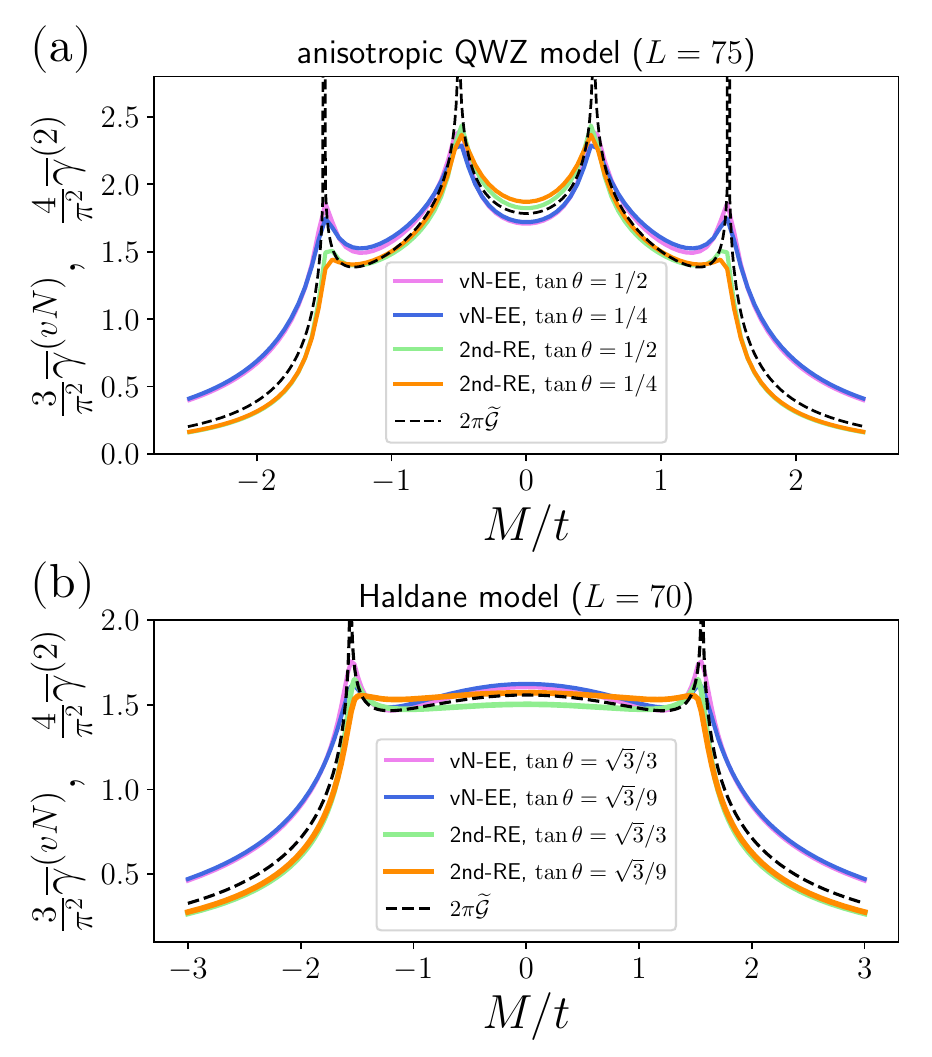}
    \caption{Corner entanglement entropies in various Chern insulator models as a function of the sublattice mass $M$ (rescaled by hopping $t$). The average corner coefficient for both the von-Neumann and second R\'enyi EEs, $\overline{\gamma}^{(vN,2)} \equiv \frac{1}{2}(\gamma^{(vN,2)}_x +\gamma^{(vN,2)}_y)$, closely follow the trace of integrated quantum metric $2\pi \widetilde{\mathcal{G}}$ upon rescaling by the leading cumulant expansion coefficient.}
    \label{fig:numerics_EE}
\end{figure}

In the following, we focus on band insulators and verify \cref{eq: key result_band insulator} in various lattice models. We first analyze a \textit{compact obstructed atomic insulator} on a square lattice \cite{bradlyn2017topological, Schindler2021,Jonah2023}, where $\mathcal{C}^{(Q)}(\theta)$ can be evaluated beyond the small-angle-limit (details provided in \cite{supp}). This model is constructed with four coinciding orbitals per unit cell on a square lattice, and the occupied eigenstate is specified as
\begin{equation}
    \widetilde{U}(\bk) = \frac{1}{4}\sum_{m=0}^3 e^{i\frac{\bk}{2}\cdot (1-C_4^m)(\hat{x}+\hat{y})}\mathcal{D}[C_4]^m  \bpm 1 \\ 1 \\1 \\1 \epm,
\end{equation}
where $C_4^m$ represents a counter-clockwise rotation by $m\pi/2$, and $\mathcal{D}[C_4] = \text{diag}(1,-1,-i,i)$ is the rotation operator in the orbital space. Consequently, $\widetilde{\mathcal{G}}_{xx} = \widetilde{\mathcal{G}}_{yy} = 1/4$. The Wannier orbitals are \textit{compactly} supported on the four corners of each plaquette, which allows us to find exactly that $\mathcal{C}^{(Q)}(\theta) = \cot{\theta}/4$ for $\cot{\theta}\in 2\mathbb{N}$, and $\mathcal{C}^{(Q)}(\theta) = 1/8$ for $\tan{\theta}\in 2\mathbb{N}$. Hence $\gamma^{(Q)}(\theta\ll 1) = \pi$ as promised, while in the opposite limit $\gamma^{(Q)}(\pi/2) = \pi^2/4$. This analysis highlights the importance of the small-angle limit for probing quantum geometry on lattices.

\noindent {\color{blue}\emph{Lattice simulation.}} We further substantiate our result for band insulators with variable quantum geometry by simulating Chern insulator models on lattices of $L\times L$ sites with open boundary conditions. Details of implementation are provided in \cite{supp}. 
%
%
We study the square-lattice Qi-Wu-Zhang (QWZ) model \cite{QWZ2006} and the Haldane's honeycomb model \cite{Haldane1988}, following the partition scheme stipulated below Eq. \eqref{eq: fluctuation_simplified}. The \textit{average} corner coefficient $\overline{\gamma}^{(Q)}(\theta) \equiv \frac{1}{2}[\gamma^{(Q)}_x (\theta) +\gamma^{(Q)}_y (\theta)]$ is calculated for various angles $\theta$, and compared with the trace of integrated metric $\widetilde{\mathcal{G}}=\widetilde{\mathcal{G}}_{xx}+\widetilde{\mathcal{G}}_{yy}$ (computed using the band projector defined below Eq. \eqref{eq:relating S to G}, for one occupied band). The results are summarized in Fig. \ref{fig:numerics_Fluc}. For the QWZ model, we have investigated the case with anisotropic nearest-neighbor inter-orbital hopping $t_x=2t_y=t$. For the Haldane model, we consider nearest-neighbor hopping $t$ and next-nearest-neighbor hopping $t_2=0.3 t$ with the phase parameter $\phi=\pi/2$. 
Varying the sublattice mass $M$ we access both trivial and topological phases with varying quantum geometry. Both $2\pi\widetilde{\mathcal{G}}$ and the closely matched small-angle $\overline{\gamma}^{(Q)}$ are lower bounded by the Chern number. While our prediction is made for $\theta \ll 1$, the numerics show an exceptional match between $\overline{\gamma}^{(Q)}$ and $2\pi \widetilde{\mathcal{G}}$ already for intermediate $\theta$. 
Recent realizations of the discussed models in ultracold Fermi gases \cite{jotzu2014experimental,Liang2023_QWZrealization}  encourage near-term experimental observation of quantum geometry with the aid of quantum gas microscopy, which offers site-resolved imaging for measuring $\mathcal{C}^{(Q)}$ \cite{Cheuk2015, Haller2015, Parsons2015, Edge2015, Omran2015, Bakr_review}.

\noindent {\color{blue}\emph{Corner entanglement entropies.}} Motivated by the established connection between quantum geometry and corner charge fluctuation, we now explore quantum geometric effects in quantum entanglement. For free fermions, it is well known that the entanglement entropies (EEs) are determined by the full counting statistics composed of charge cumulants \cite{klich2009, song2012, calabrese2012exact}. We focus on the von-Neumann (vN) and the second R\'enyi entropies, which satisfy
\begin{equation}\label{eq:cumulant_expansion}
\begin{split}
    S^{(vN)}_A  &= \frac{\pi^2}{3} \braket{Q^2_A}_c +  \frac{\pi^4}{45} \braket{Q^4_A}_c+  \frac{2\pi^6}{945} \braket{Q^6_A}_c+ ... \\
    S^{(2)}_A  &= \frac{\pi^2}{4} \braket{Q^2_A}_c -  \frac{\pi^4}{192} \braket{Q^4_A}_c+  \frac{\pi^6}{23040} \braket{Q^6_A}_c+ ...
\end{split}
\end{equation}
The EEs are also known to scale generically as Eq. \eqref{eq: def_corner_fluc_continuum}, and their corner terms have been studied extensively in conformal field theories \cite{klebanov2012shape, Krempa2015corner, faulkner2016shape, Melko2017, Melko2019, Estienne2021Dirac}, and in connection to holographic duality \cite{bueno2015corner, seminara2017corner}. Here we discover new connections for the quantum geometry in band insulators. 
The corner entanglement entropies $\mathcal{C}^{(vN,2)}$ are defined similar to Eq. \eqref{eq: def_corner_fluc_lattice}, replacing $\cc{Q_A^2}$ by $S^{(vN,2)}_A$ and similarly for other regions, and the corner coefficients $\gamma^{(vN,2)}$ are defined as in Eq. \eqref{eq:def_corner_coefficient}. Figure \ref{fig:numerics_EE} shows the comparison between the \textit{average} corner EE coefficients $\overline{\gamma}^{(vN,2)} \equiv \frac{1}{2}(\gamma^{(vN,2)}_x+\gamma^{(vN,2)}_y)$ and $2\pi \widetilde{\mathcal{G}}$ for small $\theta$, with EEs computed exactly using the correlation matrix method \cite{Peschel2001, peschel2003calculation, Cheong2004,supp}. The corner EEs are found to closely follow the trend of variation in quantum geometry. Particularly, they peak at gap-closing transitions where the corner EEs are known to diverge logarithmically with the system size \cite{Krempa2015corner}, in consonance with the logarithmic divergence of 2D quantum metric \cite{Thonhauser2006,onishi2024fundamental}.

Rescaling $\overline{\gamma}^{(vN,2)}$ by the leading coefficient of the cumulant expansion in Eq. \eqref{eq:cumulant_expansion} (see also the vertical axis of Fig. \ref{fig:numerics_EE}), a close quantitative match with $2\pi \widetilde{\mathcal{G}}$ is observed. Notably, the second R\'enyi entropy tracks the quantum geometry more closely than the von-Neumann entropy, which can be understood from the suppressed higher-order cumulant coefficients in Eq. \eqref{eq:cumulant_expansion}. In general, higher order charge cumulants have their own corner terms \cite{berthiere2023full}, which is the origin of the slight mismatch between EEs and quantum metric as seen in Fig. \ref{fig:numerics_EE}. The analytical understanding of this mismatch is left for future investigations. Yet, we have provided numerical evidence for the geometric effects in the entanglement entropy as probed by the corner term.

\noindent {\color{blue}\emph{Conclusion.}} We have demonstrated, both analytically and numerically, that the bipartite charge fluctuation contains a \textit{corner} term that captures the quantum geometry of 2D lattice insulators. We propose a practical scheme to isolate the corner term and introduce a new observable for quantum geometry, applicable to generic interacting lattice systems and readily measurable under quantum gas microscopes. For free fermions, we further provide numerical evidence for the quantum geometric effect in the corner entanglement entropies. For future investigations, it is important to investigate interaction effects for corner entanglement entropies, generalize our results for both charge fluctuation and entanglement entropies to three-dimensional insulators, and explore physical observables for higher order quantum geometric tensors \cite{PhysRevA.108.032218}.


\begin{acknowledgments}
We are grateful to Shinsei Ryu, Andrei Bernevig, Jie Wang, Hongchao Li and Hyunsoo Ha for inspiring discussions, and especially to Gilles Parez for bringing Ref. \cite{estienne2022cornering} to our attention. P.M.T. also appreciates discussions with Duncan Haldane, Ramanjit Sohal, Ruihua Fan, Zhehao Dai, Liang Fu, and particularly Charles Kane for a motivating conversation about Ref. \cite{onishi2024quantum}. P.M.T. is supported by a postdoctoral research fellowship at
the Princeton Center for Theoretical Science and a Croucher Fellowship.
J.H.-A. is supported by a Hertz Fellowship.
J.Y acknowledges the support of the Gordon and Betty
Moore Foundation at Princeton University, and is supported by the startup fund at the University of Florida.

%
\noindent {\color{blue}\emph{Notes added:} } Close to the completion of this updated version of the manuscript, we became aware of an upcoming work Ref. \cite{wu2024corner}, which also studies the relation between corner charge fluctuations and the many-body quantum geometry. We thank Meng Cheng and collaborators for communicating their unpublished work with us, and for coordinating submissions of our papers. Besides, a recent work Ref. \cite{kruchkov2024entanglemententropylatticemodels} discusses geometric contributions to the lower bound of entanglement entropy for partitions without corners. 

\end{acknowledgments}

\bibliographystyle{apsrev4-1.bst}
\bibliography{reference}

\begin{thebibliography}{98}%
\makeatletter
\providecommand \@ifxundefined [1]{%
 \@ifx{#1\undefined}
}%
\providecommand \@ifnum [1]{%
 \ifnum #1\expandafter \@firstoftwo
 \else \expandafter \@secondoftwo
 \fi
}%
\providecommand \@ifx [1]{%
 \ifx #1\expandafter \@firstoftwo
 \else \expandafter \@secondoftwo
 \fi
}%
\providecommand \natexlab [1]{#1}%
\providecommand \enquote  [1]{``#1''}%
\providecommand \bibnamefont  [1]{#1}%
\providecommand \bibfnamefont [1]{#1}%
\providecommand \citenamefont [1]{#1}%
\providecommand \href@noop [0]{\@secondoftwo}%
\providecommand \href [0]{\begingroup \@sanitize@url \@href}%
\providecommand \@href[1]{\@@startlink{#1}\@@href}%
\providecommand \@@href[1]{\endgroup#1\@@endlink}%
\providecommand \@sanitize@url [0]{\catcode `\\12\catcode `\$12\catcode
  `\&12\catcode `\#12\catcode `\^12\catcode `\_12\catcode `\%12\relax}%
\providecommand \@@startlink[1]{}%
\providecommand \@@endlink[0]{}%
\providecommand \url  [0]{\begingroup\@sanitize@url \@url }%
\providecommand \@url [1]{\endgroup\@href {#1}{\urlprefix }}%
\providecommand \urlprefix  [0]{URL }%
\providecommand \Eprint [0]{\href }%
\providecommand \doibase [0]{http://dx.doi.org/}%
\providecommand \selectlanguage [0]{\@gobble}%
\providecommand \bibinfo  [0]{\@secondoftwo}%
\providecommand \bibfield  [0]{\@secondoftwo}%
\providecommand \translation [1]{[#1]}%
\providecommand \BibitemOpen [0]{}%
\providecommand \bibitemStop [0]{}%
\providecommand \bibitemNoStop [0]{.\EOS\space}%
\providecommand \EOS [0]{\spacefactor3000\relax}%
\providecommand \BibitemShut  [1]{\csname bibitem#1\endcsname}%
\let\auto@bib@innerbib\@empty
\bibitem [{\citenamefont {T\"orm\"a}(2023)}]{Torma2023}%
  \BibitemOpen
  \bibfield  {author} {\bibinfo {author} {\bibfnamefont {P.}~\bibnamefont
  {T\"orm\"a}},\ }\href {\doibase 10.1103/PhysRevLett.131.240001} {\bibfield
  {journal} {\bibinfo  {journal} {Phys. Rev. Lett.}\ }\textbf {\bibinfo
  {volume} {131}},\ \bibinfo {pages} {240001} (\bibinfo {year}
  {2023})}\BibitemShut {NoStop}%
\bibitem [{\citenamefont {Resta}(2011)}]{resta2011insulating}%
  \BibitemOpen
  \bibfield  {author} {\bibinfo {author} {\bibfnamefont {R.}~\bibnamefont
  {Resta}},\ }\href
  {https://link.springer.com/article/10.1140/epjb/e2010-10874-4} {\bibfield
  {journal} {\bibinfo  {journal} {The European Physical Journal B}\ }\textbf
  {\bibinfo {volume} {79}},\ \bibinfo {pages} {121} (\bibinfo {year}
  {2011})}\BibitemShut {NoStop}%
\bibitem [{\citenamefont {Provost}\ and\ \citenamefont
  {Vallee}(1980)}]{provost1980riemannian}%
  \BibitemOpen
  \bibfield  {author} {\bibinfo {author} {\bibfnamefont {J.}~\bibnamefont
  {Provost}}\ and\ \bibinfo {author} {\bibfnamefont {G.}~\bibnamefont
  {Vallee}},\ }\href {https://link.springer.com/article/10.1007/bf02193559}
  {\bibfield  {journal} {\bibinfo  {journal} {Communications in Mathematical
  Physics}\ }\textbf {\bibinfo {volume} {76}},\ \bibinfo {pages} {289}
  (\bibinfo {year} {1980})}\BibitemShut {NoStop}%
\bibitem [{\citenamefont {Berry}(1984)}]{berry1984quantal}%
  \BibitemOpen
  \bibfield  {author} {\bibinfo {author} {\bibfnamefont {M.~V.}\ \bibnamefont
  {Berry}},\ }\href
  {https://royalsocietypublishing.org/doi/10.1098/rspa.1984.0023} {\bibfield
  {journal} {\bibinfo  {journal} {Proceedings of the Royal Society of London.
  A. Mathematical and Physical Sciences}\ }\textbf {\bibinfo {volume} {392}},\
  \bibinfo {pages} {45} (\bibinfo {year} {1984})}\BibitemShut {NoStop}%
\bibitem [{\citenamefont {Thouless}\ \emph {et~al.}(1982)\citenamefont
  {Thouless}, \citenamefont {Kohmoto}, \citenamefont {Nightingale},\ and\
  \citenamefont {den Nijs}}]{TKNN1982}%
  \BibitemOpen
  \bibfield  {author} {\bibinfo {author} {\bibfnamefont {D.~J.}\ \bibnamefont
  {Thouless}}, \bibinfo {author} {\bibfnamefont {M.}~\bibnamefont {Kohmoto}},
  \bibinfo {author} {\bibfnamefont {M.~P.}\ \bibnamefont {Nightingale}}, \ and\
  \bibinfo {author} {\bibfnamefont {M.}~\bibnamefont {den Nijs}},\ }\href
  {\doibase 10.1103/PhysRevLett.49.405} {\bibfield  {journal} {\bibinfo
  {journal} {Phys. Rev. Lett.}\ }\textbf {\bibinfo {volume} {49}},\ \bibinfo
  {pages} {405} (\bibinfo {year} {1982})}\BibitemShut {NoStop}%
\bibitem [{\citenamefont {Souza}\ \emph {et~al.}(2000)\citenamefont {Souza},
  \citenamefont {Wilkens},\ and\ \citenamefont {Martin}}]{SWM2000}%
  \BibitemOpen
  \bibfield  {author} {\bibinfo {author} {\bibfnamefont {I.}~\bibnamefont
  {Souza}}, \bibinfo {author} {\bibfnamefont {T.}~\bibnamefont {Wilkens}}, \
  and\ \bibinfo {author} {\bibfnamefont {R.~M.}\ \bibnamefont {Martin}},\
  }\href {\doibase 10.1103/PhysRevB.62.1666} {\bibfield  {journal} {\bibinfo
  {journal} {Phys. Rev. B}\ }\textbf {\bibinfo {volume} {62}},\ \bibinfo
  {pages} {1666} (\bibinfo {year} {2000})}\BibitemShut {NoStop}%
\bibitem [{\citenamefont {Kivelson}(1982)}]{Kivelson1982}%
  \BibitemOpen
  \bibfield  {author} {\bibinfo {author} {\bibfnamefont {S.}~\bibnamefont
  {Kivelson}},\ }\href {\doibase 10.1103/PhysRevB.26.4269} {\bibfield
  {journal} {\bibinfo  {journal} {Phys. Rev. B}\ }\textbf {\bibinfo {volume}
  {26}},\ \bibinfo {pages} {4269} (\bibinfo {year} {1982})}\BibitemShut
  {NoStop}%
\bibitem [{\citenamefont {Marzari}\ and\ \citenamefont
  {Vanderbilt}(1997)}]{Marzari1997}%
  \BibitemOpen
  \bibfield  {author} {\bibinfo {author} {\bibfnamefont {N.}~\bibnamefont
  {Marzari}}\ and\ \bibinfo {author} {\bibfnamefont {D.}~\bibnamefont
  {Vanderbilt}},\ }\href {\doibase 10.1103/PhysRevB.56.12847} {\bibfield
  {journal} {\bibinfo  {journal} {Phys. Rev. B}\ }\textbf {\bibinfo {volume}
  {56}},\ \bibinfo {pages} {12847} (\bibinfo {year} {1997})}\BibitemShut
  {NoStop}%
\bibitem [{\citenamefont {Resta}\ and\ \citenamefont
  {Sorella}(1999)}]{resta1999}%
  \BibitemOpen
  \bibfield  {author} {\bibinfo {author} {\bibfnamefont {R.}~\bibnamefont
  {Resta}}\ and\ \bibinfo {author} {\bibfnamefont {S.}~\bibnamefont
  {Sorella}},\ }\href {\doibase 10.1103/PhysRevLett.82.370} {\bibfield
  {journal} {\bibinfo  {journal} {Phys. Rev. Lett.}\ }\textbf {\bibinfo
  {volume} {82}},\ \bibinfo {pages} {370} (\bibinfo {year} {1999})}\BibitemShut
  {NoStop}%
\bibitem [{\citenamefont {Roy}(2014)}]{Roy2014}%
  \BibitemOpen
  \bibfield  {author} {\bibinfo {author} {\bibfnamefont {R.}~\bibnamefont
  {Roy}},\ }\href {\doibase 10.1103/PhysRevB.90.165139} {\bibfield  {journal}
  {\bibinfo  {journal} {Phys. Rev. B}\ }\textbf {\bibinfo {volume} {90}},\
  \bibinfo {pages} {165139} (\bibinfo {year} {2014})}\BibitemShut {NoStop}%
\bibitem [{\citenamefont {Peotta}\ and\ \citenamefont
  {T{\"o}rm{\"a}}(2015)}]{peotta2015superfluidity}%
  \BibitemOpen
  \bibfield  {author} {\bibinfo {author} {\bibfnamefont {S.}~\bibnamefont
  {Peotta}}\ and\ \bibinfo {author} {\bibfnamefont {P.}~\bibnamefont
  {T{\"o}rm{\"a}}},\ }\href@noop {} {\bibfield  {journal} {\bibinfo  {journal}
  {Nature communications}\ }\textbf {\bibinfo {volume} {6}},\ \bibinfo {pages}
  {8944} (\bibinfo {year} {2015})}\BibitemShut {NoStop}%
\bibitem [{\citenamefont {Xie}\ \emph {et~al.}(2020)\citenamefont {Xie},
  \citenamefont {Song}, \citenamefont {Lian},\ and\ \citenamefont
  {Bernevig}}]{Xie2020}%
  \BibitemOpen
  \bibfield  {author} {\bibinfo {author} {\bibfnamefont {F.}~\bibnamefont
  {Xie}}, \bibinfo {author} {\bibfnamefont {Z.}~\bibnamefont {Song}}, \bibinfo
  {author} {\bibfnamefont {B.}~\bibnamefont {Lian}}, \ and\ \bibinfo {author}
  {\bibfnamefont {B.~A.}\ \bibnamefont {Bernevig}},\ }\href {\doibase
  10.1103/PhysRevLett.124.167002} {\bibfield  {journal} {\bibinfo  {journal}
  {Phys. Rev. Lett.}\ }\textbf {\bibinfo {volume} {124}},\ \bibinfo {pages}
  {167002} (\bibinfo {year} {2020})}\BibitemShut {NoStop}%
\bibitem [{\citenamefont {Ozawa}\ and\ \citenamefont {Mera}(2021)}]{ozawa2021}%
  \BibitemOpen
  \bibfield  {author} {\bibinfo {author} {\bibfnamefont {T.}~\bibnamefont
  {Ozawa}}\ and\ \bibinfo {author} {\bibfnamefont {B.}~\bibnamefont {Mera}},\
  }\href {\doibase 10.1103/PhysRevB.104.045103} {\bibfield  {journal} {\bibinfo
   {journal} {Phys. Rev. B}\ }\textbf {\bibinfo {volume} {104}},\ \bibinfo
  {pages} {045103} (\bibinfo {year} {2021})}\BibitemShut {NoStop}%
\bibitem [{\citenamefont {Yu}\ \emph {et~al.}(2022)\citenamefont {Yu},
  \citenamefont {Chen},\ and\ \citenamefont {Das~Sarma}}]{Yu2022}%
  \BibitemOpen
  \bibfield  {author} {\bibinfo {author} {\bibfnamefont {J.}~\bibnamefont
  {Yu}}, \bibinfo {author} {\bibfnamefont {Y.-A.}\ \bibnamefont {Chen}}, \ and\
  \bibinfo {author} {\bibfnamefont {S.}~\bibnamefont {Das~Sarma}},\ }\href
  {\doibase 10.1103/PhysRevB.105.104515} {\bibfield  {journal} {\bibinfo
  {journal} {Phys. Rev. B}\ }\textbf {\bibinfo {volume} {105}},\ \bibinfo
  {pages} {104515} (\bibinfo {year} {2022})}\BibitemShut {NoStop}%
\bibitem [{\citenamefont {Herzog-Arbeitman}\ \emph
  {et~al.}(2022{\natexlab{a}})\citenamefont {Herzog-Arbeitman}, \citenamefont
  {Peri}, \citenamefont {Schindler}, \citenamefont {Huber},\ and\ \citenamefont
  {Bernevig}}]{Jonah2022b}%
  \BibitemOpen
  \bibfield  {author} {\bibinfo {author} {\bibfnamefont {J.}~\bibnamefont
  {Herzog-Arbeitman}}, \bibinfo {author} {\bibfnamefont {V.}~\bibnamefont
  {Peri}}, \bibinfo {author} {\bibfnamefont {F.}~\bibnamefont {Schindler}},
  \bibinfo {author} {\bibfnamefont {S.~D.}\ \bibnamefont {Huber}}, \ and\
  \bibinfo {author} {\bibfnamefont {B.~A.}\ \bibnamefont {Bernevig}},\ }\href
  {\doibase 10.1103/PhysRevLett.128.087002} {\bibfield  {journal} {\bibinfo
  {journal} {Phys. Rev. Lett.}\ }\textbf {\bibinfo {volume} {128}},\ \bibinfo
  {pages} {087002} (\bibinfo {year} {2022}{\natexlab{a}})}\BibitemShut
  {NoStop}%
\bibitem [{\citenamefont {Onishi}\ and\ \citenamefont
  {Fu}(2024{\natexlab{a}})}]{onishi2024topologicalboundstructurefactor}%
  \BibitemOpen
  \bibfield  {author} {\bibinfo {author} {\bibfnamefont {Y.}~\bibnamefont
  {Onishi}}\ and\ \bibinfo {author} {\bibfnamefont {L.}~\bibnamefont {Fu}},\
  }\href {https://arxiv.org/abs/2406.18654} {\enquote {\bibinfo {title}
  {Topological bound on structure factor},}\ } (\bibinfo {year}
  {2024}{\natexlab{a}}),\ \Eprint {http://arxiv.org/abs/2406.18654}
  {arXiv:2406.18654 [cond-mat.str-el]} \BibitemShut {NoStop}%
\bibitem [{\citenamefont {Parameswaran}\ \emph {et~al.}(2012)\citenamefont
  {Parameswaran}, \citenamefont {Roy},\ and\ \citenamefont
  {Sondhi}}]{Parameswaran2012}%
  \BibitemOpen
  \bibfield  {author} {\bibinfo {author} {\bibfnamefont {S.~A.}\ \bibnamefont
  {Parameswaran}}, \bibinfo {author} {\bibfnamefont {R.}~\bibnamefont {Roy}}, \
  and\ \bibinfo {author} {\bibfnamefont {S.~L.}\ \bibnamefont {Sondhi}},\
  }\href {\doibase 10.1103/PhysRevB.85.241308} {\bibfield  {journal} {\bibinfo
  {journal} {Phys. Rev. B}\ }\textbf {\bibinfo {volume} {85}},\ \bibinfo
  {pages} {241308} (\bibinfo {year} {2012})}\BibitemShut {NoStop}%
\bibitem [{\citenamefont {Dobard\ifmmode \check{z}\else
  \v{z}\fi{}i\ifmmode~\acute{c}\else \'{c}\fi{}}\ \emph
  {et~al.}(2013)\citenamefont {Dobard\ifmmode \check{z}\else
  \v{z}\fi{}i\ifmmode~\acute{c}\else \'{c}\fi{}}, \citenamefont
  {Milovanovi\ifmmode~\acute{c}\else \'{c}\fi{}},\ and\ \citenamefont
  {Regnault}}]{Regnault2013}%
  \BibitemOpen
  \bibfield  {author} {\bibinfo {author} {\bibfnamefont {E.}~\bibnamefont
  {Dobard\ifmmode \check{z}\else \v{z}\fi{}i\ifmmode~\acute{c}\else
  \'{c}\fi{}}}, \bibinfo {author} {\bibfnamefont {M.~V.}\ \bibnamefont
  {Milovanovi\ifmmode~\acute{c}\else \'{c}\fi{}}}, \ and\ \bibinfo {author}
  {\bibfnamefont {N.}~\bibnamefont {Regnault}},\ }\href {\doibase
  10.1103/PhysRevB.88.115117} {\bibfield  {journal} {\bibinfo  {journal} {Phys.
  Rev. B}\ }\textbf {\bibinfo {volume} {88}},\ \bibinfo {pages} {115117}
  (\bibinfo {year} {2013})}\BibitemShut {NoStop}%
\bibitem [{\citenamefont {Jackson}\ \emph {et~al.}(2015)\citenamefont
  {Jackson}, \citenamefont {M{\"o}ller},\ and\ \citenamefont
  {Roy}}]{jackson2015geometric}%
  \BibitemOpen
  \bibfield  {author} {\bibinfo {author} {\bibfnamefont {T.~S.}\ \bibnamefont
  {Jackson}}, \bibinfo {author} {\bibfnamefont {G.}~\bibnamefont {M{\"o}ller}},
  \ and\ \bibinfo {author} {\bibfnamefont {R.}~\bibnamefont {Roy}},\
  }\href@noop {} {\bibfield  {journal} {\bibinfo  {journal} {Nature
  communications}\ }\textbf {\bibinfo {volume} {6}},\ \bibinfo {pages} {8629}
  (\bibinfo {year} {2015})}\BibitemShut {NoStop}%
\bibitem [{\citenamefont {Claassen}\ \emph {et~al.}(2015)\citenamefont
  {Claassen}, \citenamefont {Lee}, \citenamefont {Thomale}, \citenamefont
  {Qi},\ and\ \citenamefont {Devereaux}}]{Claassen2015}%
  \BibitemOpen
  \bibfield  {author} {\bibinfo {author} {\bibfnamefont {M.}~\bibnamefont
  {Claassen}}, \bibinfo {author} {\bibfnamefont {C.~H.}\ \bibnamefont {Lee}},
  \bibinfo {author} {\bibfnamefont {R.}~\bibnamefont {Thomale}}, \bibinfo
  {author} {\bibfnamefont {X.-L.}\ \bibnamefont {Qi}}, \ and\ \bibinfo {author}
  {\bibfnamefont {T.~P.}\ \bibnamefont {Devereaux}},\ }\href {\doibase
  10.1103/PhysRevLett.114.236802} {\bibfield  {journal} {\bibinfo  {journal}
  {Phys. Rev. Lett.}\ }\textbf {\bibinfo {volume} {114}},\ \bibinfo {pages}
  {236802} (\bibinfo {year} {2015})}\BibitemShut {NoStop}%
\bibitem [{\citenamefont {Wang}\ \emph
  {et~al.}(2021{\natexlab{a}})\citenamefont {Wang}, \citenamefont {Cano},
  \citenamefont {Millis}, \citenamefont {Liu},\ and\ \citenamefont
  {Yang}}]{Wang2021}%
  \BibitemOpen
  \bibfield  {author} {\bibinfo {author} {\bibfnamefont {J.}~\bibnamefont
  {Wang}}, \bibinfo {author} {\bibfnamefont {J.}~\bibnamefont {Cano}}, \bibinfo
  {author} {\bibfnamefont {A.~J.}\ \bibnamefont {Millis}}, \bibinfo {author}
  {\bibfnamefont {Z.}~\bibnamefont {Liu}}, \ and\ \bibinfo {author}
  {\bibfnamefont {B.}~\bibnamefont {Yang}},\ }\href {\doibase
  10.1103/PhysRevLett.127.246403} {\bibfield  {journal} {\bibinfo  {journal}
  {Phys. Rev. Lett.}\ }\textbf {\bibinfo {volume} {127}},\ \bibinfo {pages}
  {246403} (\bibinfo {year} {2021}{\natexlab{a}})}\BibitemShut {NoStop}%
\bibitem [{\citenamefont {Ledwith}\ \emph {et~al.}(2023)\citenamefont
  {Ledwith}, \citenamefont {Vishwanath},\ and\ \citenamefont
  {Parker}}]{Ledwith2023}%
  \BibitemOpen
  \bibfield  {author} {\bibinfo {author} {\bibfnamefont {P.~J.}\ \bibnamefont
  {Ledwith}}, \bibinfo {author} {\bibfnamefont {A.}~\bibnamefont {Vishwanath}},
  \ and\ \bibinfo {author} {\bibfnamefont {D.~E.}\ \bibnamefont {Parker}},\
  }\href {\doibase 10.1103/PhysRevB.108.205144} {\bibfield  {journal} {\bibinfo
   {journal} {Phys. Rev. B}\ }\textbf {\bibinfo {volume} {108}},\ \bibinfo
  {pages} {205144} (\bibinfo {year} {2023})}\BibitemShut {NoStop}%
\bibitem [{\citenamefont {Liu}\ and\ \citenamefont
  {Bergholtz}(2024)}]{LIU2024515}%
  \BibitemOpen
  \bibfield  {author} {\bibinfo {author} {\bibfnamefont {Z.}~\bibnamefont
  {Liu}}\ and\ \bibinfo {author} {\bibfnamefont {E.~J.}\ \bibnamefont
  {Bergholtz}},\ }in\ \href {\doibase
  https://doi.org/10.1016/B978-0-323-90800-9.00136-0} {\emph {\bibinfo
  {booktitle} {Encyclopedia of Condensed Matter Physics (Second Edition)}}},\
  \bibinfo {editor} {edited by\ \bibinfo {editor} {\bibfnamefont
  {T.}~\bibnamefont {Chakraborty}}}\ (\bibinfo  {publisher} {Academic Press},\
  \bibinfo {address} {Oxford},\ \bibinfo {year} {2024})\ \bibinfo {edition}
  {second edition}\ ed.,\ pp.\ \bibinfo {pages} {515--538}\BibitemShut
  {NoStop}%
\bibitem [{\citenamefont {T{\"o}rm{\"a}}\ \emph {et~al.}(2022)\citenamefont
  {T{\"o}rm{\"a}}, \citenamefont {Peotta},\ and\ \citenamefont
  {Bernevig}}]{torma2022superconductivity}%
  \BibitemOpen
  \bibfield  {author} {\bibinfo {author} {\bibfnamefont {P.}~\bibnamefont
  {T{\"o}rm{\"a}}}, \bibinfo {author} {\bibfnamefont {S.}~\bibnamefont
  {Peotta}}, \ and\ \bibinfo {author} {\bibfnamefont {B.~A.}\ \bibnamefont
  {Bernevig}},\ }\href@noop {} {\bibfield  {journal} {\bibinfo  {journal}
  {Nature Reviews Physics}\ }\textbf {\bibinfo {volume} {4}},\ \bibinfo {pages}
  {528} (\bibinfo {year} {2022})}\BibitemShut {NoStop}%
\bibitem [{\citenamefont {Huhtinen}\ \emph {et~al.}(2022)\citenamefont
  {Huhtinen}, \citenamefont {Herzog-Arbeitman}, \citenamefont {Chew},
  \citenamefont {Bernevig},\ and\ \citenamefont {T\"orm\"a}}]{Jonah2022}%
  \BibitemOpen
  \bibfield  {author} {\bibinfo {author} {\bibfnamefont {K.-E.}\ \bibnamefont
  {Huhtinen}}, \bibinfo {author} {\bibfnamefont {J.}~\bibnamefont
  {Herzog-Arbeitman}}, \bibinfo {author} {\bibfnamefont {A.}~\bibnamefont
  {Chew}}, \bibinfo {author} {\bibfnamefont {B.~A.}\ \bibnamefont {Bernevig}},
  \ and\ \bibinfo {author} {\bibfnamefont {P.}~\bibnamefont {T\"orm\"a}},\
  }\href {\doibase 10.1103/PhysRevB.106.014518} {\bibfield  {journal} {\bibinfo
   {journal} {Phys. Rev. B}\ }\textbf {\bibinfo {volume} {106}},\ \bibinfo
  {pages} {014518} (\bibinfo {year} {2022})}\BibitemShut {NoStop}%
\bibitem [{\citenamefont {Herzog-Arbeitman}\ \emph
  {et~al.}(2022{\natexlab{b}})\citenamefont {Herzog-Arbeitman}, \citenamefont
  {Chew}, \citenamefont {Huhtinen}, \citenamefont {T{\"o}rm{\"a}},\ and\
  \citenamefont {Bernevig}}]{2022arXiv220900007H}%
  \BibitemOpen
  \bibfield  {author} {\bibinfo {author} {\bibfnamefont {J.}~\bibnamefont
  {Herzog-Arbeitman}}, \bibinfo {author} {\bibfnamefont {A.}~\bibnamefont
  {Chew}}, \bibinfo {author} {\bibfnamefont {K.-E.}\ \bibnamefont {Huhtinen}},
  \bibinfo {author} {\bibfnamefont {P.}~\bibnamefont {T{\"o}rm{\"a}}}, \ and\
  \bibinfo {author} {\bibfnamefont {B.~A.}\ \bibnamefont {Bernevig}},\ }\href
  {https://arxiv.org/pdf/2209.00007} {\bibfield  {journal} {\bibinfo  {journal}
  {arXiv preprint arXiv:2209.00007}\ } (\bibinfo {year}
  {2022}{\natexlab{b}})}\BibitemShut {NoStop}%
\bibitem [{\citenamefont {Hofmann}\ \emph {et~al.}(2020)\citenamefont
  {Hofmann}, \citenamefont {Berg},\ and\ \citenamefont
  {Chowdhury}}]{PhysRevB.102.201112}%
  \BibitemOpen
  \bibfield  {author} {\bibinfo {author} {\bibfnamefont {J.~S.}\ \bibnamefont
  {Hofmann}}, \bibinfo {author} {\bibfnamefont {E.}~\bibnamefont {Berg}}, \
  and\ \bibinfo {author} {\bibfnamefont {D.}~\bibnamefont {Chowdhury}},\ }\href
  {\doibase 10.1103/PhysRevB.102.201112} {\bibfield  {journal} {\bibinfo
  {journal} {Phys. Rev. B}\ }\textbf {\bibinfo {volume} {102}},\ \bibinfo
  {pages} {201112} (\bibinfo {year} {2020})}\BibitemShut {NoStop}%
\bibitem [{\citenamefont {Hofmann}\ \emph {et~al.}(2023)\citenamefont
  {Hofmann}, \citenamefont {Berg},\ and\ \citenamefont
  {Chowdhury}}]{PhysRevLett.130.226001}%
  \BibitemOpen
  \bibfield  {author} {\bibinfo {author} {\bibfnamefont {J.~S.}\ \bibnamefont
  {Hofmann}}, \bibinfo {author} {\bibfnamefont {E.}~\bibnamefont {Berg}}, \
  and\ \bibinfo {author} {\bibfnamefont {D.}~\bibnamefont {Chowdhury}},\ }\href
  {\doibase 10.1103/PhysRevLett.130.226001} {\bibfield  {journal} {\bibinfo
  {journal} {Phys. Rev. Lett.}\ }\textbf {\bibinfo {volume} {130}},\ \bibinfo
  {pages} {226001} (\bibinfo {year} {2023})}\BibitemShut {NoStop}%
\bibitem [{\citenamefont {Chen}\ and\ \citenamefont {Law}(2024)}]{KTLaw2024}%
  \BibitemOpen
  \bibfield  {author} {\bibinfo {author} {\bibfnamefont {S.~A.}\ \bibnamefont
  {Chen}}\ and\ \bibinfo {author} {\bibfnamefont {K.~T.}\ \bibnamefont {Law}},\
  }\href {\doibase 10.1103/PhysRevLett.132.026002} {\bibfield  {journal}
  {\bibinfo  {journal} {Phys. Rev. Lett.}\ }\textbf {\bibinfo {volume} {132}},\
  \bibinfo {pages} {026002} (\bibinfo {year} {2024})}\BibitemShut {NoStop}%
\bibitem [{\citenamefont {Gao}\ \emph {et~al.}(2023)\citenamefont {Gao},
  \citenamefont {Liu}, \citenamefont {Qiu}, \citenamefont {Ghosh},
  \citenamefont {V.~Trevisan}, \citenamefont {Onishi}, \citenamefont {Hu},
  \citenamefont {Qian}, \citenamefont {Tien}, \citenamefont {Chen} \emph
  {et~al.}}]{gao2023quantum}%
  \BibitemOpen
  \bibfield  {author} {\bibinfo {author} {\bibfnamefont {A.}~\bibnamefont
  {Gao}}, \bibinfo {author} {\bibfnamefont {Y.-F.}\ \bibnamefont {Liu}},
  \bibinfo {author} {\bibfnamefont {J.-X.}\ \bibnamefont {Qiu}}, \bibinfo
  {author} {\bibfnamefont {B.}~\bibnamefont {Ghosh}}, \bibinfo {author}
  {\bibfnamefont {T.}~\bibnamefont {V.~Trevisan}}, \bibinfo {author}
  {\bibfnamefont {Y.}~\bibnamefont {Onishi}}, \bibinfo {author} {\bibfnamefont
  {C.}~\bibnamefont {Hu}}, \bibinfo {author} {\bibfnamefont {T.}~\bibnamefont
  {Qian}}, \bibinfo {author} {\bibfnamefont {H.-J.}\ \bibnamefont {Tien}},
  \bibinfo {author} {\bibfnamefont {S.-W.}\ \bibnamefont {Chen}},  \emph
  {et~al.},\ }\href@noop {} {\bibfield  {journal} {\bibinfo  {journal}
  {Science}\ }\textbf {\bibinfo {volume} {381}},\ \bibinfo {pages} {181}
  (\bibinfo {year} {2023})}\BibitemShut {NoStop}%
\bibitem [{\citenamefont {Kaplan}\ \emph {et~al.}(2024)\citenamefont {Kaplan},
  \citenamefont {Holder},\ and\ \citenamefont {Yan}}]{Kaplan2024}%
  \BibitemOpen
  \bibfield  {author} {\bibinfo {author} {\bibfnamefont {D.}~\bibnamefont
  {Kaplan}}, \bibinfo {author} {\bibfnamefont {T.}~\bibnamefont {Holder}}, \
  and\ \bibinfo {author} {\bibfnamefont {B.}~\bibnamefont {Yan}},\ }\href
  {\doibase 10.1103/PhysRevLett.132.026301} {\bibfield  {journal} {\bibinfo
  {journal} {Phys. Rev. Lett.}\ }\textbf {\bibinfo {volume} {132}},\ \bibinfo
  {pages} {026301} (\bibinfo {year} {2024})}\BibitemShut {NoStop}%
\bibitem [{\citenamefont {Fang}\ \emph {et~al.}(2023)\citenamefont {Fang},
  \citenamefont {Cano},\ and\ \citenamefont
  {Ghorashi}}]{fang2023quantumgeometryinducednonlinear}%
  \BibitemOpen
  \bibfield  {author} {\bibinfo {author} {\bibfnamefont {Y.}~\bibnamefont
  {Fang}}, \bibinfo {author} {\bibfnamefont {J.}~\bibnamefont {Cano}}, \ and\
  \bibinfo {author} {\bibfnamefont {S.~A.~A.}\ \bibnamefont {Ghorashi}},\
  }\href {https://arxiv.org/abs/2310.11489} {\enquote {\bibinfo {title}
  {Quantum geometry induced nonlinear transport in altermagnets},}\ } (\bibinfo
  {year} {2023}),\ \Eprint {http://arxiv.org/abs/2310.11489} {arXiv:2310.11489
  [cond-mat.mes-hall]} \BibitemShut {NoStop}%
\bibitem [{\citenamefont {Verma}\ \emph {et~al.}(2024)\citenamefont {Verma},
  \citenamefont {Guerci},\ and\ \citenamefont
  {Queiroz}}]{PhysRevLett.132.236001}%
  \BibitemOpen
  \bibfield  {author} {\bibinfo {author} {\bibfnamefont {N.}~\bibnamefont
  {Verma}}, \bibinfo {author} {\bibfnamefont {D.}~\bibnamefont {Guerci}}, \
  and\ \bibinfo {author} {\bibfnamefont {R.}~\bibnamefont {Queiroz}},\ }\href
  {\doibase 10.1103/PhysRevLett.132.236001} {\bibfield  {journal} {\bibinfo
  {journal} {Phys. Rev. Lett.}\ }\textbf {\bibinfo {volume} {132}},\ \bibinfo
  {pages} {236001} (\bibinfo {year} {2024})}\BibitemShut {NoStop}%
\bibitem [{\citenamefont {{Kang}}\ \emph {et~al.}(2024)\citenamefont {{Kang}},
  \citenamefont {{Oh}}, \citenamefont {{Lee}},\ and\ \citenamefont
  {{Yang}}}]{2024arXiv240207171K}%
  \BibitemOpen
  \bibfield  {author} {\bibinfo {author} {\bibfnamefont {J.}~\bibnamefont
  {{Kang}}}, \bibinfo {author} {\bibfnamefont {T.}~\bibnamefont {{Oh}}},
  \bibinfo {author} {\bibfnamefont {J.}~\bibnamefont {{Lee}}}, \ and\ \bibinfo
  {author} {\bibfnamefont {B.-J.}\ \bibnamefont {{Yang}}},\ }\href {\doibase
  10.48550/arXiv.2402.07171} {\bibfield  {journal} {\bibinfo  {journal} {arXiv
  e-prints}\ ,\ \bibinfo {eid} {arXiv:2402.07171}} (\bibinfo {year} {2024})},\
  \Eprint {http://arxiv.org/abs/2402.07171} {arXiv:2402.07171
  [cond-mat.str-el]} \BibitemShut {NoStop}%
\bibitem [{\citenamefont {Yu}\ \emph {et~al.}(2024)\citenamefont {Yu},
  \citenamefont {Ciccarino}, \citenamefont {Bianco}, \citenamefont {Errea},
  \citenamefont {Narang},\ and\ \citenamefont {Bernevig}}]{Yu2024EPC}%
  \BibitemOpen
  \bibfield  {author} {\bibinfo {author} {\bibfnamefont {J.}~\bibnamefont
  {Yu}}, \bibinfo {author} {\bibfnamefont {C.~J.}\ \bibnamefont {Ciccarino}},
  \bibinfo {author} {\bibfnamefont {R.}~\bibnamefont {Bianco}}, \bibinfo
  {author} {\bibfnamefont {I.}~\bibnamefont {Errea}}, \bibinfo {author}
  {\bibfnamefont {P.}~\bibnamefont {Narang}}, \ and\ \bibinfo {author}
  {\bibfnamefont {B.~A.}\ \bibnamefont {Bernevig}},\ }\href
  {https://doi.org/10.1038/s41567-024-02486-0} {\bibfield  {journal} {\bibinfo
  {journal} {Nature Physics}\ } (\bibinfo {year} {2024})}\BibitemShut {NoStop}%
\bibitem [{\citenamefont {Neupert}\ \emph {et~al.}(2013)\citenamefont
  {Neupert}, \citenamefont {Chamon},\ and\ \citenamefont
  {Mudry}}]{Neupert2013}%
  \BibitemOpen
  \bibfield  {author} {\bibinfo {author} {\bibfnamefont {T.}~\bibnamefont
  {Neupert}}, \bibinfo {author} {\bibfnamefont {C.}~\bibnamefont {Chamon}}, \
  and\ \bibinfo {author} {\bibfnamefont {C.}~\bibnamefont {Mudry}},\ }\href
  {\doibase 10.1103/PhysRevB.87.245103} {\bibfield  {journal} {\bibinfo
  {journal} {Phys. Rev. B}\ }\textbf {\bibinfo {volume} {87}},\ \bibinfo
  {pages} {245103} (\bibinfo {year} {2013})}\BibitemShut {NoStop}%
\bibitem [{\citenamefont {Ozawa}\ and\ \citenamefont
  {Goldman}(2018)}]{Ozawa2018}%
  \BibitemOpen
  \bibfield  {author} {\bibinfo {author} {\bibfnamefont {T.}~\bibnamefont
  {Ozawa}}\ and\ \bibinfo {author} {\bibfnamefont {N.}~\bibnamefont
  {Goldman}},\ }\href {\doibase 10.1103/PhysRevB.97.201117} {\bibfield
  {journal} {\bibinfo  {journal} {Phys. Rev. B}\ }\textbf {\bibinfo {volume}
  {97}},\ \bibinfo {pages} {201117} (\bibinfo {year} {2018})}\BibitemShut
  {NoStop}%
\bibitem [{\citenamefont {Gianfrate}\ \emph {et~al.}(2020)\citenamefont
  {Gianfrate}, \citenamefont {Bleu}, \citenamefont {Dominici}, \citenamefont
  {Ardizzone}, \citenamefont {De~Giorgi}, \citenamefont {Ballarini},
  \citenamefont {Lerario}, \citenamefont {West}, \citenamefont {Pfeiffer},
  \citenamefont {Solnyshkov} \emph {et~al.}}]{gianfrate2020measurement}%
  \BibitemOpen
  \bibfield  {author} {\bibinfo {author} {\bibfnamefont {A.}~\bibnamefont
  {Gianfrate}}, \bibinfo {author} {\bibfnamefont {O.}~\bibnamefont {Bleu}},
  \bibinfo {author} {\bibfnamefont {L.}~\bibnamefont {Dominici}}, \bibinfo
  {author} {\bibfnamefont {V.}~\bibnamefont {Ardizzone}}, \bibinfo {author}
  {\bibfnamefont {M.}~\bibnamefont {De~Giorgi}}, \bibinfo {author}
  {\bibfnamefont {D.}~\bibnamefont {Ballarini}}, \bibinfo {author}
  {\bibfnamefont {G.}~\bibnamefont {Lerario}}, \bibinfo {author} {\bibfnamefont
  {K.}~\bibnamefont {West}}, \bibinfo {author} {\bibfnamefont {L.}~\bibnamefont
  {Pfeiffer}}, \bibinfo {author} {\bibfnamefont {D.}~\bibnamefont
  {Solnyshkov}},  \emph {et~al.},\ }\href@noop {} {\bibfield  {journal}
  {\bibinfo  {journal} {Nature}\ }\textbf {\bibinfo {volume} {578}},\ \bibinfo
  {pages} {381} (\bibinfo {year} {2020})}\BibitemShut {NoStop}%
\bibitem [{\citenamefont {Ahn}\ \emph {et~al.}(2022)\citenamefont {Ahn},
  \citenamefont {Guo}, \citenamefont {Nagaosa},\ and\ \citenamefont
  {Vishwanath}}]{ahn2022riemannian}%
  \BibitemOpen
  \bibfield  {author} {\bibinfo {author} {\bibfnamefont {J.}~\bibnamefont
  {Ahn}}, \bibinfo {author} {\bibfnamefont {G.-Y.}\ \bibnamefont {Guo}},
  \bibinfo {author} {\bibfnamefont {N.}~\bibnamefont {Nagaosa}}, \ and\
  \bibinfo {author} {\bibfnamefont {A.}~\bibnamefont {Vishwanath}},\ }\href
  {https://www.nature.com/articles/s41567-021-01465-z} {\bibfield  {journal}
  {\bibinfo  {journal} {Nature Physics}\ }\textbf {\bibinfo {volume} {18}},\
  \bibinfo {pages} {290} (\bibinfo {year} {2022})}\BibitemShut {NoStop}%
\bibitem [{\citenamefont {de~Sousa}\ \emph {et~al.}(2023)\citenamefont
  {de~Sousa}, \citenamefont {Cruz},\ and\ \citenamefont {Chen}}]{deSousa2023}%
  \BibitemOpen
  \bibfield  {author} {\bibinfo {author} {\bibfnamefont {M.~S.~M.}\
  \bibnamefont {de~Sousa}}, \bibinfo {author} {\bibfnamefont {A.~L.}\
  \bibnamefont {Cruz}}, \ and\ \bibinfo {author} {\bibfnamefont
  {W.}~\bibnamefont {Chen}},\ }\href {\doibase 10.1103/PhysRevB.107.205133}
  {\bibfield  {journal} {\bibinfo  {journal} {Phys. Rev. B}\ }\textbf {\bibinfo
  {volume} {107}},\ \bibinfo {pages} {205133} (\bibinfo {year}
  {2023})}\BibitemShut {NoStop}%
\bibitem [{\citenamefont {Onishi}\ and\ \citenamefont
  {Fu}(2024{\natexlab{b}})}]{onishi2024fundamental}%
  \BibitemOpen
  \bibfield  {author} {\bibinfo {author} {\bibfnamefont {Y.}~\bibnamefont
  {Onishi}}\ and\ \bibinfo {author} {\bibfnamefont {L.}~\bibnamefont {Fu}},\
  }\href {\doibase 10.1103/PhysRevX.14.011052} {\bibfield  {journal} {\bibinfo
  {journal} {Phys. Rev. X}\ }\textbf {\bibinfo {volume} {14}},\ \bibinfo
  {pages} {011052} (\bibinfo {year} {2024}{\natexlab{b}})}\BibitemShut
  {NoStop}%
\bibitem [{\citenamefont {Komissarov}\ \emph {et~al.}(2023)\citenamefont
  {Komissarov}, \citenamefont {Holder},\ and\ \citenamefont
  {Queiroz}}]{komissarov2023quantumgeometricorigincapacitance}%
  \BibitemOpen
  \bibfield  {author} {\bibinfo {author} {\bibfnamefont {I.}~\bibnamefont
  {Komissarov}}, \bibinfo {author} {\bibfnamefont {T.}~\bibnamefont {Holder}},
  \ and\ \bibinfo {author} {\bibfnamefont {R.}~\bibnamefont {Queiroz}},\ }\href
  {https://arxiv.org/abs/2306.08035} {\enquote {\bibinfo {title} {The quantum
  geometric origin of capacitance in insulators},}\ } (\bibinfo {year}
  {2023}),\ \Eprint {http://arxiv.org/abs/2306.08035} {arXiv:2306.08035
  [cond-mat.mes-hall]} \BibitemShut {NoStop}%
\bibitem [{\citenamefont {Kruchkov}\ and\ \citenamefont
  {Ryu}(2023{\natexlab{a}})}]{kruchkov2023noise}%
  \BibitemOpen
  \bibfield  {author} {\bibinfo {author} {\bibfnamefont {A.}~\bibnamefont
  {Kruchkov}}\ and\ \bibinfo {author} {\bibfnamefont {S.}~\bibnamefont {Ryu}},\
  }\href {https://arxiv.org/abs/2309.00042} {\bibfield  {journal} {\bibinfo
  {journal} {arXiv preprint arXiv:2309.00042}\ } (\bibinfo {year}
  {2023}{\natexlab{a}})}\BibitemShut {NoStop}%
\bibitem [{\citenamefont {Kruchkov}\ and\ \citenamefont
  {Ryu}(2023{\natexlab{b}})}]{kruchkov2023spectral}%
  \BibitemOpen
  \bibfield  {author} {\bibinfo {author} {\bibfnamefont {A.}~\bibnamefont
  {Kruchkov}}\ and\ \bibinfo {author} {\bibfnamefont {S.}~\bibnamefont {Ryu}},\
  }\href {https://arxiv.org/abs/2312.17318} {\bibfield  {journal} {\bibinfo
  {journal} {arXiv preprint arXiv:2312.17318}\ } (\bibinfo {year}
  {2023}{\natexlab{b}})}\BibitemShut {NoStop}%
\bibitem [{\citenamefont {Onishi}\ and\ \citenamefont
  {Fu}(2024{\natexlab{c}})}]{onishi2024quantum}%
  \BibitemOpen
  \bibfield  {author} {\bibinfo {author} {\bibfnamefont {Y.}~\bibnamefont
  {Onishi}}\ and\ \bibinfo {author} {\bibfnamefont {L.}~\bibnamefont {Fu}},\
  }\href@noop {} {\  (\bibinfo {year} {2024}{\natexlab{c}})},\ \Eprint
  {http://arxiv.org/abs/2406.06783} {arXiv:2406.06783 [cond-mat.str-el]}
  \BibitemShut {NoStop}%
\bibitem [{\citenamefont {{Verma}}\ and\ \citenamefont
  {{Queiroz}}(2024)}]{2024arXiv240307052V}%
  \BibitemOpen
  \bibfield  {author} {\bibinfo {author} {\bibfnamefont {N.}~\bibnamefont
  {{Verma}}}\ and\ \bibinfo {author} {\bibfnamefont {R.}~\bibnamefont
  {{Queiroz}}},\ }\href {\doibase 10.48550/arXiv.2403.07052} {\bibfield
  {journal} {\bibinfo  {journal} {arXiv e-prints}\ ,\ \bibinfo {eid}
  {arXiv:2403.07052}} (\bibinfo {year} {2024})},\ \Eprint
  {http://arxiv.org/abs/2403.07052} {arXiv:2403.07052 [cond-mat.mes-hall]}
  \BibitemShut {NoStop}%
\bibitem [{\citenamefont {Ryu}\ and\ \citenamefont {Hatsugai}(2006)}]{Ryu2006}%
  \BibitemOpen
  \bibfield  {author} {\bibinfo {author} {\bibfnamefont {S.}~\bibnamefont
  {Ryu}}\ and\ \bibinfo {author} {\bibfnamefont {Y.}~\bibnamefont {Hatsugai}},\
  }\href {\doibase 10.1103/PhysRevB.73.245115} {\bibfield  {journal} {\bibinfo
  {journal} {Phys. Rev. B}\ }\textbf {\bibinfo {volume} {73}},\ \bibinfo
  {pages} {245115} (\bibinfo {year} {2006})}\BibitemShut {NoStop}%
\bibitem [{\citenamefont {Paul}(2024)}]{nisarga2024}%
  \BibitemOpen
  \bibfield  {author} {\bibinfo {author} {\bibfnamefont {N.}~\bibnamefont
  {Paul}},\ }\href {\doibase 10.1103/PhysRevB.109.085146} {\bibfield  {journal}
  {\bibinfo  {journal} {Phys. Rev. B}\ }\textbf {\bibinfo {volume} {109}},\
  \bibinfo {pages} {085146} (\bibinfo {year} {2024})}\BibitemShut {NoStop}%
\bibitem [{\citenamefont {Klich}\ and\ \citenamefont
  {Levitov}(2009)}]{klich2009}%
  \BibitemOpen
  \bibfield  {author} {\bibinfo {author} {\bibfnamefont {I.}~\bibnamefont
  {Klich}}\ and\ \bibinfo {author} {\bibfnamefont {L.}~\bibnamefont
  {Levitov}},\ }\href {\doibase 10.1103/PhysRevLett.102.100502} {\bibfield
  {journal} {\bibinfo  {journal} {Phys. Rev. Lett.}\ }\textbf {\bibinfo
  {volume} {102}},\ \bibinfo {pages} {100502} (\bibinfo {year}
  {2009})}\BibitemShut {NoStop}%
\bibitem [{\citenamefont {Calabrese}\ \emph {et~al.}(2012)\citenamefont
  {Calabrese}, \citenamefont {Mintchev},\ and\ \citenamefont
  {Vicari}}]{calabrese2012exact}%
  \BibitemOpen
  \bibfield  {author} {\bibinfo {author} {\bibfnamefont {P.}~\bibnamefont
  {Calabrese}}, \bibinfo {author} {\bibfnamefont {M.}~\bibnamefont {Mintchev}},
  \ and\ \bibinfo {author} {\bibfnamefont {E.}~\bibnamefont {Vicari}},\
  }\href@noop {} {\bibfield  {journal} {\bibinfo  {journal} {Europhysics
  Letters}\ }\textbf {\bibinfo {volume} {98}},\ \bibinfo {pages} {20003}
  (\bibinfo {year} {2012})}\BibitemShut {NoStop}%
\bibitem [{\citenamefont {Song}\ \emph {et~al.}(2012)\citenamefont {Song},
  \citenamefont {Rachel}, \citenamefont {Flindt}, \citenamefont {Klich},
  \citenamefont {Laflorencie},\ and\ \citenamefont {Le~Hur}}]{song2012}%
  \BibitemOpen
  \bibfield  {author} {\bibinfo {author} {\bibfnamefont {H.~F.}\ \bibnamefont
  {Song}}, \bibinfo {author} {\bibfnamefont {S.}~\bibnamefont {Rachel}},
  \bibinfo {author} {\bibfnamefont {C.}~\bibnamefont {Flindt}}, \bibinfo
  {author} {\bibfnamefont {I.}~\bibnamefont {Klich}}, \bibinfo {author}
  {\bibfnamefont {N.}~\bibnamefont {Laflorencie}}, \ and\ \bibinfo {author}
  {\bibfnamefont {K.}~\bibnamefont {Le~Hur}},\ }\href {\doibase
  10.1103/PhysRevB.85.035409} {\bibfield  {journal} {\bibinfo  {journal} {Phys.
  Rev. B}\ }\textbf {\bibinfo {volume} {85}},\ \bibinfo {pages} {035409}
  (\bibinfo {year} {2012})}\BibitemShut {NoStop}%
\bibitem [{\citenamefont {Rachel}\ \emph {et~al.}(2012)\citenamefont {Rachel},
  \citenamefont {Laflorencie}, \citenamefont {Song},\ and\ \citenamefont
  {Le~Hur}}]{Rachel2012}%
  \BibitemOpen
  \bibfield  {author} {\bibinfo {author} {\bibfnamefont {S.}~\bibnamefont
  {Rachel}}, \bibinfo {author} {\bibfnamefont {N.}~\bibnamefont {Laflorencie}},
  \bibinfo {author} {\bibfnamefont {H.~F.}\ \bibnamefont {Song}}, \ and\
  \bibinfo {author} {\bibfnamefont {K.}~\bibnamefont {Le~Hur}},\ }\href
  {\doibase 10.1103/PhysRevLett.108.116401} {\bibfield  {journal} {\bibinfo
  {journal} {Phys. Rev. Lett.}\ }\textbf {\bibinfo {volume} {108}},\ \bibinfo
  {pages} {116401} (\bibinfo {year} {2012})}\BibitemShut {NoStop}%
\bibitem [{\citenamefont {Herviou}\ \emph {et~al.}(2019)\citenamefont
  {Herviou}, \citenamefont {Le~Hur},\ and\ \citenamefont {Mora}}]{Herviou2019}%
  \BibitemOpen
  \bibfield  {author} {\bibinfo {author} {\bibfnamefont {L.}~\bibnamefont
  {Herviou}}, \bibinfo {author} {\bibfnamefont {K.}~\bibnamefont {Le~Hur}}, \
  and\ \bibinfo {author} {\bibfnamefont {C.}~\bibnamefont {Mora}},\ }\href
  {\doibase 10.1103/PhysRevB.99.075133} {\bibfield  {journal} {\bibinfo
  {journal} {Phys. Rev. B}\ }\textbf {\bibinfo {volume} {99}},\ \bibinfo
  {pages} {075133} (\bibinfo {year} {2019})}\BibitemShut {NoStop}%
\bibitem [{\citenamefont {Wu}\ \emph {et~al.}(2021)\citenamefont {Wu},
  \citenamefont {Jian},\ and\ \citenamefont {Xu}}]{wu2021universal}%
  \BibitemOpen
  \bibfield  {author} {\bibinfo {author} {\bibfnamefont {X.-C.}\ \bibnamefont
  {Wu}}, \bibinfo {author} {\bibfnamefont {C.-M.}\ \bibnamefont {Jian}}, \ and\
  \bibinfo {author} {\bibfnamefont {C.}~\bibnamefont {Xu}},\ }\href@noop {}
  {\bibfield  {journal} {\bibinfo  {journal} {SciPost Physics}\ }\textbf
  {\bibinfo {volume} {11}},\ \bibinfo {pages} {033} (\bibinfo {year}
  {2021})}\BibitemShut {NoStop}%
\bibitem [{\citenamefont {Wang}\ \emph
  {et~al.}(2021{\natexlab{b}})\citenamefont {Wang}, \citenamefont {Cheng},\
  and\ \citenamefont {Meng}}]{Meng2021}%
  \BibitemOpen
  \bibfield  {author} {\bibinfo {author} {\bibfnamefont {Y.-C.}\ \bibnamefont
  {Wang}}, \bibinfo {author} {\bibfnamefont {M.}~\bibnamefont {Cheng}}, \ and\
  \bibinfo {author} {\bibfnamefont {Z.~Y.}\ \bibnamefont {Meng}},\ }\href
  {\doibase 10.1103/PhysRevB.104.L081109} {\bibfield  {journal} {\bibinfo
  {journal} {Phys. Rev. B}\ }\textbf {\bibinfo {volume} {104}},\ \bibinfo
  {pages} {L081109} (\bibinfo {year} {2021}{\natexlab{b}})}\BibitemShut
  {NoStop}%
\bibitem [{\citenamefont {Estienne}\ \emph {et~al.}(2022)\citenamefont
  {Estienne}, \citenamefont {St{\'e}phan},\ and\ \citenamefont
  {Witczak-Krempa}}]{estienne2022cornering}%
  \BibitemOpen
  \bibfield  {author} {\bibinfo {author} {\bibfnamefont {B.}~\bibnamefont
  {Estienne}}, \bibinfo {author} {\bibfnamefont {J.-M.}\ \bibnamefont
  {St{\'e}phan}}, \ and\ \bibinfo {author} {\bibfnamefont {W.}~\bibnamefont
  {Witczak-Krempa}},\ }\href
  {https://www.nature.com/articles/s41467-021-27727-1} {\bibfield  {journal}
  {\bibinfo  {journal} {Nature Communications}\ }\textbf {\bibinfo {volume}
  {13}},\ \bibinfo {pages} {287} (\bibinfo {year} {2022})}\BibitemShut
  {NoStop}%
\bibitem [{\citenamefont {Wu}(2024)}]{wu2024bipartite}%
  \BibitemOpen
  \bibfield  {author} {\bibinfo {author} {\bibfnamefont {X.-C.}\ \bibnamefont
  {Wu}},\ }\href {https://arxiv.org/abs/2404.04331} {\bibfield  {journal}
  {\bibinfo  {journal} {arXiv preprint arXiv:2404.04331}\ } (\bibinfo {year}
  {2024})}\BibitemShut {NoStop}%
\bibitem [{\citenamefont {Tam}\ \emph {et~al.}(2022)\citenamefont {Tam},
  \citenamefont {Claassen},\ and\ \citenamefont {Kane}}]{tamkane2022}%
  \BibitemOpen
  \bibfield  {author} {\bibinfo {author} {\bibfnamefont {P.~M.}\ \bibnamefont
  {Tam}}, \bibinfo {author} {\bibfnamefont {M.}~\bibnamefont {Claassen}}, \
  and\ \bibinfo {author} {\bibfnamefont {C.~L.}\ \bibnamefont {Kane}},\ }\href
  {\doibase 10.1103/PhysRevX.12.031022} {\bibfield  {journal} {\bibinfo
  {journal} {Phys. Rev. X}\ }\textbf {\bibinfo {volume} {12}},\ \bibinfo
  {pages} {031022} (\bibinfo {year} {2022})}\BibitemShut {NoStop}%
\bibitem [{\citenamefont {Tam}\ and\ \citenamefont {Kane}(2024)}]{tamkane2024}%
  \BibitemOpen
  \bibfield  {author} {\bibinfo {author} {\bibfnamefont {P.~M.}\ \bibnamefont
  {Tam}}\ and\ \bibinfo {author} {\bibfnamefont {C.~L.}\ \bibnamefont {Kane}},\
  }\href {\doibase 10.1103/PhysRevB.109.035413} {\bibfield  {journal} {\bibinfo
   {journal} {Phys. Rev. B}\ }\textbf {\bibinfo {volume} {109}},\ \bibinfo
  {pages} {035413} (\bibinfo {year} {2024})}\BibitemShut {NoStop}%
\bibitem [{\citenamefont {Haldane}(2014)}]{haldane2014attachment}%
  \BibitemOpen
  \bibfield  {author} {\bibinfo {author} {\bibfnamefont {F.}~\bibnamefont
  {Haldane}},\ }\href {https://arxiv.org/abs/1401.0529} {\bibfield  {journal}
  {\bibinfo  {journal} {arXiv preprint arXiv:1401.0529}\ } (\bibinfo {year}
  {2014})}\BibitemShut {NoStop}%
\bibitem [{\citenamefont {Lim}\ \emph {et~al.}(2015)\citenamefont {Lim},
  \citenamefont {Fuchs},\ and\ \citenamefont {Montambaux}}]{Lim2015}%
  \BibitemOpen
  \bibfield  {author} {\bibinfo {author} {\bibfnamefont {L.-K.}\ \bibnamefont
  {Lim}}, \bibinfo {author} {\bibfnamefont {J.-N.}\ \bibnamefont {Fuchs}}, \
  and\ \bibinfo {author} {\bibfnamefont {G.}~\bibnamefont {Montambaux}},\
  }\href {\doibase 10.1103/PhysRevA.92.063627} {\bibfield  {journal} {\bibinfo
  {journal} {Phys. Rev. A}\ }\textbf {\bibinfo {volume} {92}},\ \bibinfo
  {pages} {063627} (\bibinfo {year} {2015})}\BibitemShut {NoStop}%
\bibitem [{\citenamefont {Simon}\ and\ \citenamefont
  {Rudner}(2020)}]{Simon2020}%
  \BibitemOpen
  \bibfield  {author} {\bibinfo {author} {\bibfnamefont {S.~H.}\ \bibnamefont
  {Simon}}\ and\ \bibinfo {author} {\bibfnamefont {M.~S.}\ \bibnamefont
  {Rudner}},\ }\href {\doibase 10.1103/PhysRevB.102.165148} {\bibfield
  {journal} {\bibinfo  {journal} {Phys. Rev. B}\ }\textbf {\bibinfo {volume}
  {102}},\ \bibinfo {pages} {165148} (\bibinfo {year} {2020})}\BibitemShut
  {NoStop}%
\bibitem [{Note1()}]{Note1}%
  \BibitemOpen
  \bibinfo {note} {Another commonly used orbital embedding is to Fourier
  transform as $\rho _{\protect \bf q}= \DOTSB \sum@ \slimits@ _{{\protect \bf
  R},\sigma } e^{-i{\protect \bf q}\cdot ({\protect \bf R}+{\protect \bf
  r}_\sigma )} c^\dagger _{{\protect \bf R},\sigma } c_{{\protect \bf R},\sigma
  }$, where ${\protect \bf R}+{\protect \bf r}_\sigma $ is the physical
  position of the orbital. This is not used in the main text, so we shall
  discuss it only in the supplementary.}\BibitemShut {Stop}%
\bibitem [{Note2()}]{Note2}%
  \BibitemOpen
  \bibinfo {note} {To be concrete, the twisted periodic boundary condition
  corresponds to $\Psi _{\protect \bf k}({\protect \bf r}_1,..., {\protect \bf
  r}_a+{\protect \bf L}, ..., {\protect \bf r}_N) = e^{i{\protect \bf k}\cdot
  {\protect \bf L}} \Psi _{\protect \bf k}({\protect \bf r}_1,..., {\protect
  \bf r}_a, ..., {\protect \bf r}_N)$, with $k_i \in [0,2\pi /L_i)$ and $L_i$
  is the period of the system in the $i$-th direction.}\BibitemShut {Stop}%
\bibitem [{\citenamefont {Kohn}(1964)}]{Kohn1964}%
  \BibitemOpen
  \bibfield  {author} {\bibinfo {author} {\bibfnamefont {W.}~\bibnamefont
  {Kohn}},\ }\href {\doibase 10.1103/PhysRev.133.A171} {\bibfield  {journal}
  {\bibinfo  {journal} {Phys. Rev.}\ }\textbf {\bibinfo {volume} {133}},\
  \bibinfo {pages} {A171} (\bibinfo {year} {1964})}\BibitemShut {NoStop}%
\bibitem [{sup()}]{supp}%
  \BibitemOpen
  \href@noop {} {}\bibinfo {note} {Supplementary Materials, which (1) review
  the relation between static structure factor and quantum geometry, (2)
  explain the analytical calculation for the obstructed atomic insulator model,
  (3) justifies the applicability of our work for any Bravais lattice, (4)
  discuss specific examples where quantum metric of the physical orbital
  embedding can be extracted, and (5) provide implementation details for the
  numerical studies.}\BibitemShut {Stop}%
\bibitem [{Note3()}]{Note3}%
  \BibitemOpen
  \bibinfo {note} {A simple proof of this statement is included in the
  Supplementary \cite {supp} for completeness. See also Refs. \cite
  {Regnault2013,onishi2024quantum}}\BibitemShut {NoStop}%
\bibitem [{\citenamefont {Bradlyn}\ \emph {et~al.}(2017)\citenamefont
  {Bradlyn}, \citenamefont {Elcoro}, \citenamefont {Cano}, \citenamefont
  {Vergniory}, \citenamefont {Wang}, \citenamefont {Felser}, \citenamefont
  {Aroyo},\ and\ \citenamefont {Bernevig}}]{bradlyn2017topological}%
  \BibitemOpen
  \bibfield  {author} {\bibinfo {author} {\bibfnamefont {B.}~\bibnamefont
  {Bradlyn}}, \bibinfo {author} {\bibfnamefont {L.}~\bibnamefont {Elcoro}},
  \bibinfo {author} {\bibfnamefont {J.}~\bibnamefont {Cano}}, \bibinfo {author}
  {\bibfnamefont {M.~G.}\ \bibnamefont {Vergniory}}, \bibinfo {author}
  {\bibfnamefont {Z.}~\bibnamefont {Wang}}, \bibinfo {author} {\bibfnamefont
  {C.}~\bibnamefont {Felser}}, \bibinfo {author} {\bibfnamefont {M.~I.}\
  \bibnamefont {Aroyo}}, \ and\ \bibinfo {author} {\bibfnamefont {B.~A.}\
  \bibnamefont {Bernevig}},\ }\href
  {https://www.nature.com/articles/nature23268} {\bibfield  {journal} {\bibinfo
   {journal} {Nature}\ }\textbf {\bibinfo {volume} {547}},\ \bibinfo {pages}
  {298} (\bibinfo {year} {2017})}\BibitemShut {NoStop}%
\bibitem [{\citenamefont {Schindler}\ and\ \citenamefont
  {Bernevig}(2021)}]{Schindler2021}%
  \BibitemOpen
  \bibfield  {author} {\bibinfo {author} {\bibfnamefont {F.}~\bibnamefont
  {Schindler}}\ and\ \bibinfo {author} {\bibfnamefont {B.~A.}\ \bibnamefont
  {Bernevig}},\ }\href {\doibase 10.1103/PhysRevB.104.L201114} {\bibfield
  {journal} {\bibinfo  {journal} {Phys. Rev. B}\ }\textbf {\bibinfo {volume}
  {104}},\ \bibinfo {pages} {L201114} (\bibinfo {year} {2021})}\BibitemShut
  {NoStop}%
\bibitem [{\citenamefont {Herzog-Arbeitman}\ \emph {et~al.}(2023)\citenamefont
  {Herzog-Arbeitman}, \citenamefont {Song}, \citenamefont {Elcoro},\ and\
  \citenamefont {Bernevig}}]{Jonah2023}%
  \BibitemOpen
  \bibfield  {author} {\bibinfo {author} {\bibfnamefont {J.}~\bibnamefont
  {Herzog-Arbeitman}}, \bibinfo {author} {\bibfnamefont {Z.-D.}\ \bibnamefont
  {Song}}, \bibinfo {author} {\bibfnamefont {L.}~\bibnamefont {Elcoro}}, \ and\
  \bibinfo {author} {\bibfnamefont {B.~A.}\ \bibnamefont {Bernevig}},\ }\href
  {\doibase 10.1103/PhysRevLett.130.236601} {\bibfield  {journal} {\bibinfo
  {journal} {Phys. Rev. Lett.}\ }\textbf {\bibinfo {volume} {130}},\ \bibinfo
  {pages} {236601} (\bibinfo {year} {2023})}\BibitemShut {NoStop}%
\bibitem [{\citenamefont {Qi}\ \emph {et~al.}(2006)\citenamefont {Qi},
  \citenamefont {Wu},\ and\ \citenamefont {Zhang}}]{QWZ2006}%
  \BibitemOpen
  \bibfield  {author} {\bibinfo {author} {\bibfnamefont {X.-L.}\ \bibnamefont
  {Qi}}, \bibinfo {author} {\bibfnamefont {Y.-S.}\ \bibnamefont {Wu}}, \ and\
  \bibinfo {author} {\bibfnamefont {S.-C.}\ \bibnamefont {Zhang}},\ }\href
  {\doibase 10.1103/PhysRevB.74.085308} {\bibfield  {journal} {\bibinfo
  {journal} {Phys. Rev. B}\ }\textbf {\bibinfo {volume} {74}},\ \bibinfo
  {pages} {085308} (\bibinfo {year} {2006})}\BibitemShut {NoStop}%
\bibitem [{\citenamefont {Haldane}(1988)}]{Haldane1988}%
  \BibitemOpen
  \bibfield  {author} {\bibinfo {author} {\bibfnamefont {F.~D.~M.}\
  \bibnamefont {Haldane}},\ }\href {\doibase 10.1103/PhysRevLett.61.2015}
  {\bibfield  {journal} {\bibinfo  {journal} {Phys. Rev. Lett.}\ }\textbf
  {\bibinfo {volume} {61}},\ \bibinfo {pages} {2015} (\bibinfo {year}
  {1988})}\BibitemShut {NoStop}%
\bibitem [{\citenamefont {Jotzu}\ \emph {et~al.}(2014)\citenamefont {Jotzu},
  \citenamefont {Messer}, \citenamefont {Desbuquois}, \citenamefont {Lebrat},
  \citenamefont {Uehlinger}, \citenamefont {Greif},\ and\ \citenamefont
  {Esslinger}}]{jotzu2014experimental}%
  \BibitemOpen
  \bibfield  {author} {\bibinfo {author} {\bibfnamefont {G.}~\bibnamefont
  {Jotzu}}, \bibinfo {author} {\bibfnamefont {M.}~\bibnamefont {Messer}},
  \bibinfo {author} {\bibfnamefont {R.}~\bibnamefont {Desbuquois}}, \bibinfo
  {author} {\bibfnamefont {M.}~\bibnamefont {Lebrat}}, \bibinfo {author}
  {\bibfnamefont {T.}~\bibnamefont {Uehlinger}}, \bibinfo {author}
  {\bibfnamefont {D.}~\bibnamefont {Greif}}, \ and\ \bibinfo {author}
  {\bibfnamefont {T.}~\bibnamefont {Esslinger}},\ }\href
  {https://www.nature.com/articles/nature13915} {\bibfield  {journal} {\bibinfo
   {journal} {Nature}\ }\textbf {\bibinfo {volume} {515}},\ \bibinfo {pages}
  {237} (\bibinfo {year} {2014})}\BibitemShut {NoStop}%
\bibitem [{\citenamefont {Liang}\ \emph {et~al.}(2023)\citenamefont {Liang},
  \citenamefont {Wei}, \citenamefont {Zhang}, \citenamefont {Wang},
  \citenamefont {Zhang}, \citenamefont {Wang}, \citenamefont {Qi},
  \citenamefont {Liu},\ and\ \citenamefont {Zhang}}]{Liang2023_QWZrealization}%
  \BibitemOpen
  \bibfield  {author} {\bibinfo {author} {\bibfnamefont {M.-C.}\ \bibnamefont
  {Liang}}, \bibinfo {author} {\bibfnamefont {Y.-D.}\ \bibnamefont {Wei}},
  \bibinfo {author} {\bibfnamefont {L.}~\bibnamefont {Zhang}}, \bibinfo
  {author} {\bibfnamefont {X.-J.}\ \bibnamefont {Wang}}, \bibinfo {author}
  {\bibfnamefont {H.}~\bibnamefont {Zhang}}, \bibinfo {author} {\bibfnamefont
  {W.-W.}\ \bibnamefont {Wang}}, \bibinfo {author} {\bibfnamefont
  {W.}~\bibnamefont {Qi}}, \bibinfo {author} {\bibfnamefont {X.-J.}\
  \bibnamefont {Liu}}, \ and\ \bibinfo {author} {\bibfnamefont
  {X.}~\bibnamefont {Zhang}},\ }\href {\doibase
  10.1103/PhysRevResearch.5.L012006} {\bibfield  {journal} {\bibinfo  {journal}
  {Phys. Rev. Res.}\ }\textbf {\bibinfo {volume} {5}},\ \bibinfo {pages}
  {L012006} (\bibinfo {year} {2023})}\BibitemShut {NoStop}%
\bibitem [{\citenamefont {Cheuk}\ \emph {et~al.}(2015)\citenamefont {Cheuk},
  \citenamefont {Nichols}, \citenamefont {Okan}, \citenamefont {Gersdorf},
  \citenamefont {Ramasesh}, \citenamefont {Bakr}, \citenamefont {Lompe},\ and\
  \citenamefont {Zwierlein}}]{Cheuk2015}%
  \BibitemOpen
  \bibfield  {author} {\bibinfo {author} {\bibfnamefont {L.~W.}\ \bibnamefont
  {Cheuk}}, \bibinfo {author} {\bibfnamefont {M.~A.}\ \bibnamefont {Nichols}},
  \bibinfo {author} {\bibfnamefont {M.}~\bibnamefont {Okan}}, \bibinfo {author}
  {\bibfnamefont {T.}~\bibnamefont {Gersdorf}}, \bibinfo {author}
  {\bibfnamefont {V.~V.}\ \bibnamefont {Ramasesh}}, \bibinfo {author}
  {\bibfnamefont {W.~S.}\ \bibnamefont {Bakr}}, \bibinfo {author}
  {\bibfnamefont {T.}~\bibnamefont {Lompe}}, \ and\ \bibinfo {author}
  {\bibfnamefont {M.~W.}\ \bibnamefont {Zwierlein}},\ }\href {\doibase
  10.1103/PhysRevLett.114.193001} {\bibfield  {journal} {\bibinfo  {journal}
  {Phys. Rev. Lett.}\ }\textbf {\bibinfo {volume} {114}},\ \bibinfo {pages}
  {193001} (\bibinfo {year} {2015})}\BibitemShut {NoStop}%
\bibitem [{\citenamefont {Haller}\ \emph {et~al.}(2015)\citenamefont {Haller},
  \citenamefont {Hudson}, \citenamefont {Kelly}, \citenamefont {Cotta},
  \citenamefont {Peaudecerf}, \citenamefont {Bruce},\ and\ \citenamefont
  {Kuhr}}]{Haller2015}%
  \BibitemOpen
  \bibfield  {author} {\bibinfo {author} {\bibfnamefont {E.}~\bibnamefont
  {Haller}}, \bibinfo {author} {\bibfnamefont {J.}~\bibnamefont {Hudson}},
  \bibinfo {author} {\bibfnamefont {A.}~\bibnamefont {Kelly}}, \bibinfo
  {author} {\bibfnamefont {D.~A.}\ \bibnamefont {Cotta}}, \bibinfo {author}
  {\bibfnamefont {B.}~\bibnamefont {Peaudecerf}}, \bibinfo {author}
  {\bibfnamefont {G.~D.}\ \bibnamefont {Bruce}}, \ and\ \bibinfo {author}
  {\bibfnamefont {S.}~\bibnamefont {Kuhr}},\ }\href {\doibase
  10.1038/nphys3403} {\bibfield  {journal} {\bibinfo  {journal} {Nature
  Physics}\ }\textbf {\bibinfo {volume} {11}},\ \bibinfo {pages} {738}
  (\bibinfo {year} {2015})}\BibitemShut {NoStop}%
\bibitem [{\citenamefont {Parsons}\ \emph {et~al.}(2015)\citenamefont
  {Parsons}, \citenamefont {Huber}, \citenamefont {Mazurenko}, \citenamefont
  {Chiu}, \citenamefont {Setiawan}, \citenamefont {Wooley-Brown}, \citenamefont
  {Blatt},\ and\ \citenamefont {Greiner}}]{Parsons2015}%
  \BibitemOpen
  \bibfield  {author} {\bibinfo {author} {\bibfnamefont {M.~F.}\ \bibnamefont
  {Parsons}}, \bibinfo {author} {\bibfnamefont {F.}~\bibnamefont {Huber}},
  \bibinfo {author} {\bibfnamefont {A.}~\bibnamefont {Mazurenko}}, \bibinfo
  {author} {\bibfnamefont {C.~S.}\ \bibnamefont {Chiu}}, \bibinfo {author}
  {\bibfnamefont {W.}~\bibnamefont {Setiawan}}, \bibinfo {author}
  {\bibfnamefont {K.}~\bibnamefont {Wooley-Brown}}, \bibinfo {author}
  {\bibfnamefont {S.}~\bibnamefont {Blatt}}, \ and\ \bibinfo {author}
  {\bibfnamefont {M.}~\bibnamefont {Greiner}},\ }\href {\doibase
  10.1103/PhysRevLett.114.213002} {\bibfield  {journal} {\bibinfo  {journal}
  {Phys. Rev. Lett.}\ }\textbf {\bibinfo {volume} {114}},\ \bibinfo {pages}
  {213002} (\bibinfo {year} {2015})}\BibitemShut {NoStop}%
\bibitem [{\citenamefont {Edge}\ \emph {et~al.}(2015)\citenamefont {Edge},
  \citenamefont {Anderson}, \citenamefont {Jervis}, \citenamefont {McKay},
  \citenamefont {Day}, \citenamefont {Trotzky},\ and\ \citenamefont
  {Thywissen}}]{Edge2015}%
  \BibitemOpen
  \bibfield  {author} {\bibinfo {author} {\bibfnamefont {G.~J.~A.}\
  \bibnamefont {Edge}}, \bibinfo {author} {\bibfnamefont {R.}~\bibnamefont
  {Anderson}}, \bibinfo {author} {\bibfnamefont {D.}~\bibnamefont {Jervis}},
  \bibinfo {author} {\bibfnamefont {D.~C.}\ \bibnamefont {McKay}}, \bibinfo
  {author} {\bibfnamefont {R.}~\bibnamefont {Day}}, \bibinfo {author}
  {\bibfnamefont {S.}~\bibnamefont {Trotzky}}, \ and\ \bibinfo {author}
  {\bibfnamefont {J.~H.}\ \bibnamefont {Thywissen}},\ }\href {\doibase
  10.1103/PhysRevA.92.063406} {\bibfield  {journal} {\bibinfo  {journal} {Phys.
  Rev. A}\ }\textbf {\bibinfo {volume} {92}},\ \bibinfo {pages} {063406}
  (\bibinfo {year} {2015})}\BibitemShut {NoStop}%
\bibitem [{\citenamefont {Omran}\ \emph {et~al.}(2015)\citenamefont {Omran},
  \citenamefont {Boll}, \citenamefont {Hilker}, \citenamefont {Kleinlein},
  \citenamefont {Salomon}, \citenamefont {Bloch},\ and\ \citenamefont
  {Gross}}]{Omran2015}%
  \BibitemOpen
  \bibfield  {author} {\bibinfo {author} {\bibfnamefont {A.}~\bibnamefont
  {Omran}}, \bibinfo {author} {\bibfnamefont {M.}~\bibnamefont {Boll}},
  \bibinfo {author} {\bibfnamefont {T.~A.}\ \bibnamefont {Hilker}}, \bibinfo
  {author} {\bibfnamefont {K.}~\bibnamefont {Kleinlein}}, \bibinfo {author}
  {\bibfnamefont {G.}~\bibnamefont {Salomon}}, \bibinfo {author} {\bibfnamefont
  {I.}~\bibnamefont {Bloch}}, \ and\ \bibinfo {author} {\bibfnamefont
  {C.}~\bibnamefont {Gross}},\ }\href {\doibase 10.1103/PhysRevLett.115.263001}
  {\bibfield  {journal} {\bibinfo  {journal} {Phys. Rev. Lett.}\ }\textbf
  {\bibinfo {volume} {115}},\ \bibinfo {pages} {263001} (\bibinfo {year}
  {2015})}\BibitemShut {NoStop}%
\bibitem [{\citenamefont {Gross}\ and\ \citenamefont
  {Bakr}(2021)}]{Bakr_review}%
  \BibitemOpen
  \bibfield  {author} {\bibinfo {author} {\bibfnamefont {C.}~\bibnamefont
  {Gross}}\ and\ \bibinfo {author} {\bibfnamefont {W.~S.}\ \bibnamefont
  {Bakr}},\ }\href {\doibase 10.1038/s41567-021-01370-5} {\bibfield  {journal}
  {\bibinfo  {journal} {Nature Physics}\ }\textbf {\bibinfo {volume} {17}},\
  \bibinfo {pages} {1316} (\bibinfo {year} {2021})}\BibitemShut {NoStop}%
\bibitem [{\citenamefont {Klebanov}\ \emph {et~al.}(2012)\citenamefont
  {Klebanov}, \citenamefont {Nishioka}, \citenamefont {Pufu},\ and\
  \citenamefont {Safdi}}]{klebanov2012shape}%
  \BibitemOpen
  \bibfield  {author} {\bibinfo {author} {\bibfnamefont {I.~R.}\ \bibnamefont
  {Klebanov}}, \bibinfo {author} {\bibfnamefont {T.}~\bibnamefont {Nishioka}},
  \bibinfo {author} {\bibfnamefont {S.~S.}\ \bibnamefont {Pufu}}, \ and\
  \bibinfo {author} {\bibfnamefont {B.~R.}\ \bibnamefont {Safdi}},\ }\href@noop
  {} {\bibfield  {journal} {\bibinfo  {journal} {Journal of High Energy
  Physics}\ }\textbf {\bibinfo {volume} {2012}},\ \bibinfo {pages} {1}
  (\bibinfo {year} {2012})}\BibitemShut {NoStop}%
\bibitem [{\citenamefont {Bueno}\ \emph {et~al.}(2015)\citenamefont {Bueno},
  \citenamefont {Myers},\ and\ \citenamefont
  {Witczak-Krempa}}]{Krempa2015corner}%
  \BibitemOpen
  \bibfield  {author} {\bibinfo {author} {\bibfnamefont {P.}~\bibnamefont
  {Bueno}}, \bibinfo {author} {\bibfnamefont {R.~C.}\ \bibnamefont {Myers}}, \
  and\ \bibinfo {author} {\bibfnamefont {W.}~\bibnamefont {Witczak-Krempa}},\
  }\href {\doibase 10.1103/PhysRevLett.115.021602} {\bibfield  {journal}
  {\bibinfo  {journal} {Phys. Rev. Lett.}\ }\textbf {\bibinfo {volume} {115}},\
  \bibinfo {pages} {021602} (\bibinfo {year} {2015})}\BibitemShut {NoStop}%
\bibitem [{\citenamefont {Faulkner}\ \emph {et~al.}(2016)\citenamefont
  {Faulkner}, \citenamefont {Leigh},\ and\ \citenamefont
  {Parrikar}}]{faulkner2016shape}%
  \BibitemOpen
  \bibfield  {author} {\bibinfo {author} {\bibfnamefont {T.}~\bibnamefont
  {Faulkner}}, \bibinfo {author} {\bibfnamefont {R.~G.}\ \bibnamefont {Leigh}},
  \ and\ \bibinfo {author} {\bibfnamefont {O.}~\bibnamefont {Parrikar}},\
  }\href@noop {} {\bibfield  {journal} {\bibinfo  {journal} {Journal of High
  Energy Physics}\ }\textbf {\bibinfo {volume} {2016}},\ \bibinfo {pages} {1}
  (\bibinfo {year} {2016})}\BibitemShut {NoStop}%
\bibitem [{\citenamefont {Hayward~Sierens}\ \emph {et~al.}(2017)\citenamefont
  {Hayward~Sierens}, \citenamefont {Bueno}, \citenamefont {Singh},
  \citenamefont {Myers},\ and\ \citenamefont {Melko}}]{Melko2017}%
  \BibitemOpen
  \bibfield  {author} {\bibinfo {author} {\bibfnamefont {L.~E.}\ \bibnamefont
  {Hayward~Sierens}}, \bibinfo {author} {\bibfnamefont {P.}~\bibnamefont
  {Bueno}}, \bibinfo {author} {\bibfnamefont {R.~R.~P.}\ \bibnamefont {Singh}},
  \bibinfo {author} {\bibfnamefont {R.~C.}\ \bibnamefont {Myers}}, \ and\
  \bibinfo {author} {\bibfnamefont {R.~G.}\ \bibnamefont {Melko}},\ }\href
  {\doibase 10.1103/PhysRevB.96.035117} {\bibfield  {journal} {\bibinfo
  {journal} {Phys. Rev. B}\ }\textbf {\bibinfo {volume} {96}},\ \bibinfo
  {pages} {035117} (\bibinfo {year} {2017})}\BibitemShut {NoStop}%
\bibitem [{\citenamefont {Bednik}\ \emph {et~al.}(2019)\citenamefont {Bednik},
  \citenamefont {Hayward~Sierens}, \citenamefont {Guo}, \citenamefont {Myers},\
  and\ \citenamefont {Melko}}]{Melko2019}%
  \BibitemOpen
  \bibfield  {author} {\bibinfo {author} {\bibfnamefont {G.}~\bibnamefont
  {Bednik}}, \bibinfo {author} {\bibfnamefont {L.~E.}\ \bibnamefont
  {Hayward~Sierens}}, \bibinfo {author} {\bibfnamefont {M.}~\bibnamefont
  {Guo}}, \bibinfo {author} {\bibfnamefont {R.~C.}\ \bibnamefont {Myers}}, \
  and\ \bibinfo {author} {\bibfnamefont {R.~G.}\ \bibnamefont {Melko}},\ }\href
  {\doibase 10.1103/PhysRevB.99.155153} {\bibfield  {journal} {\bibinfo
  {journal} {Phys. Rev. B}\ }\textbf {\bibinfo {volume} {99}},\ \bibinfo
  {pages} {155153} (\bibinfo {year} {2019})}\BibitemShut {NoStop}%
\bibitem [{\citenamefont {Cr\'epel}\ \emph {et~al.}(2021)\citenamefont
  {Cr\'epel}, \citenamefont {Hackenbroich}, \citenamefont {Regnault},\ and\
  \citenamefont {Estienne}}]{Estienne2021Dirac}%
  \BibitemOpen
  \bibfield  {author} {\bibinfo {author} {\bibfnamefont {V.}~\bibnamefont
  {Cr\'epel}}, \bibinfo {author} {\bibfnamefont {A.}~\bibnamefont
  {Hackenbroich}}, \bibinfo {author} {\bibfnamefont {N.}~\bibnamefont
  {Regnault}}, \ and\ \bibinfo {author} {\bibfnamefont {B.}~\bibnamefont
  {Estienne}},\ }\href {\doibase 10.1103/PhysRevB.103.235108} {\bibfield
  {journal} {\bibinfo  {journal} {Phys. Rev. B}\ }\textbf {\bibinfo {volume}
  {103}},\ \bibinfo {pages} {235108} (\bibinfo {year} {2021})}\BibitemShut
  {NoStop}%
\bibitem [{\citenamefont {Bueno}\ and\ \citenamefont
  {Myers}(2015)}]{bueno2015corner}%
  \BibitemOpen
  \bibfield  {author} {\bibinfo {author} {\bibfnamefont {P.}~\bibnamefont
  {Bueno}}\ and\ \bibinfo {author} {\bibfnamefont {R.~C.}\ \bibnamefont
  {Myers}},\ }\href@noop {} {\bibfield  {journal} {\bibinfo  {journal} {Journal
  of High Energy Physics}\ }\textbf {\bibinfo {volume} {2015}},\ \bibinfo
  {pages} {1} (\bibinfo {year} {2015})}\BibitemShut {NoStop}%
\bibitem [{\citenamefont {Seminara}\ \emph {et~al.}(2017)\citenamefont
  {Seminara}, \citenamefont {Sisti},\ and\ \citenamefont
  {Tonni}}]{seminara2017corner}%
  \BibitemOpen
  \bibfield  {author} {\bibinfo {author} {\bibfnamefont {D.}~\bibnamefont
  {Seminara}}, \bibinfo {author} {\bibfnamefont {J.}~\bibnamefont {Sisti}}, \
  and\ \bibinfo {author} {\bibfnamefont {E.}~\bibnamefont {Tonni}},\
  }\href@noop {} {\bibfield  {journal} {\bibinfo  {journal} {Journal of High
  Energy Physics}\ }\textbf {\bibinfo {volume} {2017}},\ \bibinfo {pages} {1}
  (\bibinfo {year} {2017})}\BibitemShut {NoStop}%
\bibitem [{\citenamefont {Chung}\ and\ \citenamefont
  {Peschel}(2001)}]{Peschel2001}%
  \BibitemOpen
  \bibfield  {author} {\bibinfo {author} {\bibfnamefont {M.-C.}\ \bibnamefont
  {Chung}}\ and\ \bibinfo {author} {\bibfnamefont {I.}~\bibnamefont
  {Peschel}},\ }\href {\doibase 10.1103/PhysRevB.64.064412} {\bibfield
  {journal} {\bibinfo  {journal} {Phys. Rev. B}\ }\textbf {\bibinfo {volume}
  {64}},\ \bibinfo {pages} {064412} (\bibinfo {year} {2001})}\BibitemShut
  {NoStop}%
\bibitem [{\citenamefont {Peschel}(2003)}]{peschel2003calculation}%
  \BibitemOpen
  \bibfield  {author} {\bibinfo {author} {\bibfnamefont {I.}~\bibnamefont
  {Peschel}},\ }\href
  {https://iopscience.iop.org/article/10.1088/0305-4470/36/14/101/meta?casa_token=YtmlnEA7i7AAAAAA:6iwAUOzAxlqsCVgfaAuNRyH-hz-MRiXOgaFW90i7-uZFb8_Fcn8MyziLHD3oe9o2D0pDnZ9n6fj-Oubk2EX9E9TUpQ}
  {\bibfield  {journal} {\bibinfo  {journal} {Journal of Physics A:
  Mathematical and General}\ }\textbf {\bibinfo {volume} {36}},\ \bibinfo
  {pages} {L205} (\bibinfo {year} {2003})}\BibitemShut {NoStop}%
\bibitem [{\citenamefont {Cheong}\ and\ \citenamefont
  {Henley}(2004)}]{Cheong2004}%
  \BibitemOpen
  \bibfield  {author} {\bibinfo {author} {\bibfnamefont {S.-A.}\ \bibnamefont
  {Cheong}}\ and\ \bibinfo {author} {\bibfnamefont {C.~L.}\ \bibnamefont
  {Henley}},\ }\href {\doibase 10.1103/PhysRevB.69.075111} {\bibfield
  {journal} {\bibinfo  {journal} {Phys. Rev. B}\ }\textbf {\bibinfo {volume}
  {69}},\ \bibinfo {pages} {075111} (\bibinfo {year} {2004})}\BibitemShut
  {NoStop}%
\bibitem [{\citenamefont {Thonhauser}\ and\ \citenamefont
  {Vanderbilt}(2006)}]{Thonhauser2006}%
  \BibitemOpen
  \bibfield  {author} {\bibinfo {author} {\bibfnamefont {T.}~\bibnamefont
  {Thonhauser}}\ and\ \bibinfo {author} {\bibfnamefont {D.}~\bibnamefont
  {Vanderbilt}},\ }\href {\doibase 10.1103/PhysRevB.74.235111} {\bibfield
  {journal} {\bibinfo  {journal} {Phys. Rev. B}\ }\textbf {\bibinfo {volume}
  {74}},\ \bibinfo {pages} {235111} (\bibinfo {year} {2006})}\BibitemShut
  {NoStop}%
\bibitem [{\citenamefont {Berthiere}\ \emph {et~al.}(2023)\citenamefont
  {Berthiere}, \citenamefont {Estienne}, \citenamefont {St\'ephan},\ and\
  \citenamefont {Witczak-Krempa}}]{berthiere2023full}%
  \BibitemOpen
  \bibfield  {author} {\bibinfo {author} {\bibfnamefont {C.}~\bibnamefont
  {Berthiere}}, \bibinfo {author} {\bibfnamefont {B.}~\bibnamefont {Estienne}},
  \bibinfo {author} {\bibfnamefont {J.-M.}\ \bibnamefont {St\'ephan}}, \ and\
  \bibinfo {author} {\bibfnamefont {W.}~\bibnamefont {Witczak-Krempa}},\ }\href
  {\doibase 10.1103/PhysRevB.108.L201109} {\bibfield  {journal} {\bibinfo
  {journal} {Phys. Rev. B}\ }\textbf {\bibinfo {volume} {108}},\ \bibinfo
  {pages} {L201109} (\bibinfo {year} {2023})}\BibitemShut {NoStop}%
\bibitem [{\citenamefont {Het\'enyi}\ and\ \citenamefont
  {L\'evay}(2023)}]{PhysRevA.108.032218}%
  \BibitemOpen
  \bibfield  {author} {\bibinfo {author} {\bibfnamefont {B.}~\bibnamefont
  {Het\'enyi}}\ and\ \bibinfo {author} {\bibfnamefont {P.}~\bibnamefont
  {L\'evay}},\ }\href {\doibase 10.1103/PhysRevA.108.032218} {\bibfield
  {journal} {\bibinfo  {journal} {Phys. Rev. A}\ }\textbf {\bibinfo {volume}
  {108}},\ \bibinfo {pages} {032218} (\bibinfo {year} {2023})}\BibitemShut
  {NoStop}%
\bibitem [{\citenamefont {Wu}\ \emph {et~al.}(2024)\citenamefont {Wu},
  \citenamefont {Cai}, \citenamefont {Cheng},\ and\ \citenamefont
  {Kumar}}]{wu2024corner}%
  \BibitemOpen
  \bibfield  {author} {\bibinfo {author} {\bibfnamefont {X.-C.}\ \bibnamefont
  {Wu}}, \bibinfo {author} {\bibfnamefont {K.-L.}\ \bibnamefont {Cai}},
  \bibinfo {author} {\bibfnamefont {M.}~\bibnamefont {Cheng}}, \ and\ \bibinfo
  {author} {\bibfnamefont {P.}~\bibnamefont {Kumar}},\ }\href
  {https://arxiv.org/abs/2408.16057} {\enquote {\bibinfo {title} {Corner charge
  fluctuations and many-body quantum geometry},}\ } (\bibinfo {year} {2024}),\
  \Eprint {http://arxiv.org/abs/2408.16057} {arXiv:2408.16057
  [cond-mat.str-el]} \BibitemShut {NoStop}%
\bibitem [{\citenamefont {Kruchkov}\ and\ \citenamefont
  {Ryu}(2024)}]{kruchkov2024entanglemententropylatticemodels}%
  \BibitemOpen
  \bibfield  {author} {\bibinfo {author} {\bibfnamefont {A.}~\bibnamefont
  {Kruchkov}}\ and\ \bibinfo {author} {\bibfnamefont {S.}~\bibnamefont {Ryu}},\
  }\href {https://arxiv.org/abs/2408.10314} {\enquote {\bibinfo {title}
  {Entanglement entropy in lattice models with quantum metric},}\ } (\bibinfo
  {year} {2024}),\ \Eprint {http://arxiv.org/abs/2408.10314} {arXiv:2408.10314
  [cond-mat.str-el]} \BibitemShut {NoStop}%
\bibitem [{\citenamefont {Harper}(1955)}]{harper1955single}%
  \BibitemOpen
  \bibfield  {author} {\bibinfo {author} {\bibfnamefont {P.~G.}\ \bibnamefont
  {Harper}},\ }\href
  {https://iopscience.iop.org/article/10.1088/0370-1298/68/10/304} {\bibfield
  {journal} {\bibinfo  {journal} {Proceedings of the Physical Society. Section
  A}\ }\textbf {\bibinfo {volume} {68}},\ \bibinfo {pages} {874} (\bibinfo
  {year} {1955})}\BibitemShut {NoStop}%
\bibitem [{\citenamefont {Hofstadter}(1976)}]{Hofstadter1976}%
  \BibitemOpen
  \bibfield  {author} {\bibinfo {author} {\bibfnamefont {D.~R.}\ \bibnamefont
  {Hofstadter}},\ }\href {\doibase 10.1103/PhysRevB.14.2239} {\bibfield
  {journal} {\bibinfo  {journal} {Phys. Rev. B}\ }\textbf {\bibinfo {volume}
  {14}},\ \bibinfo {pages} {2239} (\bibinfo {year} {1976})}\BibitemShut
  {NoStop}%
\end{thebibliography}%

\clearpage
\newpage
\widetext
\begin{center}
\textbf{\large Supplementary Materials for ``Corner Charge Fluctuation as an Observable for Quantum Geometry and Entanglement
in Two-dimensional Insulators"}\\
\vspace{0.5cm}
\text{Pok Man Tam, Jonah Herzog-Arbeitman, and Jiabin Yu}
\end{center}
\onecolumngrid
\setcounter{secnumdepth}{2}
\renewcommand{\theequation}{\thesection.\arabic{equation}}
\renewcommand{\theHequation}{\theHsection.\arabic{equation}}
\renewcommand{\thefigure}{\thesection.\arabic{figure}}  

The supplemental information consists of five sections. In Sec. \ref{supp_sec:S and G} we provide a detailed discussion relating the static structure factor to quantum geometry, first in the non-interacting case for band geometry, then in the interacting case for many-body quantum geometry. In particular, we provide simple proofs for Eq. \eqref{eq:relating S to G} and Eq. \eqref{eq: relating S to localization} in the main text. In Sec. \ref{supp_sec:OAI} we discuss how to analytically and exactly calculate the corner charge fluctuation in an obstructed atomic insulator with non-trivial quantum geometry. In Sec. \ref{supp_sec: partition_scheme}, we explain why our key result in Eq. \eqref{eq:key_result} applies to the triangular lattice, and more generally to any Bravais lattices. In Sec. \ref{supp_sec:orbital_embedding}, we analyze specific examples where the integrated metric $\mathcal{G}_{ii}$ of the physical embedding can be extracted from the corner flucturation, with the Harper-Hofstadter model and the Haldane model as our focus. In Sec. \ref{supp_sec:numerics}, we review the correlation matrix method for numerically computing the bipartite fluctuation and entanglement entropies exactly, and collect all the real-space Hamiltonians as well as representative partition configurations used in our numerical studies. 

\section{Static structure factor and Quantum Geometry}\label{supp_sec:S and G}
\setcounter{equation}{0}
\setcounter{figure}{0} 

In this section we review the relation between static structure factor and quantum geometry, and some relevant results are already known and discussed in Ref. \cite{Regnault2013, onishi2024fundamental,onishi2024quantum}. Nevertheless, we include the following discussion not just for reader's convenience, but also to clarify the role of orbital embedding in relating structure factor to quantum geometry. To our knowledge, the latter discussion has not been covered in the literature. 

\subsection{Band quantum geometry}
Here we provide a simple proof of Eq. \eqref{eq:relating S to G} for band insulators. We will do this in both the physical and origin orbital embedding, where the respective Fourier transformations for the density operator are expressed as 
\begin{equation}\label{eq: two kinds of FT}
    \rho_\bq = \sum_{\bR,\sigma} e^{-i\bq\cdot \bR_\sigma} \rho_\sigma(\bR) = \sum_{\bR,\sigma} e^{-i\bq\cdot \bR_\sigma} c^\dagger_{\bR,\sigma} c_{\bR,\sigma} \quad\text{and}\quad
    \widetilde{\rho}_\bq  = \sum_{\bR} e^{-i\bq\cdot \bR} \rho(\bR) = \sum_{\bR} e^{-i\bq\cdot \bR} \Big(\sum_\sigma c^\dagger_{\bR,\sigma} c_{\bR,\sigma} \Big),
\end{equation}
while for the electron annihilation operator we have
\begin{equation}\label{eq:FT_Bloch}
    c_{\bR, \sigma} = \frac{1}{\sqrt{N_c}}\sum_{\bk,n} e^{i\bk\cdot \bR_\sigma} U_{\sigma,n}(\bk) c_{\bk,n} \quad \text{and} \quad c_{\bR, \sigma} = \frac{1}{\sqrt{N_c}}\sum_{\bk,n} e^{i\bk\cdot \bR} \widetilde{U}_{\sigma,n}(\bk) c_{\bk,n}, 
\end{equation}
where $N_c$ is the number of unit cells, $c_{\bR, \sigma}$ annihilates the electron at the $\sigma$th orbital with absolute position $\bR_\sigma=\bR+\br_\sigma$ ($\bR$ is the unit cell position and $\br_\sigma$ the relative intra-cell position), and $c_{\bk,n}$ is the electron annihilation operator in the $n$th band. The above defines the $n$-th band eigenvector $U_n(\bk)$ in the physical embedding and $\widetilde{U}_n(\bk)$ in the origin embedding. The Fourier-transformed density operators in the two embeddings can thus be expressed as 
\begin{equation}
    \rho_\bq = \sum_{\bk,\sigma mn} U^\dagger_{m,\sigma}(\bk)U_{\sigma,n}(\bk+\bq) c^\dag_{\bk,m} c_{\bk+\bq,n}\quad \text{and}\quad \widetilde{\rho}_\bq = \sum_{\bk,\sigma mn} \widetilde{U}^\dagger_{m,\sigma}(\bk)\widetilde{U}_{\sigma,n}(\bk+\bq) c^\dag_{\bk,m} c_{\bk+\bq,n}.
\end{equation}
For band insulators, the connected correlator $\langle \rho_\bq \rho_{-\bq}\rangle_c \equiv \langle \rho_\bq \rho_{-\bq}\rangle - \langle \rho_\bq \rangle \langle\rho_{-\bq}\rangle$ can be evaluated by Wick's contraction: $\braket{c^\dagger_{\bk,m}c_{\bk+\bq,n}c^\dagger_{\bk',m'}c_{\bk'-\bq,n'}}_c = \delta_{\bk,\bk'-\bq} \bar{\delta}_{m,n'}(\delta_{n,m'}-\bar{\delta}_{n,m'})$, where $\delta_{n,m'}$ is the usual Kronecker delta and $\bar{\delta}_{m,n'}$ is the Kronecker delta when $m, n'$ are occupied and zero otherwise. Thus, the static structure factor is
\begin{equation}
\begin{split}
    S_\bq &=\frac{1}{\mathcal{A}} \braket{\rho_\bq \rho_{-\bq}}_c = \int_{\text{BZ}} [d\bk] \tr[P(\bk) (1-P(\bk+\bq))] = q^i q^j \int_{\text{BZ}} [d\bk] \frac{1}{2} \tr[\partial_i P(\bk) \partial_j P(\bk)] + O(q^3), 
\end{split}
\end{equation}
where $\mathcal{A}$ is the area of the system (for 3D it would be the volume) and $P_{\sigma,\sigma'}(\bk) = \sum_{n\in\text{occ.}}U_{\sigma,n}(\bk)U^\dagger_{n,\sigma'}(\bk)$ is the single-particle projector onto the occupied bands. While the above is done for the physical embedding, the derivation for the case of origin embedding with $\widetilde{S}_{\bq}= \mathcal{A}^{-1} \cc{\widetilde{\rho}_{\bq}\widetilde{\rho}_{-\bq}}$ follows the exact same way by replacing $P(\bk)$ by $\widetilde{P}(\bk) = \sum_{m\in occ} \widetilde{U}_m(\bk)\widetilde{U}^\dagger_m(\bk)$. Altogether, we conclude
\begin{equation}
    \frac{1}{2}\partial_i\partial_j S_\bq \vert_{\bq=0} = \int_{\text{BZ}} [d\bk] \frac{1}{2} \tr[\partial_i P(\bk)\partial_j P(\bk)] = \mathcal{G}_{ij} \quad \text{and}\quad\frac{1}{2}\partial_i\partial_j \widetilde{S}_\bq \vert_{\bq=0} = \int_{\text{BZ}} [d\bk] \frac{1}{2} \tr[\partial_i \widetilde{P}(\bk)\partial_j \widetilde{P}(\bk)] = \widetilde{\mathcal{G}}_{ij},
\end{equation}
which proves Eq. \eqref{eq:relating S to G} for band quantum geometry.

\subsection{Many-body quantum geometry}
Now we provide the details for proving Eq. \eqref{eq: relating S to localization} and then establishing Eq. \eqref{eq:relating S to G} for the many-body quantum geometry in interacting systems. Recently the concept of ``quantum weight" was introduced in Refs. \cite{onishi2024fundamental, onishi2024quantum}, which connects the structure factor to the negative-first-moment of optical conductivity known as the Souza-Wilkens-Martin (SWM) sum rule \cite{SWM2000}. In what follows, we take a more direct path to relate the structure factor to the localization tensor $\cc{X_i X_j}$ (also known as the polarization fluctuation), which can then be directly related to the many-body quantum geometry (defined via twisted boundary conditions) following the analysis in Ref. \cite{SWM2000}. Again, we will emphasize the role of orbital embedding in establishing such a relation. \\

\subsubsection{Simple proof of Eq. \eqref{eq: relating S to localization}}
Let us first focus on the origin orbital embedding, in which case we directly evaluate using Eq. \eqref{eq: two kinds of FT}:
\begin{equation}
\begin{split}
    \widetilde{S}_\bq &= \frac{1}{\mathcal{A}} \cc{\widetilde{\rho}_\bq \widetilde{\rho}_{-\bq}} = \frac{1}{\mathcal{V}} \sum_{\bR}\sum_{\bR'} e^{-i\bq\cdot(\bR-\bR')} \cc{\rho(\bR) \rho(\bR')} \\
    \implies \partial_{q_i} \partial_{q_j} \widetilde{S}_\bq \vert_{\bq=0} &= - \frac{1}{\mathcal{A}} \sum_{\bR}\sum_{\bR'} (\bR-\bR')_i (\bR-\bR')_j \cc{\rho(\bR) \rho(\bR')} \\
    &= - \frac{1}{\mathcal{A}} \sum_{\bR}\sum_{\bR'} [ \bR_{i} \bR_{j} + \bR'_{i} \bR'_{j} - \bR_{i} \bR'_{j}-\bR'_{i} \bR_{j}] \cc{\rho(\bR) \rho(\bR')}.
\end{split}
\end{equation}
Notice the first two terms vanish after summation over unit cell positions due to the total conservation of charge (i.e., $\sum_\bR \rho(\bR)$ is conserved, hence the associated contribution to the connected correlator is rendered zero). Thus
\begin{equation}\label{eq:q^2_coeff_of_S_q_origin_embedding}
    \partial_{i} \partial_{j} \widetilde{S}_\bq \vert_{\bq=0}  =  \frac{2}{\mathcal{A}} \cc{\Big( \sum_\bR \bR_i \rho(\bR) \Big) \Big( \sum_{\bR'} \bR'_j \rho(\bR') \Big)} = \frac{2}{\mathcal{A}} \cc{\widetilde{X}_i\widetilde{X}_j}.
\end{equation}
Notice that $\sum_\bR \bR_i \rho(\bR) $ is the second quantized version of the \textit{center-of-cell} position operator, whose first-quantized form for an $N$-particle system is $\widetilde{\bX} = \sum_{a=1}^N \widetilde{\bx}_{a}$, where $\widetilde{\bx}_a$ measures the position of the \textit{unit-cell} to which the $a$-th particle belongs. This proves Eq. \eqref{eq: relating S to localization} in the main text.\\

For the physical orbital embedding, the analysis is almost identical. We have
\begin{equation}
    \partial_{i} \partial_{j} S_\bq \vert_{\bq=0} = \frac{2}{\mathcal{A}} \cc{ \Big( \sum_{\bR,\sigma} \bR_{\sigma,i} \rho_\sigma(\bR)\Big) \Big( \sum_{\bR',\sigma'} \bR'_{\sigma',j} \rho_{\sigma'}(\bR')\Big)} = \frac{2}{\mathcal{A}} \cc{X_iX_j}.
\end{equation}
Here we notice that $\sum_{\bR,\sigma} \bR_{\sigma,i} \rho_\sigma(\bR)$ is the second-quantized version of the $N$-particle center-of-mass position operator $\bX = \sum_{a=1}^N \bx_{a}$, where $\bx_a$ measures the \textit{absolute} position of the $a$-th particle.\\

\subsubsection{From localization tensor to quantum geometry}

The seminal work by Souza-Wilkens-Martin (SWM) in Ref. \cite{SWM2000} directly connects the localization tensor $ \cc{X_iX_j}$ to many-body quantum geometry under the physical orbital embedding. Below we carry out a similar analysis, geared towards the origin orbital embedding specifically, to relate $\cc{\widetilde{X}_i\widetilde{X}_j}$ and the many-body quantum geometry. Our analysis is simplified as compared to SWM in the sense that we do not invoke Kohn's conjecture (see Ref. \cite{Kohn1964}) about wavefunction localization in insulating states.\\

Suppose $\ket{\Psi_{\bk}}$ is a many-body state with $\bk$ specifying its boundary condition:
\eq{
T_{\bM}\ket{\Psi_{\bk}} = e^{i \bk\cdot \bR_\bM } \ket{\Psi_{\bk}}\ ,
}
where $k_i \in [0, 2\pi/L_i)$,  $T_{\bM}$ is the many-body translation operator along $\bR_\bM = \sum_{i=1}^D M_i \bL_i$ ($M_i$ integers), with $D$ the dimension and $L_i=\abs{\bL_i}$ the length of the sample along the $i$th direction.
Just like the Bloch's theorem, we can define
\eq{
\ket{\widetilde{\Phi}_{\bk}}=e^{-i\bk\cdot \widetilde{\bX}}\ket{\Psi_{\bk}}\ .
}
where $\widetilde{\bX}$ is the \textit{center-of-cell} position operator that we have introduced above, whose fluctuation is related to the structure factor in the origin orbital embedding. Under the action of $T_\bM$, we have $T_\bM \widetilde{\bX} T_\bM^{-1} = \widetilde{\bX}+\bR_{\bM}$.\\

Let us now consider $ \bra{\Psi_{\bk=0}} e^{-i  \bsl{\alpha}\cdot  \widetilde{\bX} }  \ket{\Psi_{\bk=0}} $.
We note that $ \bra{\Psi_{\bk=0}} e^{-i  \bsl{\alpha}\cdot  \widetilde{\bX}  }  \ket{\Psi_{\bk=0}} $ is only nonzero for
\eq{
\label{eq:alpha_values}
\bsl{\alpha} \in \otimes_{i} \dsZ \frac{2\pi}{L_i} 
}
due to the aforementioned boundary condition.
Therefore, we have to choose $\bsl{\alpha}$ according to \cref{eq:alpha_values} for nonzero $ \bra{\Psi_{\bk=0}} e^{-i  \bsl{\alpha}\cdot  \widetilde{\bX} }  \ket{\Psi_{\bk=0}} $.
Clearly, the intervals of $\bsl{\alpha}$ go to zero as we go to the thermodynamic limit, and thus it will not prevent us from taking derivatives in $\bsl{\alpha}$---just like how we take derivatives in the Bloch momentum of single-particle Bloch states though the Bloch momentum are in $\otimes_{i} \dsZ \frac{2\pi}{L_i} $.
We note that $\bq$ in the structure factor also has the intervals of $2\pi/L_i$ due to the periodic boundary condition.
%
From the cumulant generating function we have
\eqa{
\left. -\partial_{\alpha_i}\partial_{\alpha_j} \log  \bra{\Psi_{\bk=0}} e^{-i  \bsl{\alpha}\cdot  \widetilde{\bX} }  \ket{\Psi_{\bk=0}} \right|_{\bsl{\alpha}= 0} = \cc{\widetilde{X}_i \widetilde{X}_j} \ .
}
On the other hand,
\eqa{
\left. -\partial_{\alpha_i}\partial_{\alpha_j} \log  \bra{\Psi_{\bk=0}} e^{-i  \bsl{\alpha}\cdot  \widetilde{\bX} }  \ket{\Psi_{\bk=0}} \right|_{\bsl{\alpha}= 0} &=\left. -\partial_{\alpha_i}\partial_{\alpha_j} \log  \braket{\widetilde{\Phi}_{0}|\widetilde{\Phi}_{\bsl{\alpha}}} \right|_{\bsl{\alpha}= 0} \\
&= -\braket{\widetilde{\Phi}_{0}|\partial_{\alpha_i} \partial_{\alpha_j} \widetilde{\Phi}_{\bsl{\alpha}}}\Big|_{\bsl{\alpha}= 0}+\braket{\widetilde{\Phi}_0|\partial_{\alpha_i}\widetilde{\Phi}_{\bsl{\alpha}}}\braket{\widetilde{\Phi}_0| \partial_{\alpha_j}\widetilde{\Phi}_{\bsl{\alpha}}}\Big|_{\bsl{\alpha}= 0}\\
&= -\braket{\widetilde{\Phi}_{\bk}|\partial_{k_i} \partial_{k_j} \widetilde{\Phi}_{\bk}}\Big|_{\bk= 0}+\braket{\widetilde{\Phi}_\bk|\partial_{k_i}\widetilde{\Phi}_{\bk}}\braket{\widetilde{\Phi}_\bk| \partial_{k_j}\widetilde{\Phi}_{\bk}}\Big|_{\bk= 0}\\
& \equiv I_{ij}(\bk=0) \ .
}
Let us define $I_{ij}(\bk) \equiv -\braket{\widetilde{\Phi}_{\bk}|\partial_{k_i} \partial_{k_j} \widetilde{\Phi}_{\bk}}+\braket{\widetilde{\Phi}_\bk|\partial_{k_i}\widetilde{\Phi}_{\bk}}\braket{\widetilde{\Phi}_\bk| \partial_{k_j}\widetilde{\Phi}_{\bk}}$, then from the finiteness of $\cc{\widetilde{X}_i \widetilde{X}_j}/\mathcal{A}$ (correspondingly the finiteness of $\partial_i \partial_j S_\bq|_{\bq=0}$) which is taken as the characteristic property of an insulating matter, we have
\begin{equation}
\label{eq:tylor}
    \frac{1}{\mathcal{A}}I_{ij}(\bk) = \frac{1}{\mathcal{A}}I_{ij}(0) + O(\frac{1}{L}), \quad \text{for}\quad k_i \in [0,\frac{2\pi}{L_i}),
\end{equation}
thus in the thermodynamic limit, we have
\begin{equation}
\label{eq:final}
\begin{split}
    \frac{1}{\mathcal{A}}\cc{\widetilde{X}_i\widetilde{X}_j} = \int [d\bk]\;I_{ij} (\bk) &= \int [d\bk] \Big(\frac{1}{2}\braket{\partial_{k_i}\widetilde{\Phi}_{\bk}| \partial_{k_j} \widetilde{\Phi}_{\bk}} + \frac{1}{2}\braket{\partial_{k_j}\widetilde{\Phi}_{\bk}| \partial_{k_i} \widetilde{\Phi}_{\bk}}-\braket{\partial_{k_i}\widetilde{\Phi}_\bk|\widetilde{\Phi}_{\bk}}\braket{\widetilde{\Phi}_\bk| \partial_{k_j}\widetilde{\Phi}_{\bk}}\Big) \\
    & = \int [d\bk] \frac{1}{2} \tr[\partial_i \widetilde{P}(\bk)\partial_j \widetilde{P}(\bk)] \equiv \widetilde{\mathcal{G}}_{ij},
\end{split}
\end{equation}
with $\widetilde{P}(\bk)= \ket{\widetilde{\Phi}_\bk}\bra{\widetilde{\Phi}_\bk}$. Combining this with Eq. \eqref{eq: relating S to localization}, we arrive at Eq. \eqref{eq:relating S to G}.


\section{Compact obstructed atomic insulator}\label{supp_sec:OAI}
\setcounter{equation}{0}
\setcounter{figure}{0}  

\begin{figure}[h!]
    \centering
    \includegraphics[width=0.8\columnwidth]{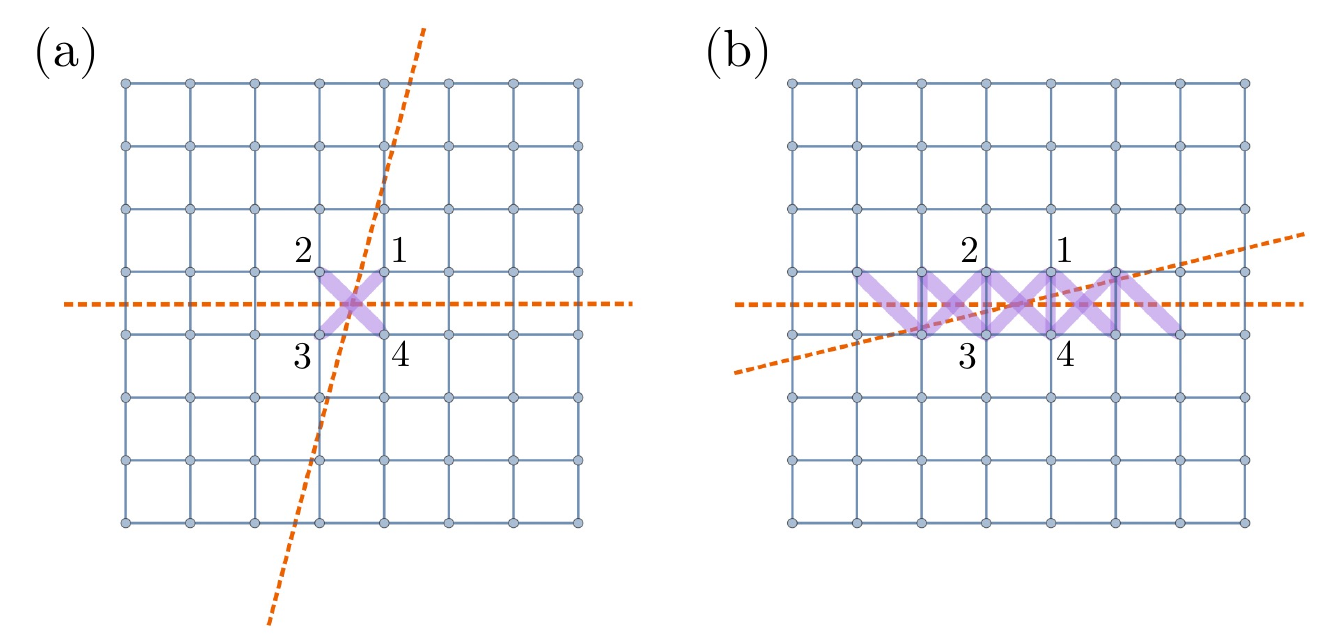}
    \caption{Corner charge fluctuation in OAI. (a) Partition scheme with $\tan\theta = 4$. This is representative for all $\tan\theta \in 2\mathbb{N}$, where only two diagonal bonds contribute to Eq. \eqref{supp_eq:corner_fluc}. (b) Partition scheme with $\cot\theta = 4$. For small angle, the number of bonds contributing to Eq. \eqref{supp_eq:corner_fluc} is proportional to $\cot\theta$. }
    \label{supp_fig:OAI}
\end{figure}

In this section, we study a compact obstructed atomic insulator (OAI) to compute the corner contribution to the charge fluctuation analytically. This analysis complements the other solvable models, the Dirac fermion and Landau levels, where the corner contribution can be analytically calculated by virtue of isotropy as in Ref. \cite{estienne2022cornering}. We build a compact OAI following Ref. \cite{Jonah2023} using a four-orbital model on the square lattice (with primitive vectors $\hat{x}$ and $\hat{y}$), where $s,d, p_x,p_y$ orbitals are placed at each site with $C_4$ representation $D[C_4] = \text{diag}(1, -1, -i ,i)$. As intra-cell orbitals are coinciding, the physical embedding and the origin embedding coincide, so for simplicity we drop the tilde notation in this discussion. The orthonormal eigenstates ($j=0, 1,2,3$) are introduced as follows,
\bea
\label{supp_eq:complete}
U_j(\mbf{k}) = \frac{1}{4} \bpm 1 \\ 1 \\1 \\1 \epm +\frac{e^{-i \frac{2\pi}{4}j}}{4} \bpm 1 \\-1 \\-i \\i \epm e^{i \mbf{k} \cdot \hat{x}}+\frac{e^{-i \frac{2\pi}{4}2j}}{4} \bpm 1\\ 1\\ -1\\ -1 \epm e^{i \mbf{k} \cdot (\hat{x}+\hat{y})}+\frac{e^{-i \frac{2\pi}{4}3j}}{4} \bpm 1\\ -1\\ i \\-i \epm e^{i \mbf{k} \cdot \hat{y}}.
\eea
Each has a Berry connection
\bea
\mbf{A}_j(\mbf{k}) = U_j^\dag (i \pmb{\nabla}) U_j = -\frac{1}{2} \hat{x} - \frac{1}{2} \hat{y}
\eea
indicating that the Wannier states built from $U_j(\mbf{k})$ are centered on the plaquette, where there are no atoms. This is a defining feature of an obstructed atomic insulator \cite{bradlyn2017topological,Schindler2021, Jonah2023}. Below, we consider a ground state with the $j=0$ band completely occupied and all other bands empty. The parent Hamiltonian of this state can be constructed as $H(\bk) = -P_0(\bk)=-U_0(\bk) U_0^\dagger(\bk)$, and it is easy to check that it describes a tight-binding model with up to second nearest neighbor hoppings. One also easily sees that $g_{xx}(\bk)=g_{yy}(\bk) = \frac{1}{2} \tr[(\partial_i P_0(\bk))^2] = \frac{1}{4}$. The trace of integrated quantum metric is $ \mathcal{G} = \frac{1}{2}$. 

To compute the bipartite charge fluctuation, we first construct the Wannier states for the $j$-th band:
\begin{equation}
    w^\dagger_{\bR,j} = \frac{1}{\sqrt{N}} \sum_{\bk,\sigma} e^{i\bk\cdot \bR} [U_{j}(\bk)]_\sigma c^\dagger_{\bk,\sigma} = \sum_{\bd, \sigma} [W_{j}(\bd)]_\sigma c^\dagger_{\bR+\bd,\sigma}\;, \quad \text{with } W_j (\bd) = \frac{1}{N} \sum_\bk e^{-i\bk\cdot\bd} U_j (\bk).
\end{equation}
Here $c^\dagger_{\bk,\sigma}$ and $c^\dagger_{\bR,\sigma}$ are the fermionic creation operators in the momentum and real spaces, respectively, and $N$ is the number of sites. The inverse transform that we need is 
\begin{equation}
    c^\dagger_{\bR,\sigma} = \sum_{\bd,j} [V_j(\bd)]_\sigma w^\dagger_{\bR-\bd,j}\;, \quad \text{with } V_j(\bd) = \frac{1}{N} \sum_\bk e^{i\bk\cdot \bd} U^*_j(\bk).
\end{equation}
It is clear from our construction of the compact OAI that $V_j(\bd)$ and $W_j(\bd)$ are non-zero only for $\bd= 0, \hat{x}, \hat{y}, \hat{x}+\hat{y}$, so each Wannier orbital is \textit{compactly} supported on four corners of a plaquette. The many-body ground state of our choice corresponds to filling up all Wannier orbitals of the $j=0$ band:
\begin{equation}
    \ket{GS}  = \prod_{\text{all } \bR} w^\dagger_{\bR,0} \ket{0},
\end{equation}
and hence
\begin{equation}
    \braket{c^\dagger_{\bR,\sigma} c_{\bR',\sigma'}}=
    \sum_{\bd} [V_{0}(\bd)]_\sigma [V_0(\bd-\bR+\bR')]^*_{\sigma'}.
\end{equation}
Now, noting that the bipartite fluctuation can be computed as $\braket{Q^2_A}_c = -\braket{Q_A Q_{\bar{A}}}$, so upon substituting into the simplified expression of corner charge fluctuation in Eq. \eqref{eq:simplified_CQ}, we find
\begin{equation}\label{supp_eq:corner_fluc}
    \mathcal{C}^{(Q)}(\theta) = \Big(\sum_{\substack{\bR\in B \\ \bR'\in D}} + \sum_{\substack{\bR\in A \\ \bR'\in C}} \Big)\sum_{\sigma,\sigma'} \sum_{\bd,\bd'}\; [V_{0}(\bd)]_\sigma [V_0(\bd-\bR+\bR')]^*_{\sigma'} [V_{0}(\bd')]^*_\sigma [V_0(\bd'-\bR+\bR')]_{\sigma'}.
\end{equation}

Let us now compute $\mathcal{C}^{(Q)}(\theta)$ in two cases: (I) for large angles with $\tan\theta \in 2\mathbb{N}$, and (II) for small angles with $\cot\theta \in 2\mathbb{N}$. For case (I), it is obvious from Fig. \ref{supp_fig:OAI}(a) that only the two diagonal bonds crossing in the center of the figure contribute. Bond $\{13\}$ corresponds to $\bR-\bR'=\hat{x}+\hat{y}$ and $\bd=\bd'=\hat{x}+\hat{y}$, and contributes $1/16$ to $\mathcal{C}^{(Q)}(\theta)$. By $C_4$, one deduces that bond $\{24\}$ also contributes $1/16$, so altogether $\mathcal{C}^{(Q)}=1/8$ in case (I). For case (II), it is obvious from Fig. \ref{supp_fig:OAI}(b) that three types of bonds contribute: the vertical ones like bond $\{14\}$ and bond $\{23\}$ (altogether $\cot\theta$ of these), the diagonal ones like bond $\{24\}$ (altogether $\cot\theta +1$ of these), and the diagonal ones like bond $\{13\}$ (altogether $\cot\theta -1$ of these). Here, bond $\{14\}$ (and its alike) contributes $1/8$ (as one can choose $\bd=\bd'=\hat{y}$ and $\bd=\bd'=\hat{x}+\hat{y}$), and the diagonal bonds again just contribute $1/16$. Altogether we conclude $\mathcal{C}^{(Q)} = \frac{1}{4}\cot\theta$.

\section{Lattice partition scheme}\label{supp_sec: partition_scheme}
\setcounter{equation}{0}
\setcounter{figure}{0}  
\begin{figure}[t!]
    \centering
    \includegraphics[width=\columnwidth]{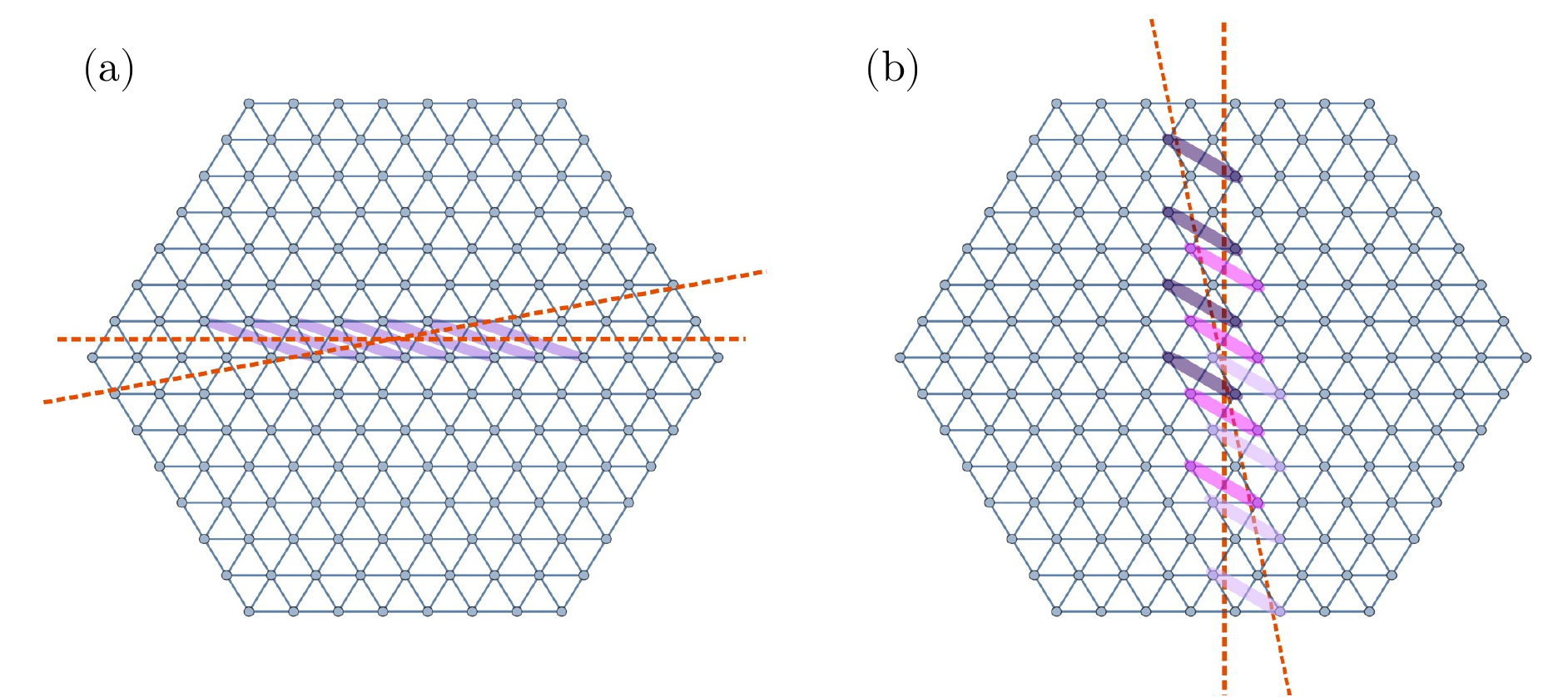}
    \caption{Partition scheme for the triangular lattice with $\tan\theta = \sqrt{3}/9$. The small angle limit of (a) and (b) extracts $ \widetilde{\mathcal{G}}_{yy}$ and $\widetilde{\mathcal{G}}_{xx}$, respectively.}
    \label{supp_fig:triangular}
\end{figure}
In this section, we elaborate on how the argument presented in the main text to establish Eq. \eqref{eq:key_result} can be applied to a generic Bravias lattice.  
%
%
We only provide a \textit{sufficient} scheme of partition, which is applicable to any Bravais lattice (with the set of angles depending on the microscopic lattice geometry).

\subsection{Triangular lattice: a two-orientation scheme}
 Here we first specify the partition scheme used for a triangular lattice, and later generalize to arbitrary oblique lattices. In particular, we explain why we have chosen the set of angles $\tan\theta=\infty,\sqrt{3}, \sqrt{3}/3, \sqrt{3}/9, \sqrt{3}/15$ in our numerics for the Haldane model presented in Fig. \ref{fig:numerics_Fluc}. In short, just like in the case of a square lattice, they are so chosen such that the partition boundary never intersect any lattice site and that an exact counting similar to Eq. \eqref{eq:exact_counting} can be attained.

To extract $\partial^2_y \widetilde{S}_\bq\vert_{\bq=0}$ and $\partial^2_x \widetilde{S}_\bq\vert_{\bq=0}$ (in anticipation of obtaining the metric trace),  we need to consider two partition orientations: (I) one kind is oriented such that one of the partition boundary is pointing along $\hat{x}$, and (II) another kind is oriented such that one of the partition boundary is pointing along $\hat{y}$. For (I)/(II), we first put the horizontal/vertical boundary in the middle of two central rows/columns, and then lay down the slanted boundary such that it intersects these two central rows/columns at the mid-point of some edges. The ``central" rows/columns are picked for convenience, so that after the partition we can specify four bulk subregions that are far away enough from the physical edge of the total system to suppress spurious edge contributions. Cases (I) and (II) are illustrated in Fig. \ref{supp_fig:triangular} (a) and (b), respectively, for $\tan\theta=\sqrt{3}/9$. One can appreciate that this is the exact same rationale we used to partition the square lattice in Fig. \ref{fig:setup}. For (I), it can be seen that the allowed $\theta$ satisfies $\tan\theta = \frac{\sqrt{3}}{2n+1}$ with $n\in \mathbb{Z}$, while for (II) we require $\tan\theta = \frac{\sqrt{3}}{3(2m+1)}$ with $m \in \mathbb{Z}$. The common solutions thus give $\tan\theta = \sqrt{3}/3, \sqrt{3}/9, \sqrt{3}/15$, etc. .  Notice also that whenever $\theta$ gives an unambiguous partition, $\pi/2-\theta$ also gives an unambiguous partition, thus we have also considered $\tan\theta=\infty, \sqrt{3}$ in our simulation.

Now let us perform the same counting argument as in the main text to evaluate Eq. \eqref{eq:simplified_CQ} in the small-angle limit (with large $B$ and $D$, and small and far-separated $A$ and $C$) for the triangular lattice given the aforementioned partition scheme.  For illustration, focus on Fig. \ref{supp_fig:triangular}(a) for case (I): given any fixed $\bR-\bR'$ on the triangular lattice, the number of such bonds obeying $\bR \in B$ and $\bR' \in D$ ($B,D$ are the large regions) are counted as
\begin{equation}
     [\frac{(\bR-\bR')_y}{a}\cot\theta - \frac{(\bR-\bR')_x}{a}]\cdot \frac{(\bR-\bR')_y}{a\sqrt{3}/2},
\end{equation}
which is the same as Eq. \eqref{eq:exact_counting} upon recognizing $a^2\sqrt{3}/4=\mathcal{A}_{\text{cell}}$. For case (II) in Fig. \ref{supp_fig:triangular}(b), one again obtain the same expression with $x\leftrightarrow y$, $a\sqrt{3}/2 \mapsto a/2$ and $a \mapsto a\sqrt{3}$, which is again Eq. \eqref{eq:exact_counting}. In Fig. \ref{supp_fig:partition}, we show some representative partition configurations that we used for producing Figs. \ref{fig:numerics_Fluc} and \ref{fig:numerics_EE} of the main text for the Haldane's model. Notice that the triangular lattice represents the unit cell positions of the honeycomb, and we are using the partition scheme stipulated below Eq. \eqref{eq: fluctuation_simplified}. 

The above two-orientation partition scheme works for extracting the trace of integrated metric, $\widetilde{\mathcal{G}} = \widetilde{\mathcal{G}}_{xx}+ \widetilde{\mathcal{G}}_{yy}$, as long as the Bravais lattice contains two orthogonal lattice vectors. This, however, is not true for a generic oblique lattice, which requires a three-orientation partition scheme as described next. 

\subsection{General oblique lattice: a three-orientation scheme}
Now consider a generic oblique Bravais lattice with primitive vectors $\ba_1$, $\ba_2$ and $\ba_3 = -\ba_1-\ba_2$. Along each crystal axis $\ba_i$, we extract the small-angle corner coefficient based on the counting described above, which gives us 
\begin{equation}
    \mathcal{C}^{(Q)}_i(\theta) = -\frac{\cot\theta}{2 \mathcal{A}_{\text{cell}}} \sum_{\Delta\bR} [\Delta\bR\cdot\hat{\bb}_i]^2 \cc{\rho(\Delta\bR)\rho(0)},
\end{equation}
where $\hat{\bb}_i\perp \ba_i$ is a unit vector. Denote the angle between $\hat{\bb}_i$ and $\hat{\bb}_j$ by $\phi_{ij}$. Without loss of generality, assume $\hat{\bb}_1 =\hat{x}$, then 
\begin{subequations}
\begin{align}
[\Delta\bR\cdot\hat{\bb}_1]^2 &= \Delta\bR_x^2\\
[\Delta\bR\cdot\hat{\bb}_2]^2 &=  \Delta\bR_x^2 \cos^2\phi_{12} +\Delta\bR_y^2\sin^2\phi_{12} +\Delta\bR_x\Delta\bR_y \sin{2\phi_{12}}\\
[\Delta\bR\cdot\hat{\bb}_3]^2 &= \Delta\bR_x^2 \cos^2\phi_{13} +\Delta\bR_y^2\sin^2\phi_{13} -\Delta\bR_x\Delta\bR_y \sin{2\phi_{13}}
\end{align}
\end{subequations}
It is straighforward to check that 
\begin{equation}\label{supp_eq:three-orientation_1}
    \Delta\bR^2 = \frac{\cos\phi_{23}}{\sin\phi_{12}\sin\phi_{13}} [\Delta\bR\cdot\hat{\bb}_1]^2 + (\text{cyclic permutations of 123}).
\end{equation}
The trace of the integrated metric can be extracted as $2\pi \widetilde{\mathcal{G}} = \lim_{\theta\rightarrow 0 } \overline{\gamma}^{(Q)}(\theta)$ with
\begin{equation}\label{supp_eq:three-orientation_2}
    \overline{\gamma}^{(Q)} = \frac{1}{2}\Big[\frac{\cos\phi_{23}}{\sin\phi_{12}\sin\phi_{13}} \gamma^{(Q)}_1 + (\text{cyclic permutations of 123})\Big].
\end{equation}

\section{Orbital Embedding: Physical vs Origin}\label{supp_sec:orbital_embedding}
\setcounter{equation}{0}
\setcounter{figure}{0}  
\begin{figure}[t]
    \centering
    \includegraphics[width=0.6\columnwidth]{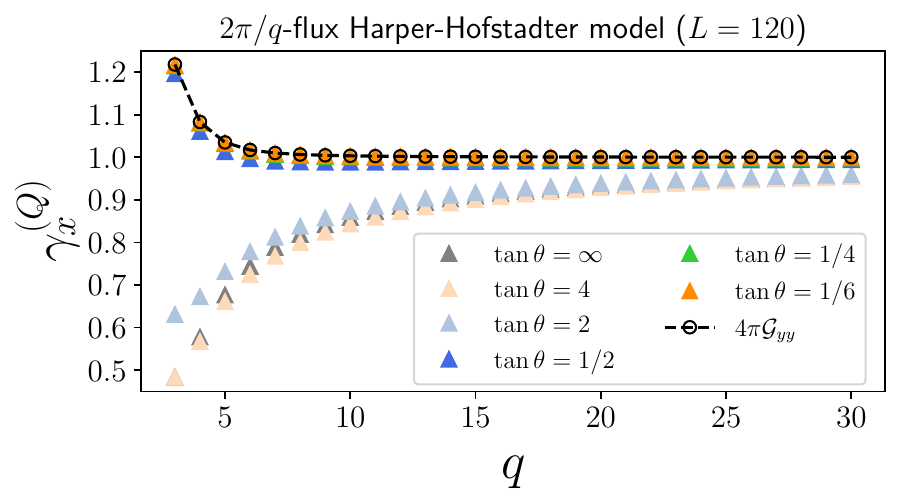}
    \caption{Corner coefficient $\gamma_x^{(Q)}$ in the Harper-Hofstadter model, and comparison with the integrated quantum metric $\mathcal{G}_{yy}$ of the physical orbital embedding.}
    \label{fig:Hofstadter}
\end{figure}

In this section we discuss how to extract the quantum metric $\mathcal{G}_{ii}$ in the physical orbital embedding even when it is different from the origin orbital embedding. The examples we focus on are the Harper-Hoftstadter model \cite{harper1955single, Hofstadter1976} and the Haldane honeycomb model \cite{Haldane1988}, and their Hamiltonians are provided in Sec. \ref{supp_sec:numerics}.

\subsection{Harper-Hofstadter model}
\begin{figure}[t]
    \centering
    \includegraphics[width=0.9\columnwidth]{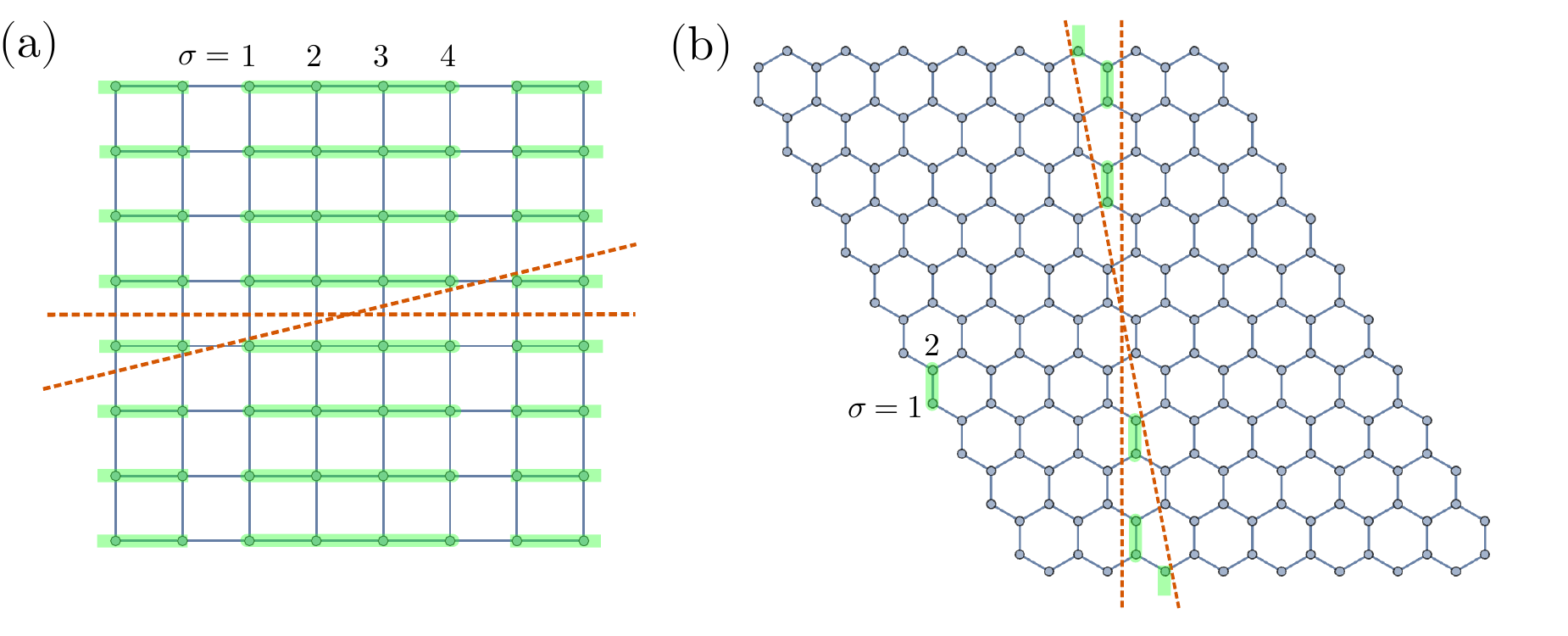}
    \caption{(a) Partition of the square lattice which supports the Harper-Hofstadter model, with $2\pi/q$-flux per plaquette. For $q=4$, the magnetic unit cell (colored in green) remain undivided by the partition scheme with a corner angle $\tan\theta=1/4$. (b) Partition of the honeycomb lattice which supports the Haldane model. The depicted partition corresponds to a corner with $\tan\theta = \sqrt{3}/9$, and importantly, the partition preserves the unit cell (colored in green) without dividing it. }
    \label{supp_fig:embedding}
\end{figure}
Here we study the Harper-Hofstadter (HH) model on the square lattice (one orbital per site) with $2\pi/q$-flux ($q \in \mathbb{Z}$) per plaquette. The ground state corresponds to occupying the lowest band. As $q \rightarrow \infty$, the lowest band gets flattened and effectively becomes the lowest LL with a uniform quantum geometry and $4\pi \mathcal{G}_{ii}=1$ \cite{peotta2015superfluidity,ozawa2021}. Note that the magnetic unit cell consists of $q$ sites. Let us use the partition exactly based on Fig. \ref{fig:setup}, instead of the stipulated partition scheme mentioned below Eq. \eqref{eq: fluctuation_simplified} (where our main results Eqs. \eqref{eq:key_result}, \eqref{eq: key result_band insulator} are based on). This partition is convenient to use as we do not need to keep track of the unit cell when doing the partition, the partition is just faithfully represented in Fig. \ref{fig:setup}. The corner charge fluctuation is compared with the integrated metric of the \textit{physical} orbital embedding, and an exceedingly nice match is obtained for small $\theta$, as shown in Fig. \ref{fig:Hofstadter}.

To explain this match, first notice for the specific case of $\tan\theta=1/q$ we can actually understand $\gamma^{(Q)}_{x} = 4\pi \mathcal{G}_{yy}$ based on Eq. \eqref{eq: key result_band insulator}. As depicted in Fig. \ref{supp_fig:embedding}(a), it clearly shows that the chosen magnetic unit cell containing $q$ orbitals are \textit{not} divided by the partition with corner angle $\theta = \arctan{q^{-1}}$. Notice that the sublattice position difference $(\br_\sigma-\br_{\sigma'}) \parallel \hat{x}$, hence the physical embedding projector $P_{\sigma,\sigma'}(\bk)$ and the origin orbital embedding projector $\widetilde{P}_{\sigma,\sigma'} (\bk) = e^{i\bk\cdot(\br_\sigma-\br_{\sigma'})} P_{\sigma,\sigma'} (\bk) $ differ only by a $k_y$-independent unitary transformation. Thus,
\begin{equation}
    \widetilde{g}_{yy} = \frac{1}{2}\tr[(\partial_y\widetilde{P}(\bk))^2] = \frac{1}{2}\tr[(\partial_y P(\bk))^2]= g_{yy}.
\end{equation}
As we have noted in the main text, this argument suffices to explain the match in Fig. \ref{fig:Hofstadter} for the specific cases with $\tan\theta=1/q$, but it is clear that the exceedingly nice match between $\gamma^{(Q)}_{x}$ and $4\pi \mathcal{G}_{yy}$ holds even more generally when the partition of square lattice can divide the magnetic unit cell. Below we explain this generic phenomenon by modifying the counting argument around Eq. \eqref{eq:exact_counting} to the current situation. 

For readers' convenience, let us recollect from the main text that the corner charge fluctuation can be expressed as 
\begin{equation}\label{supp_eq:compact_expression_corner_fluc}
    \mathcal{C}^{(Q)}(\theta)= \sum_{\sigma,\sigma'}\Big(\sum_{\substack{\bR_\sigma \in B \\ \bR'_{\sigma'}\in D}} + \sum_{\substack{\bR_\sigma \in A \\ \bR'_{\sigma'}\in C}}\Big) \mathcal{F}_{\sigma,\sigma'}(\bR_\sigma-\bR'_{\sigma'}),
\end{equation}
with
\begin{equation}
    \mathcal{F}_{\sigma,\sigma'}(\bR_\sigma-\bR'_{\sigma'}) = \mathcal{A}_{\text{cell}}^2 \int_{BZ} [d\bk] [d\bk'] e^{-i(\bk-\bk')\cdot(\bR_\sigma-\bR'_{\sigma'})} P_{\sigma',\sigma}(\bk) P_{\sigma,\sigma'}(\bk').
\end{equation}
To be generally consistent with the implementation of partition we used for the numerics, here we \textbf{do not} invoke the stipulation mentioned below Eq. \eqref{eq: fluctuation_simplified}. Whether $\bR_\sigma$ is within a subregion is solely determined by its physical position, making no reference to the unit cell position. Notice that translation symmetry in the Harper-Hofstadter model implies that $\mathcal{F}_{\sigma,\sigma'}(\bR_\sigma-\bR'_{\sigma'})$ is only explicitly dependent on the positional displacement $\bR_\sigma-\bR'_{\sigma'}$, but not on the sublattice indices, hence for the moment we can write $\mathcal{F}_{\sigma,\sigma'}(\bR_\sigma-\bR'_{\sigma'}) \equiv f(\br-\br')$. Let us also replace $\sum_{\sigma, \sigma'} \sum_{\bR_{\sigma}\in B, \bR'_{\sigma'}\in D}$ by $\sum_{\br\in B, \br'\in D}$, with $\br\;(\br')$ summed over \textbf{square lattice sites} in region $B$ ($D$). We remark that the above replacement cannot be generalized to an arbitrary multi-orbital model, which is why for an arbitrary model we need to stipulate a special kind of partition, as mentioned below Eq. \eqref{eq: fluctuation_simplified}, to arrive at a simple universal result. With our focus on the Harper-Hofstadter model, we realize that given a fixed $\br-\br'$, the number of terms that contribute to the first sum in Eq. \eqref{supp_eq:compact_expression_corner_fluc} is
\begin{equation}
        \frac{1}{\mathcal{A}_{\text{plaq.}}}\big[(\br-\br')_y \cot{\theta} - (\br-\br')_x\big] (\br-\br')_y, 
\end{equation}
where $\mathcal{A}_{\text{plaq.}}=\mathcal{A}_{\text{cell}}/q$ is the area of an elementary plaquette on the square lattice. As in the main text, we take the small-angle-limit ($\theta\rightarrow 0$), only retain the $\cot\theta$-term and neglect all of the rest, we obtain
\begin{equation}
\begin{split}
    \mathcal{C}^{(Q)}_{x} (\theta \rightarrow 0 ) &= \frac{\cot\theta}{2}\sum_{\br-\br'} \frac{(\br-\br')_y^2}{\mathcal{A}_{\text{plaq.}}} f(\br-\br')\\
    &=\frac{\cot\theta}{2} \mathcal{A}_{\text{cell}} \sum_{\sigma,\sigma'} \sum_{\substack{\bR_\sigma - \bR'_{\sigma'}}} (\bR_{\sigma}-\bR'_{\sigma'})_y^2\;\int_{BZ} [d\bk] [d\bk'] e^{-i(\bk-\bk')\cdot(\bR_\sigma-\bR'_{\sigma'})} P_{\sigma',\sigma}(\bk) P_{\sigma,\sigma'}(\bk')\\
    & = \cot{\theta} \int_{BZ}[d\bk] \frac{1}{2}\tr[(\partial_y P(\bk))^2].
\end{split}
\end{equation}
In the second equality, we have replaced $\sum_{\br-\br'}$ by $\frac{1}{q} \sum_{\sigma,\sigma'} \sum_{\bR_{\sigma}-\bR'_{\sigma'}}$. In the third equality, we have used $(\bR_\sigma-\bR'_{\sigma'})_y^2 e^{-i(\bk-\bk')\cdot(\bR_\sigma-\bR'_{\sigma'})} = \partial_y\partial_{y'} e^{-i(\bk-\bk')\cdot(\bR_\sigma-\bR'_{\sigma'})}$, and subsequently integrated by parts. Note that in the final expression we have the projector $P(\bk)$ for the physical orbital embedding. We have thus explained the general match between the corner coefficient $\gamma^{(Q)}_x$ and $4\pi\mathcal{G}_{yy}$ in Fig. \ref{fig:Hofstadter}.

\subsection{Haldane's honeycomb model}
\begin{figure}[t]
    \centering
    \includegraphics[width=\columnwidth]{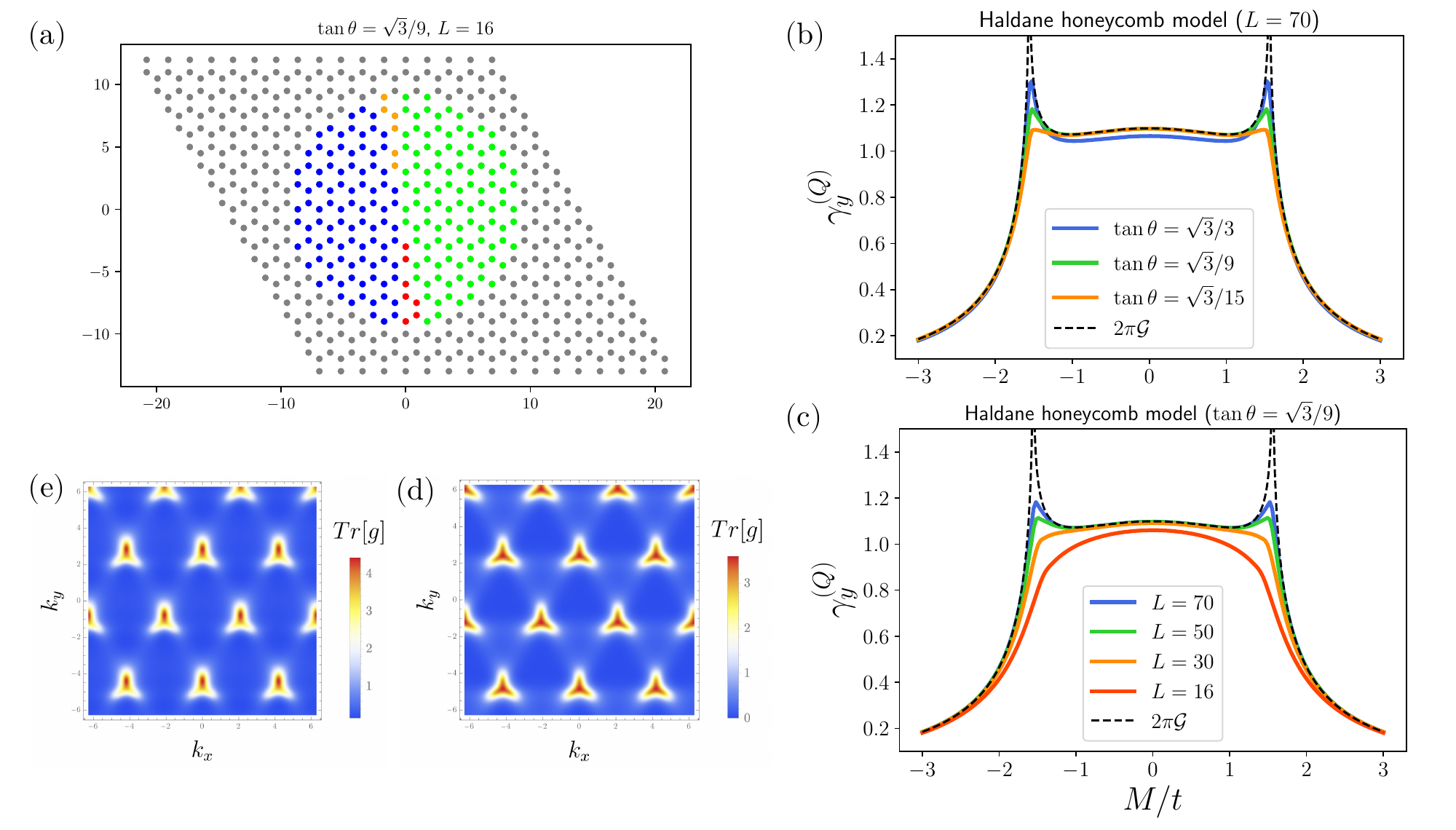}
    \caption{Extracting $\mathcal{G}$ of the Haldane honeycomb model ($t'=0.3t$, $\phi=\pi/2$). (a) shows an example of spatial partition used in our simulation to extract the corner charge fluctuation coefficient $\gamma^{(Q)}_y$. (b,c) show the comparison between $2\pi\mathcal{G}$ and $\gamma^{(Q)}_y$ for various partition angles and total system sizes. Notice that the trace of integrated metric $\mathcal{G}$ and $\widetilde{\mathcal{G}}$ are indeed different for the honeycomb lattice model, by comparing (b) here with Fig. \ref{fig:numerics_Fluc}(a) and notice the quantitative difference of the dashed lines. The momentum-space distribution of $\Tr[g]\equiv g_{xx}+g_{yy}$ (for $M=t$) is shown in (d) for the physical orbital embedding of the honeycomb lattice model, showing the presence of $C_3$ symmetry, and in (e) for the origin orbital embedding which lacks $C_3$.}
    \label{supp_fig:honeycomb}
\end{figure}
In the main text when we studied the Haldane model, we adopted the stipulated partition scheme mentioned below Eq. \eqref{eq: fluctuation_simplified}. From the corner charge fluctuation we were able to extract the trace of integrated quantum metric $\widetilde{\mathcal{G}}$ as shown in Fig. \ref{fig:numerics_Fluc}. Here we demonstrate how the metric  $\mathcal{G}$ of the physical embedding of the honeycomb model can be extracted. 

Our goal can be achieved by the kind of partition depicted in Fig. \ref{supp_fig:embedding}(b), which does not divide the unit cell (labeled in green). According to our key result in the main text, Eq. \eqref{eq:key_result}, the corner coefficient gives the integrated quantum metric evaluated with the origin orbital embedding. But notice, just like in the above analysis of the Harper-Hofstadter model, here $\widetilde{P}_{\sigma,\sigma'} (\bk) = e^{i\bk\cdot(\br_\sigma-\br_{\sigma'})} P_{\sigma,\sigma'} (\bk) $ differ from the physical embedding projector $P(\bk)$ only by a $k_x$-independent unitary transformation, as $(\br_2-\br_1)\parallel \hat{y}$. Consequently, with small $\theta$, we obtain $\gamma^{(Q)}_y  = 4\pi \widetilde{\mathcal{G}}_{xx} = 4\pi \mathcal{G}_{xx}$. To obtain the trace of integrated metric $\mathcal{G}=\mathcal{G}_{xx}+\mathcal{G}_{yy}$, we make use of the $C_3$ symmetry of the honeycomb Haldane model together with the three-orientation partition scheme based on Eqs. \eqref{supp_eq:three-orientation_1}. With $\phi_{12} = \phi_{23}=\phi_{13} = 2\pi/3$, we expect
\begin{equation}
    2\pi \mathcal{G} =\frac{4\pi ( \mathcal{G}_{xx}+ \mathcal{G}_{C_3x, C_3x}+\mathcal{G}_{C_3^2x, C_3^2x})}{3} = 4\pi\mathcal{G}_{xx} = \gamma^{(Q)}_y.
\end{equation}
This is confirmed in Fig. \ref{supp_fig:honeycomb}.

\section{Additional information for numerical studies}\label{supp_sec:numerics}
\setcounter{equation}{0}
\setcounter{figure}{0}  

\subsection{Correlation matrix method}
The central quantity we compute for a subsystem $A$ is its two-point correlation matrix $(C_A)_{ij}= \langle c^\dagger_i c_j\rangle$, where $i,j \in A$ label all the orbitals inside this subsystem. From this we calculate the bipartite particle-number fluctuation as
\begin{equation}\label{supp_eq:fluc}
    \braket{Q^2_A}_c = \sum_{i,j\in A} \braket{c^\dagger_i c_i c^\dagger_j c_j}_c  = \sum_{i\in A} \braket{c^\dagger_i c_i} - \sum_{i,j\in A} \braket{c^\dagger_i c_j}\braket{c^\dagger_j c_i} = \Tr[C_A-C_A^2],
\end{equation}
where $\Tr$ represents tracing over the orbitals in subsystem $A$. The subscript $c$ means connected correlation. More generally, 
\begin{equation}
    \langle Q_A Q_B \rangle_c  \equiv \langle Q_AQ_B \rangle - \langle Q_A \rangle \langle Q_B \rangle =\delta_{AB} \langle Q_A \rangle - \sum_{i\in A} \sum_{j \in B} \langle c^\dagger_{i} c_{j} \rangle \langle c^\dagger_{j} c_{i} \rangle.
\end{equation}

The correlation matrix also allows us to compute entanglement entropies (EEs) for free-fermion systems \cite{Peschel2001,peschel2003calculation,Cheong2004}. In this work we have focused on the von-Neumann EE $S^{(vN)}_A = -\Tr[\rho_A \log \rho_A]$, and the second R\'enyi EE $S^{(2)}_A = -\log\Tr[\rho^2_A]$. The key idea of the method is to express the reduced density matrix $\rho_A$ in an exponential form,
\begin{equation}
\rho_A = \frac{e^{-\mathcal{H}_A}}{Z_A}
\end{equation}
with $Z_A = \Tr[e^{-\mathcal{H}_A}]$, and the entanglement Hamiltonian $\mathcal{H}_A$ is chosen as a free-fermion operator
\begin{equation}\label{EntHam}
\mathcal{H}_A = \sum_{i,j \in A}(h_A)_{ij}c^\dagger_i c_j.
\end{equation}
As such, $n$-point correlation functions would factorize due to Wick's theorem, as appropriate for free-fermionic systems under our study. Matrices $h_{A}$ and $C_{A}$ are related as follows,
\begin{equation}
(C_A)_{ij}  = \Tr[\rho_A c^\dagger_i c_j] = \Big(\frac{1}{1+e^{h_A}}\Big)_{ji}, 
\end{equation}
which can be shown easily by first transforming to the basis that diagonalizes $h_A$. Next, we define a generating function
\begin{equation}
\begin{split}
Z_A(\beta) &\equiv \Tr[e^{-\beta \mathcal{H}_A}]\\
&= \det[1+(C_A^{-1}-1)^{-\beta}], 
\end{split}
\end{equation} 
which relates to the von Neumann EE by
\begin{equation}\label{supp_eq:vNEE}
S^{(vN)}_A = (1-\partial_\beta) \log Z_A(\beta) \rvert_{\beta=1} = -\Tr[C_A\log C_A+(1-C_A)\log (1-C_A)],
\end{equation}
and relates to the second R\'enyi EE by
\begin{equation}\label{supp_eq:2REE}
    S^{(2)}_A = -\log \big[\frac{Z_A(2)}{Z_A(1)^2}\big] = -\Tr \log [C_A^2 + (1-C_A)^2].
\end{equation}
\eqref{supp_eq:fluc},\eqref{supp_eq:vNEE} and \eqref{supp_eq:2REE} are the central equations used in our numerical calculation. 

\subsection{Details on lattice simulation}
\begin{figure}[h!]
    \centering
    \includegraphics[width=\columnwidth]{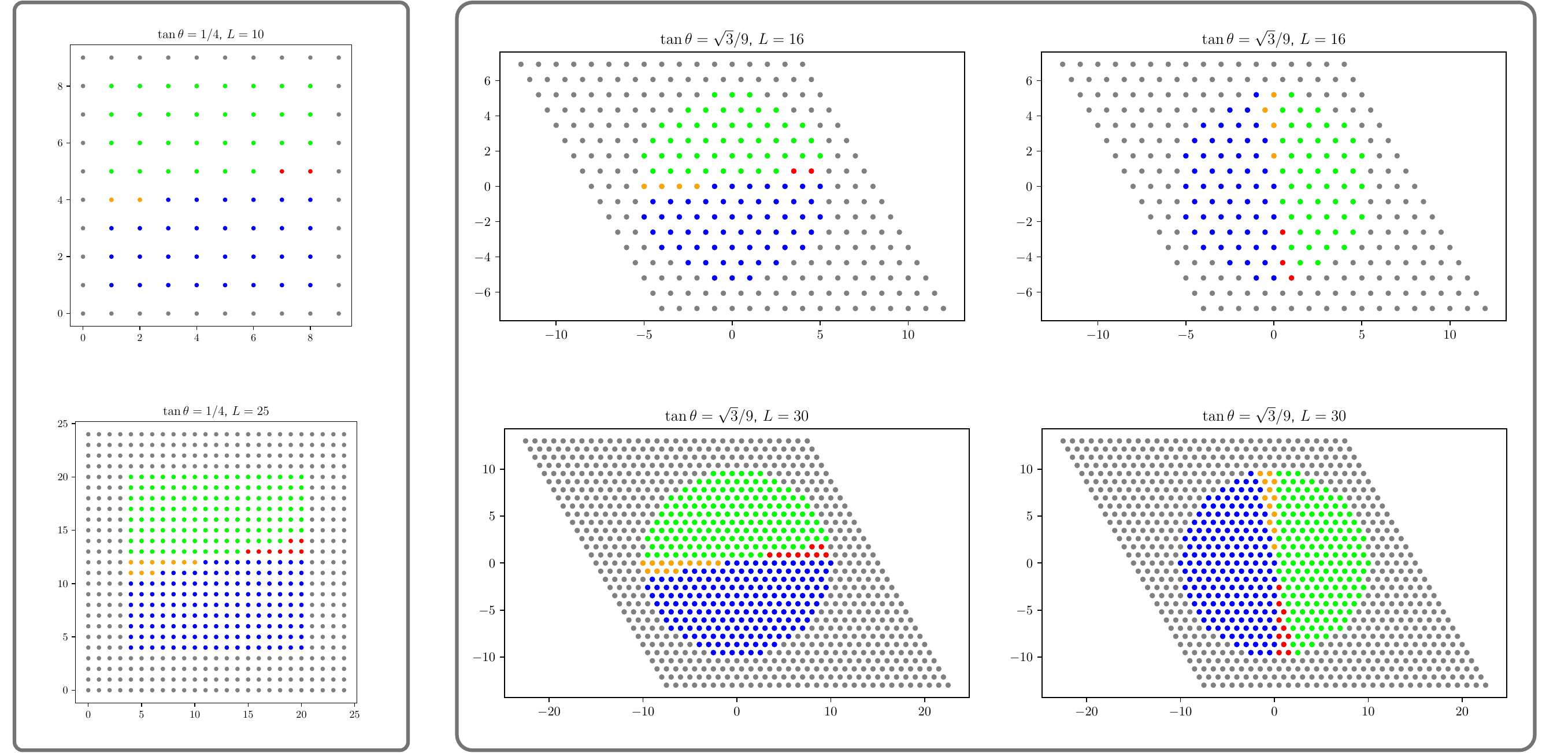}
    \caption{Depiction of representative real-space partitions used in our lattice simulation to produce Fig. \ref{fig:numerics_Fluc} and Fig. \ref{fig:numerics_EE} in the main text. Left panel for the square lattice, and right panel for the triangular lattice with both $x$- and $y$-partition shown. Sites colored in gray belong to region $E$, which are not used in the computation. The remaining four colored regions $A,B,C,D$ are used, with total linear size $2L/3$. This strategy allows us to suppress unwanted contribution from gapless boundary modes which exist in a topological phase. One-dimensional gapless modes generally contribute a logarithmic divergence (in the size of the boundary interval where it lives in), and cannot be properly canceled out in the combination in Eq. \eqref{eq: def_corner_fluc_lattice}. }
    \label{supp_fig:partition}
\end{figure}

For convenience of interested readers, here we specify explicitly the real-space lattice Hamiltonian and illustrate some representative real-space partition configurations we use for obtaining our numerical results shown in Fig. \ref{fig:numerics_Fluc} and Fig. \ref{fig:numerics_EE} of the main text. In this work we have studied three lattice models with open boundary conditions. For the Harper-Hofstadter (HH) model with $2\pi/q$-flux per plaquette \cite{harper1955single, Hofstadter1976}, we have
\begin{equation}
    H_{HH} = \sum_{\bR} \big( e^{i\frac{2\pi\bR_x}{q}}c^\dagger_{\bR+\hat{y}} c_\bR + c^\dagger_{\bR+\hat{x}} c_\bR \big) + \text{H.c.},
\end{equation}
where $c^\dagger_\bR$ is the fermionic creation operator at site $\bR$ on a square lattice. For the Qi-Wu-Zhang (QWZ) model \cite{QWZ2006} on a square lattice with two orbitals (labeled $1$ and $2$) per site, we have 
\begin{equation}
\begin{split}
    H_{QWZ} = \sum_\bR \Big\{ &-\frac{t_y}{2}(c^\dagger_{\bR+\hat{y},2}c_{\bR,1}-c^\dagger_{\bR+\hat{y},1}c_{\bR,2}+c^\dagger_{\bR+\hat{y},1}c_{\bR,1}-c^\dagger_{\bR+\hat{y},2}c_{\bR,2})\\
    & - \frac{t_x}{2} (c^\dagger_{\bR+\hat{x},1}c_{\bR,1}-c^\dagger_{\bR+\hat{x},2}c_{\bR,2} -i c^\dagger_{\bR+\hat{x},2}c_{\bR,1}-ic^\dagger_{\bR+\hat{x},1}c_{\bR,2})\\
    & +\frac{M}{2} (c^\dagger_{\bR,1}c_{\bR,1}-c^\dagger_{\bR,2}c_{\bR,2}) \Big\} + \text{H.c.}.
\end{split}
\end{equation}
We studied the anisotropic case with $t_x=2t_y=t$ in the main text. Lastly, we have the Haldane model \cite{Haldane1988} on the honeycomb lattice with two orbitals (labeled $1$ and $2$) per unit cell. 
%
Denoting the three $C_3$-related primitive vectors as $\ba_{i=1,2,3}$, we have
\begin{equation}
\begin{split}
    H_{H} = &\sum_{\bR} \Big\{\;t( c^\dagger_{\bR,2} c_{\bR,1}+c^\dagger_{\bR-\ba_3,2}c_{\bR,1}+c^\dagger_{\bR+\ba_2,2}c_{\bR,1} ) \\
    & + t'\Big[\big(e^{-i\phi}\sum_{i=1}^3 c^\dagger_{\bR+\ba_i,1}c_{\bR,1}\big) + \big(1\rightarrow 2, \; \phi\rightarrow -\phi\big)\Big]+\frac{M}{2}(c^\dagger_{\bR,1}c_{\bR,1}- c^\dagger_{\bR,2}c_{\bR,2})\;\Big\} + \text{H.c.}. 
\end{split}
\end{equation}
In this work, we have focused on $t'=0.3 t$ and $\phi=\pi/2$. 

Finally, we have shown in Fig. \ref{supp_fig:partition} some of the real-space partition configurations that we have used for the numerical simulation of corner fluctuation and corner entanglement entropies.

\end{document}